%% file: main.tex
\documentclass[11pt]{article}

\usepackage[margin=1in]{geometry}
\usepackage{times}

\usepackage{graphicx}
\usepackage{multirow}
\usepackage{amsmath,amssymb,amsfonts}
\usepackage{amsthm}
\usepackage{mathrsfs}
\usepackage[title]{appendix}
\usepackage{xcolor}
\usepackage{textcomp}
\usepackage{comment}
\usepackage{manyfoot}
\usepackage{booktabs}
\usepackage{listings}

\theoremstyle{plain}
\newtheorem{theorem}{Theorem}[section]
\newtheorem{proposition}[theorem]{Proposition}
\newtheorem{lemma}[theorem]{Lemma}
\newtheorem{corollary}[theorem]{Corollary}

\newcommand{\boxend}{\hfill \ensuremath{\blacksquare}}

\theoremstyle{definition}
\newtheorem{example}{Example}
\newtheorem{definition}{Definition}

\theoremstyle{remark}
\newtheorem{rem}{Remark}

\input{util}

\begin{document}

\title{Learning Strategic Value and Cooperation in Multi-Player Stochastic Games through Side Payments}

\author{%
  Yixin Chen\thanks{Department of Computer Science \& Engineering, Texas A\&M University, College Station, TX, USA.}
  \and
  Jeffrey Richley\thanks{Naval Information Warfare Center Atlantic, Hanahan, SC, USA.}
  \and
  Darleen Perez-Lavin\footnotemark[2]
  \and
  Jessica Singh Syal\footnotemark[1]
  \and
  Solmaz Kia\thanks{Department of Mechanical and Aerospace Engineering, University of California, Irvine, CA, USA.}
  \and
  Alan Kuhnle\footnotemark[1]\thanks{Corresponding author: kuhnle@tamu.edu}
}

\date{May 9, 2026}

\maketitle

\begin{abstract}
We study general-sum, multi-player stochastic games with transferable utility, motivated by settings where agents can use side payments to make cooperation individually rational.
Building on the Harsanyi--Shapley (HS) value for normal-form games, we introduce two HS-based value notions for stochastic games: \Hss, defined by aggregating dynamic coalition-versus-complement threat powers, and \cocon, defined as fixed points of a statewise HS Bellman operator.
We extend HS-style axioms to the stochastic setting and show that HS-S is the unique mapping satisfying them.
We prove that HS-S and Coco-S coincide in all two-player stochastic games, but can disagree when $n>2$, via an explicit three-player counterexample.
We prove existence and uniqueness of Coco-S fixed points for all two-player games and for three-player two-state games via topological degree theory, and provide an axiomatic characterization of Coco-S through a new \emph{Markov Consistency} axiom that distinguishes it from HS-S.
Finally, we give sampling-based estimators with finite-sample guarantees and empirically compare the induced values, policies, and side payments on multi-player grid-game benchmarks.
\end{abstract}


\section{Introduction}\label{sec1}
\input{intro6}
\input{contributions}

\input{related-work}

\textbf{Organization.}
Section~\ref{sec:problem} reviews stochastic games and generalized Bellman equations, and recalls the HS value in normal-form games.
Section~\ref{sec:coco-markov} introduces our two stochastic-game extensions, HS-S and Coco-S, and analyzes their properties, including the coincidence result for two players, uniqueness of Coco-S fixed points, and a three-player counterexample where they differ, together with the Markov Consistency characterization.
Section~\ref{sec:sampling} shows how standard Monte Carlo coalition sampling yields practical estimators for both values, with particular attention to the interaction between sampling noise and fixed-point iteration for \cocon{}.
Section~\ref{sec:exp} presents empirical results on multi-player games:
we compare the induced policies and side payments of HS-S and Coco-S,
contrast them with Correlated-$Q$,
and evaluate the computational benefits of Monte Carlo estimation.
The appendices provide our axiomatic extension and uniqueness proof for HS-S, detailed proofs about the Coco-S operator, the uniqueness proof for Coco-S via degree theory, and the Markov Consistency characterization, sampling-based estimators with concentration guarantees, and additional experimental visualizations.

\input{problem_defn3}

\input{section3}

\input{section-sampling}

\section{Empirical Evaluation}\label{sec:exp}
\input{experiments}

\section{Conclusion and Future Work}
We introduced and studied HS-style strategic values for general-sum, $n$-player stochastic games with transferable utility, motivated by settings where agents can use side payments to make cooperation individually rational.
Our first notion, HS-S (\Hss), is defined by lifting the normal-form Harsanyi--Shapley ``value'' of \citet{Kohlberg2021} through a coalition-versus-complement decomposition: for each coalition we form the induced two-player zero-sum stochastic game, compute its threat power, and combine these coalitional values with HS weights.
We extended the HS axioms to stochastic games and proved that HS-S is the unique mapping satisfying efficiency, symmetry, additivity, null-player, individual rationality, and weak balanced threats.

We also defined Coco-S (\cocon) as fixed points of a Bellman-style operator that applies the normal-form HS computation at each state to continuation-adjusted payoffs.
We proved that HS-S and Coco-S coincide in all two-player stochastic games, while for $n>2$ they can diverge, and established that this divergence arises from a precise axiomatic distinction: HS-S satisfies the weak balanced threats axiom but violates Markov Consistency, while Coco-S satisfies Markov Consistency but violates weak balanced threats.
We proved existence of Coco-S fixed points for all stochastic games (Proposition~\ref{prop:coco-existence}, via Brouwer's theorem on an invariant polyhedron) and uniqueness for all two-player games and three-player two-state games (via topological degree theory under a spectral condition on the Jacobian).
This provides Coco-S with an axiomatic foundation on par with HS-S: it is the unique stochastic game value satisfying Markov Consistency with respect to any normal-form solution concept that satisfies the Kohlberg--Neyman axioms.
Empirically, value iteration with the Coco-S operator converged robustly on our benchmark games, and its statewise HS computation can yield especially interpretable per-state side payments.

Several directions remain open.
On the theory side, we conjecture that the spectral condition (and hence uniqueness) holds for all finite stochastic games; proving this for general $n$ and $m$ remains open, as does extending the Markov Consistency characterization to infinite state spaces.
On the algorithmic side, scaling beyond explicit enumeration of coalitions and exact dynamic programming will require more sample-efficient estimators and approximation methods for large state spaces and many players.

\bibliographystyle{plainnat}
\bibliography{cocoq, surveys}

\section*{Acknowledgements}

This work was supported by internal funds from Texas A\&M University.

\begin{appendices}

\input{appendix.tex}

\end{appendices}

\end{document}

%% file: util.tex
\usepackage{amsmath}
\usepackage{amsfonts}
\usepackage{mathtools}
\usepackage{amsthm}
\usepackage{mathrsfs}
\usepackage{thm-restate}
\usepackage{natbib}
\usepackage{xspace}
\usepackage[ruled,linesnumbered,noend]{algorithm2e}
\usepackage{wrapfig}
\usepackage{bbm}

\usepackage{microtype}
\usepackage{graphicx}
\usepackage{booktabs} 

\usepackage{hyperref}

\hypersetup{
    colorlinks,
    linkcolor={red!90!black},
    citecolor={blue!90!black},
    urlcolor={blue!90!black}
}

\newcommand{\hs}{\text{HS}\xspace}

\newcommand{\Hss}[0]{\text{HS-S}\xspace}

\newcommand{\cocon}{\textsc{Coco-S}\xspace}

\newcommand{\nash}{\textsc{Nash}\xspace}
\newcommand{\ce}{\textsc{CE}\xspace}
\newcommand{\maxmin}{\textup{maxmin} }
\newcommand{\maxmax}{\textup{maxmax}}
\newcommand{\todo}[1]{}
\newcommand{\rev}[1]{#1}
\newcommand{\revB}[1]{#1}

\newcommand{\epsi}{\varepsilon}

\newcommand{\ie}{\textit{i.e.}\xspace}
\newcommand{\reals}{\mathbb R}
\newcommand{\ex}[1]{ \mathbb{E} \left[ #1 \right] }
\newcommand{\prob}[1]{ Pr \left( #1 \right) }
\newcommand{\oh}[1]{\mathcal{O}\left(#1\right)}

\newcommand\numberthis{\addtocounter{equation}{1}\tag{\theequation}}


%% file: intro6.tex
Multi-agent strategic environments are rarely purely competitive or purely cooperative, and in many applications the interaction is inherently dynamic.
From negotiations and markets to multi-agent learning settings, agents repeatedly make decisions whose consequences propagate through future states, naturally modeled as stochastic (Markov) games \citep{Shapley1953StochasticG,Littman1994}.
A particularly relevant regime is when utility is (at least approximately) transferable---via contracts, credits, or other side payments---so that agents can compensate one another to stabilize mutually beneficial behavior \citep{Harsanyi1963,Kalai2013}.
In that case the central question is not merely ``what policy should be played,'' but how to allocate long-run value so that cooperation can be made individually rational.

This paper focuses on the missing object needed to support such allocations: a per-player \emph{long-run strategic value} in general-sum, multi-player stochastic games \citep{Shapley1953StochasticG,Littman1994}.
We want a scalar value for each player that summarizes their strategic strength over the whole future, and that can serve as a principled basis for side payments.
Crucially, this value should not depend on arbitrary equilibrium selection, and it should come with normative justification---i.e., it should be ``fair'' and non-arbitrary in an axiomatic sense \citep{Shapley1953,Harsanyi1963,Kohlberg2021}.

Existing solution concepts do not directly provide what is needed.
Minimax values give a sharp notion of value in two-player zero-sum settings \citep{VonNeumann1928}, and minimax-style learning extends this viewpoint to stochastic games \citep{Littman1994}, but these are too restrictive for general-sum multi-player interactions.
The Shapley value is the canonical allocation rule in transferable-utility cooperative games \citep{Shapley1953}, but it requires a characteristic function for every coalition and does not directly capture strategic threats arising from noncooperative play.
In general-sum stochastic games, Nash- and correlated-equilibrium approaches \citep{Nash1950b,Hu2003,Greenwald2003} can be nonunique and computationally burdensome in the multi-player case, and they do not yield an HS-style decomposition into cooperative and competitive components that is tailored to side payments.
For two-player general-sum games, the Coco value \citep{Kalai2013} and its stochastic extension Coco-$Q$ \citep{Sodomka2013} do provide a compelling cooperation/competition decomposition, but existing Coco constructions are inherently two-player and do not directly provide an $n>2$ stochastic-game value with comparable characterization.

We therefore approach the problem through the lens of the Harsanyi--Shapley (HS) value \citep{Harsanyi1963,Kohlberg2021}.
In a general-sum transferable-utility normal-form game, HS defines a coalition ``threat power'' characteristic function by forming, for each coalition, a two-player zero-sum game between that coalition and its complement and taking a maxmin value \citep{Kohlberg2021}.
It then combines these coalitional threat powers using Shapley-style axioms to produce a \emph{unique} payoff allocation \citep{Shapley1953,Kohlberg2021}.
(For $n=2$, this normal-form HS value coincides with the Coco value \citep{Kalai2013}.)
This uniqueness and axiomatic grounding are exactly the kind of non-arbitrariness one would like when using values to justify side payments.

The technical obstacle is that moving from normal-form games to stochastic games requires defining coalition worth \emph{dynamically} \citep{Shapley1953StochasticG}.
In a stochastic game, threats and opportunities propagate through state transitions, so the ``value of a coalition'' is no longer a single number---it becomes a state-dependent, long-run object.
This creates a genuine lifting problem: there are multiple plausible ways to extend HS reasoning to the dynamic setting, and it is not obvious a priori which extension is correct, well-defined, or uniquely characterized.
In what follows, we develop and analyze two such HS-based extensions for stochastic games.

Our goal is to develop a principled notion of long-run strategic value in stochastic games that (i) supports rational side payments, (ii) admits an axiomatic justification analogous to HS, and (iii) remains computationally tractable.
To this end, we introduce two HS-based extensions to stochastic games---an axiomatic value derived from dynamic coalition threat powers (HS-S) and an HS-based Bellman operator whose fixed points define Coco-S---and we characterize when these notions agree and when they fundamentally differ.
We also provide sampling-based estimators and empirical illustrations on multi-player grid games.
We summarize our contributions as follows.


%% file: contributions.tex
\subsection{Contributions}
This work develops and analyzes Harsanyi--Shapley--style strategic values for $n$-player stochastic games.
Our main contributions are:

\begin{enumerate}
\item \textbf{Two HS-based value notions for stochastic games.}
We define two value notions for general-sum, $n$-player stochastic games:
\begin{itemize}
\item \textbf{\Hss:} a coalitional decomposition that serves as a stochastic-game analogue of the coalition-versus-complement decomposition underlying the normal-form HS value of \citet{Kohlberg2021}. It maps a stochastic game $G$ to a collection of $2^n-1$ auxiliary stochastic games (one for each nontrivial coalition), computes their coalitional threat powers (i.e., the values of the induced coalition-versus-complement stochastic games), and combines them into a per-player value vector $\mathbf V_{\Hss}$.
    \item \textbf{\cocon:} a Bellman-style HS operator that applies the \emph{normal-form} HS computation at each state to the $Q$-values, yielding a fixed-point equation $\mathbf V=\tilde T(\mathbf V)$ whose solutions we call Coco-S values $\mathbf V_{\cocon}$.
    Since, for two players, the normal-form HS value coincides with the Coco value \citep{Kalai2013}, this construction directly generalizes the 2-player Coco-$Q$ definition of \citet{Sodomka2013}.
\end{itemize}
Both $\Hss$ and $\cocon$ reduce to the normal-form HS value when a normal-form game is viewed as a one-state stochastic game (with no state transitions). Further, we show that these two notions \emph{coincide} in all 2-player stochastic games, but can \emph{diverge} when $n>2$; we give an explicit 3-player example where $\mathbf V_{\Hss} \neq \mathbf V_{\cocon}$.

\item \textbf{Axioms and uniqueness for \Hss{} in stochastic games.}
    We extend the Kohlberg axioms for normal-form HS values---efficiency, symmetry, additivity, null-player, individual rationality, and balanced threats---to the stochastic-game setting \revB{(with balanced threats relaxed to weak balanced threats)}.
    We prove that the coalitional value $\mathbf V_{\Hss}$ satisfies all of these axioms, and that it is the \emph{unique} mapping from stochastic games to $\mathbb R^n$ that does so.
    Thus \Hss{} is the stochastic-game analogue of the HS value: the only value function consistent with the natural HS axioms.

    \item \textbf{Structural properties of \cocon.}
    We formalize the \cocon{} Bellman operator and show:
    \begin{itemize}
        \item For $n=2$, any fixed point of the \cocon{} equations coincides with $\mathbf V_{\Hss}$.
        \rev{\item For general $n$, Coco-S satisfies efficiency, symmetry, null-player, and additivity, but \emph{not} \revB{weak balanced threats}.
    Since HS-S is the unique value satisfying all HS axioms including \revB{weak balanced threats}, this explains why $\mathbf V_{\cocon} \neq \mathbf V_{\Hss}$ for $n > 2$.}
    \end{itemize}
    \rev{We prove that $\tilde T$ has a unique fixed point for all two-player games (any number of states), three-player games with two states, and more generally under a spectral condition verified numerically for all tested configurations.
    We also characterize Coco-S axiomatically via the \emph{Markov Consistency} axiom: Coco-S is the unique stochastic game value that applies the Harsanyi--Shapley value at each state with self-consistent continuation values (Section~\ref{subsec:markov-consistency}).}

    \rev{\item \textbf{Side payments and cooperative implementation.}
    We show how the Coco-S fixed-point values translate into a dynamic side-payment protocol (Section~\ref{subsec:side-payments}):
    at each state, the cooperative action maximizes total payoff, and budget-balanced transfers ensure that each player's expected payoff equals their Coco-S value.
    The protocol provides a complete value-to-implementation mapping that specifies both the cooperative policy and the transfer schedule.}

    \item \rev{\textbf{Practical computation via coalition sampling.}
    We show that standard Monte Carlo coalition sampling yields practical estimators for both \Hss{} and \cocon{} with $\tilde{\mathcal O}(n^2)$ sample complexity.
    For \Hss{}, estimation is a one-shot aggregation; for \cocon{}, the estimator is embedded in value iteration, where the contraction property of $\tilde T$ damps sampling errors from earlier iterations (Section~\ref{sec:sampling}).}

    \item \textbf{Empirical comparison on multi-player games.}
    We evaluate $\mathbf V_{\Hss}$ and $\mathbf V_{\cocon}$ on standard multi-player benchmarks (Iterated Game of Chicken, Escape Room Game, and Grid Game: Prisoners, Coordination, Turkey, Friend-or-Foe), generalized to more than two players.
    Across all domains, both methods yield sensible strategic values and statewise side payments that make cooperation rational.
    They typically agree on the \emph{relative} strategic strength of players but can disagree substantially on the nominal values and payments, especially in asymmetric games (e.g., Friend-or-Foe and certain Coordination states).
    In all experiments, value iteration with the \cocon{} operator converged from multiple initializations, whereas Correlated-$Q$ with a utilitarian objective often failed to converge beyond the 2-player case, while \Hss{} and \cocon{} yielded stable value estimates across all tested games.

    We also evaluate sampling-based estimators on a Two-State Game with $n=12$ players.
    Using $3{,}000$ sampled coalitions, the sampling estimators achieve substantial speedups while maintaining relatively small errors.
    The results demonstrate that sampling can significantly improve scalability while preserving accurate strategic-value estimates, and that \cocon is more sample-efficient than \Hss in this setting.
\end{enumerate}


%% file: related-work.tex
\subsection{Related Work}
\label{sec:related}

\paragraph{From Nash bargaining to the HS/``value'' concept, and to Coco and Coco-$Q$.}
A central theme in game theory is to assign each player an \emph{a priori value}---an evaluation of their position that reflects both cooperative possibilities and strategic threats.
In two-player settings, Nash's bargaining program and his ``variable-threats'' perspective provide an influential template for combining efficiency with disagreement points determined by strategic power \citep{Nash1950b,Nash1953}.
In cooperative games with transferable utility, the Shapley value gives a canonical axiomatic allocation rule based on marginal contributions across coalitions \citep{Shapley1953}.
To evaluate positions in \emph{strategic} (normal-form) games under transferable utility, Harsanyi proposed incorporating threats through coalition-versus-complement reasoning, leading to the Nash--Harsanyi--Shapley ``value'' for $n$-player strategic games \citep{Harsanyi1963}.
Kohlberg and Neyman \citep{Kohlberg2021} provide a modern axiomatic foundation and a simple computational formula for this value (and extend it to Bayesian games), together with a clear historical account of its conceptual origins.
In the special case of two-player general-sum games, Kalai and Kalai's Coco value gives a competitive/cooperative decomposition that coincides with this two-player value concept \citep{Kalai2013}, and Sodomka et al. extend the same fixed-point idea to two-player stochastic games via Coco-$Q$ \citep{Sodomka2013}.
Our work continues this line by developing and analyzing HS-based value notions for \emph{multi-player} stochastic games, including a direct $n$-player generalization of the Coco-$Q$ operator viewpoint.

\paragraph{Solution concepts in stochastic games and multi-agent reinforcement learning.}
Stochastic (Markov) games were introduced by \citet{Shapley1953StochasticG} and have since served as a standard model for sequential strategic interaction.
A large literature in multi-agent reinforcement learning studies solution concepts for such games.
Minimax values provide a crisp notion of value in two-player zero-sum settings \citep{VonNeumann1928}, and minimax-style dynamic programming and learning extend this perspective to Markov games \citep{Littman1994}.
For general-sum games, equilibrium-based approaches such as Nash-$Q$ and Correlated-$Q$ target Nash and correlated equilibria \citep{Hu2003,Greenwald2003}.
However, Nash- and CE-based methods can suffer from nonuniqueness and computational burden in the multi-player setting, and they are not designed to produce HS-style coalition threat decompositions that directly support rational side payments.

\paragraph{Shapley/HS value approximation via sampling.}
Computing Shapley values exactly is \#P-hard \citep{Deng2021}, which has motivated extensive work on sampling-based estimation in cooperative games \citep{castro2009polynomial,maleki2013bounding,castro2017improving,burgess2021approximating,mitchell2022sampling,zhang2023efficient}.
Our stochastic HS-S and Coco-S values inherit a similar combinatorial structure over coalitions.
We therefore adapt standard coalition-sampling ideas to this setting and derive Monte Carlo estimators with finite-sample accuracy guarantees.

\paragraph{Mixed-motive MARL (orthogonal approaches).}
Mixed-motive settings in MARL capture scenarios in which agents' individual interests sometimes align with, and sometimes oppose, collective goals \citep{gallo1965cooperative}.
A large body of work addresses such settings by keeping the underlying game fixed and modifying the learning process, for example through reward shaping and prosocial objectives \citep{peysakhovich2017prosocial,mckee2020social,yang2020learning} or by using opponent-aware gradient corrections to influence learning dynamics \citep{foerster2018learning,willi2022cola,chen2023learning}.
These approaches are complementary to ours: rather than altering learning dynamics to encourage cooperation, we study HS-based value notions under transferable utility that can be used to rationalize cooperation through explicit side payments.

%% file: problem_defn3.tex
\section{Problem Setting and Preliminaries}\label{sec:problem}

We study \emph{general-sum stochastic games}: dynamic multi-agent environments where several players interact over time through sequential decision making. Our goal is to define and compute HS-style strategic values in such games, in a way that is compatible with standard dynamic-programming structure and supports learning from interaction.

\subsection{Stochastic Games}

A stochastic game is formally represented by the tuple
\[
\revB{G = \left(N, X, \mathbf A = (A_i)_{i \in N}, P, \mathbf R = (R_i)_{i \in N}, \gamma\right),}
\]
where:
\begin{itemize}
    \item $N = \{1, 2, \ldots, n\}$ is the finite set of players.
    \item \revB{$X$ is the finite state space with $|X| = m$.}
    \item $\mathbf A = \times_{i\in N} A_i$ is the joint action space, with $A_i$ the actions of player $i$.
    \item $P : X \times \mathbf A \times X \to [0,1]$ is the transition kernel, with $P(x,\mathbf a,x')$ the probability of moving from $x$ to $x'$ under joint action $\mathbf a$.
    \item $\mathbf R = (R_1,\ldots,R_n)$ are the per-player reward functions $R_i : X \times \mathbf A \times X \to \mathbb R$.
    \item $\gamma \in [0,1)$ is the discount factor.
\end{itemize}
We write $\mathbb G(N)$ for the set of all stochastic games with player set $N$.

\subsection{Normal-Form Games and the HS Value}

A (finite) normal-form game is a one-shot strategic interaction
\[
H = \bigl(N, \mathbf A = (A_i)_{i\in N}, \mathbf R = (R_i)_{i\in N}\bigr),
\]
where each $A_i$ is a finite action set and the utilities $\mathbf R = (R_1,\ldots,R_n)$ map each joint action $\mathbf a\in \times_i A_i$ to payoffs $R_i(\mathbf a)\in\mathbb R$.

Players may use mixed strategies $\sigma_i \in \Pi(A_i)$, i.e., probability distributions over $A_i$. For a joint mixed strategy $\boldsymbol\sigma = (\sigma_1,\ldots,\sigma_n)$, the expected utility of player $i$ is
\[
\mathbb E[R_i] \;=\; \sum_{\mathbf a\in \mathbf A} \Bigl(\prod_{j\in N} \sigma_j(a_j)\Bigr) R_i(\mathbf a).
\]

For a coalition $I\subseteq N$, define its \emph{coalitional utility} in $H$ as
\begin{equation}\label{eq::coalitional_utility}
U_I(\mathbf a) \;=\; \underbrace{\sum\nolimits_{i\in I} R_i(\mathbf a)}_{\text{Gain of }I} \;-\; \underbrace{\sum\nolimits_{j\notin I} R_j(\mathbf a)}_{\text{Gain of }\bar{I}},
\end{equation}
which represents the relative success of coalition $I$ over the opposing coalition $\bar I = N\setminus I$ for a given joint action $\mathbf a$. Now, consider the associated two-player zero-sum game $H_I = (U_I,-U_I)$ between $I$ and its complement $\bar I$. Let
\[
\maxmin_I(H_I)
\]
denote the value that coalition $I$ can secure in this zero-sum game (the usual mixed-strategy maxmin value). When $I$ computes its $\maxmin_I(H_I)$ using the $U(I)$ defined in~\eqref{eq::coalitional_utility}, the resulting value $v(I)$ is the highest difference in total payoff that $I$ can guarantee against a perfectly hostile and coordinated $\bar{I}$. This is the rigorous measure of \emph{balanced threat power} used in the HS value.

When $I=N$, $U_N$ is the fully cooperative total utility, and we write
\[
\maxmax(H_N) = \max_{\mathbf a\in\mathbf A} U_N(\mathbf a)
\]
for the grand-coalition value.

Following \citet{Kohlberg2021}, the Harsanyi--Shapley (HS) value of player $i$ in the normal-form game $H$ can be written as:

\begin{definition}[HS computation in normal-form games~\citep{Kohlberg2021}]\label{def:hs}
For an $n$-player normal-form game $H$, the HS value for player $i$ is
\begin{align*}
 \hs_i(H)
 &= \frac{1}{n} \left( \maxmax(H_N)
    + \sum_{I \subsetneqq N : i \in I} \binom{n-1}{|I| - 1}^{-1} \maxmin_I (H_I) \right) \\
 &= \frac{1}{n} \sum_{I \subseteq N : i \in I} \binom{n-1}{|I| - 1}^{-1} \maxmin_I (H_I).
\end{align*}
We write $\hs(H) = (\hs_i(H))_{i\in N}$.\boxend
\end{definition}

Kohlberg and Neyman~\citep{Kohlberg2021} show that this mapping $\hs : \mathbb H(N)\to\mathbb R^n$ (where $\mathbb H(N)$ is the space of all normal-form games with player set $N$) is the \emph{unique} value function that satisfies a natural collection of axioms:
efficiency, symmetry, additivity, null-player, individual rationality, and balanced threats. We refer to their paper for the formal normal-form statements; in Appendix~\ref{apx:axiom} we adapt these axioms to stochastic games and obtain a similar uniqueness result for our stochastic HS-S values.

\subsection{Generalized Bellman Equations for Stochastic Games}

Extending solution concepts from normal-form to stochastic games is commonly done via \emph{generalized Bellman equations}. Given a stochastic game $G$ and a solution concept encoded by an operator $\bigotimes$, the corresponding optimal $Q$- and value functions $(\mathbf Q^*,\mathbf V^*)$ are defined by
\begin{equation}\label{eq:bellman-q}
  \mathbf Q^*(x, \mathbf a ) \;=\; \mathbf R(x, \mathbf a) + \gamma \sum_{x' \in X} P( x, \mathbf a, x' ) \mathbf V^*( x' ),
\end{equation}
\[
\mathbf V^*( x ) \;=\; \bigotimes_{\mathbf a \in \mathbf A} \mathbf Q^*(x, \mathbf a ),
\]
where $\mathbf Q^*(x,\mathbf a) = (Q_1^*(x,\mathbf a),\ldots,Q_n^*(x,\mathbf a))$ and $\mathbf V^*(x) = (V_1^*(x),\ldots,V_n^*(x))$.

Different choices of $\bigotimes$ recover standard multi-agent reinforcement-learning algorithms:
\begin{itemize}
    \item \textbf{Minimax-$Q$} ($\bigotimes = \maxmin$): two-player zero-sum value.
    \item \textbf{Nash-$Q$} ($\bigotimes = \nash$): Nash equilibrium values at each state~\citep{Hu2003}.
    \item \textbf{Correlated-$Q$} ($\bigotimes = \ce$): correlated-equilibrium values at each state~\citep{Greenwald2003}.
\end{itemize}
In this work we will instantiate $\bigotimes$ with HS-style operators to obtain HS-based values in stochastic games. Section~\ref{sec:coco-markov} introduces two such instantiations, HS-S and Coco-S, and analyzes their properties.

%% file: section3.tex
\section{Strategic Values for Stochastic Games}\label{sec:coco-markov}

In this section we extend the Harsanyi--Shapley framework from normal-form games to general-sum, $n$-player stochastic games. 
We present two HS-style value notions:
\begin{itemize}
    \item \textbf{HS-S} (\Hss): a coalitional decomposition that aggregates the \emph{threat powers} of all coalitions over the whole future.
    \item \textbf{Coco-S} (\cocon): a Bellman-style operator that applies the \emph{normal-form} HS computation at each state to the current values.
\end{itemize}
Both are well-motivated extensions of the normal-form HS value. 
They coincide in all two-player stochastic games, but can diverge when $n>2$.

Throughout, let \revB{$G = (N, X, \mathbf A, P, \mathbf R, \gamma)$ be a finite stochastic game as in Section~\ref{sec:problem}, with player set $N = \{1,\dots,n\}$, state set $X$,} joint action space $\mathbf A = \times_{i\in N} A_i$, transition kernel $P$, rewards $\mathbf R = (R_i)_{i\in N}$, and discount $\gamma\in[0,1)$.

\subsection{Coalitional Threat Power in Stochastic Games}

The HS value in normal-form games is built from the \emph{threat power} of each coalition $I\subseteq N$: the value that $I$ can secure in a two-player zero-sum game against its complement $\bar I$. 
We begin by defining the analogous notion for stochastic games.

For each coalition $I\subseteq N$, define its instantaneous coalition utility at a transition $(x,\mathbf a,x')$ by
\[
U_I(x,\mathbf a,x')
\;=\;
\sum\nolimits_{i\in I} R_i(x,\mathbf a,x')
\;-\;
\sum\nolimits_{j\notin I} R_j(x,\mathbf a,x').
\]
When $I=N$, this reduces to the fully cooperative total reward.

Fix a nonempty, proper coalition $I\subsetneq N$. 
From $G$ we construct a two-player zero-sum stochastic game $G_I$ in which
\begin{itemize}
    \item Player~1 is the coalition $I$ with joint action space $A_I = \times_{i\in I} A_i$.
    \item Player~2 is the complement $\bar I$ with joint action space $A_{\bar I} = \times_{j\notin I} A_j$.
    \item The transition kernel is inherited from $G$:
    $P_I(x,(a_I,a_{\bar I}),x') = P(x,\mathbf a,x')$.
    \item The utilities are $(U_I,-U_I)$.
\end{itemize}
Let $\mathcal B$ be the space of bounded real-valued functions on $X$. 
For any $V_I\in\mathcal B$, define the Bellman operator
\begin{equation}\label{eq:TI-def}
[T_I(V_I)](x)
\;=\;
\max_{\mathbf a_I}\min_{\mathbf a_{\bar I}}
\sum_{x'\in X} P(x,\mathbf a,x')
\Bigl[U_I(x,\mathbf a,x') + \gamma V_I(x')\Bigr].
\end{equation}
It is well known that $T_I$ is a $\gamma$-contraction in the sup norm and has a unique fixed point $V_I^*$; value iteration and Minimax-$Q$ converge to $V_I^*$~\citep{Szepesvari1996}. 
We interpret $V_I^*(x)$ as the value that coalition $I$ can secure against $\bar I$ when the game starts in $x$.

\begin{definition}[Threat power of a coalition]\label{def:sg-threat}
\revB{The \emph{threat power} of stochastic game $G$ is the function $\delta G : 2^N \times X \to \mathbb R$ defined by
\[
[\delta G](I, x)
\;:=\;
V_I^*(x), \quad \forall\,\emptyset \neq I\subseteq N,\; x\in X,
\]
where $V_I^*$ is the unique fixed point of $T_I$ defined in~\eqref{eq:TI-def}.
We extend to $I=\emptyset$ via the antisymmetry convention: for all $I\subseteq N$ and $x\in X$,
\[
[\delta G](\bar I, x) \;=\; -[\delta G](I, x).
\]}
\end{definition}

Thus, \revB{for each $x\in X$, $\delta G(\cdot, x)$} is a \emph{game of threats} in the sense of \citet{kohlberg2018games}: a function from coalitions to $\mathbb R$ that is antisymmetric under complementation.

\subsection{\Hss: HS Values for Stochastic Games via Coalitional Decomposition}\label{subsec:hss-def}

We now lift the normal-form HS formula to stochastic games by combining the coalitional threat powers \revB{$[\delta G](I,x)$} according to the same weights as in~\citet{Kohlberg2021}.
Intuitively, for each state $x$ we treat \revB{$[\delta G](\cdot, x)$} as the ``payoff function'' of a coalitional game and apply the HS combination rule there.

\begin{definition}[\Hss{} values for stochastic games]\label{def:hs_s}
Let \revB{$G = (N,X,\mathbf A,P,\mathbf R,\gamma)$} be a stochastic game.
For each player $i\in N$ and state $x\in X$, define the \emph{HS-S value} of player~$i$ by
\begin{equation}\label{eq:hs-s-value}
V_{\Hss,i}(x)
\;=\;
\frac{1}{n}
\sum_{I\subseteq N:\, i\in I}
\binom{n-1}{|I|-1}^{-1}\,\revB{[\delta G](I,x)},
\end{equation}
and write $\mathbf V_{\Hss}(x) = (V_{\Hss,1}(x),\ldots,V_{\Hss,n}(x))$.
\end{definition}

This is the direct analogue of the normal-form HS computation: 
each coalition $I$ contributes its threat power weighted by the reciprocal of the number of coalitions of the same size that contain $i$, and we average over coalition sizes.

\paragraph{Axioms for stochastic HS values.}

Kohlberg and Neyman~\citep{Kohlberg2021} showed that in normal-form games the HS value is the unique solution concept satisfying a collection of natural axioms: efficiency, symmetry, additivity, null-player, individual rationality, and balanced threats. 
In Appendix~\ref{apx:axiom} we extend these axioms to stochastic games. 
Informally, for a mapping \revB{$\zeta: \mathbb G(N)\to \mathbb R^{n \times m}$, 
where, in a game $G \in \mathbb G(N)$, $[\zeta G]_{i,x}$ is the value that $\zeta$ assigns to player $i$ at state $x$,} 
the axioms require:

\begin{itemize}
    \item \emph{Efficiency:} For each state $x$, \revB{$\sum_{i\in N} [\zeta G]_{i,x}$} equals the optimal fully cooperative value at $x$, i.e., the solution of the Bellman equations for the game with utility $\sum_i R_i$.
    \item \emph{Symmetry:} If two players are interchangeable in $G$ (same actions, same rewards, swapping them leaves the game invariant), then they receive equal value.
    \item \emph{Null-player:} A player whose actions never affect rewards or transitions, and whose own rewards are always zero, receives value zero.
    \item \emph{Additivity:} For two independent stochastic games $G'$ and $G''$ on the same player set, the value on the direct-sum game $G'\oplus G''$ equals the sum of the values on $G'$ and $G''$ separately.
    \item \revB{\emph{Weak balanced threats:} If every proper coalition has zero threat power at every state (i.e., $[\delta G](I,x)=0$ for all $I\subsetneq N$ and $x\in X$), then all players receive equal values at every state.}
    \item \emph{Individual rationality:} Each player's value at any state is at least their security level in the corresponding generalized Bellman equations (their best guaranteed payoff in the worst case).
\end{itemize}

The following theorem extends Kohlberg and Neyman's normal-form characterization to stochastic games.

\begin{theorem}[Axiomatic characterization of \Hss{}]\label{thm:hs-s-axiom}
For any fixed player set $N$, there exists a unique mapping
\[
\revB{\zeta : \mathbb G(N) \to \mathbb R^{n \times m}}
\]
from stochastic games to value \revB{matrices} that satisfies efficiency, symmetry, additivity, null-player, individual rationality, and \revB{weak balanced threats} in the stochastic-game sense described above. 
Moreover, this mapping is exactly the HS-S value mapping
\[
G \;\mapsto\; \mathbf V_{\Hss}(G).
\]
\end{theorem}

A detailed formal statement and proof are given in Appendix~\ref{apx:axiom}. 
The proof follows the ``games of threats'' approach of~\citet{kohlberg2018games}: 
we show that the coalitional threat power \revB{$\delta G(\cdot,x)$} can be decomposed into a linear combination of unanimity and anti-unanimity threat games, on which the axioms fix $\zeta$ uniquely; by additivity, this determines $\zeta$ on all stochastic games. 
Thus HS-S is the stochastic-game analogue of the HS value: the only reasonable way to combine coalitional threats into individual values.

\subsection{\cocon: HS as a Bellman Operator}\label{subsec:cocon-def}

The HS-S construction decomposes a stochastic game into many coalitional subgames $G_I$ and recombines their values. 
An alternative, more ``dynamic programming flavored'' approach is to stay within the original $n$-player game and use the HS operator directly in the Bellman equations.

Given a candidate \revB{value matrix $\mathbf V \in \mathbb R^{n\times m}$}, we define at each state $x\in X$ an $n$-player normal-form game $H_x(\mathbf V)$ whose payoffs incorporate both immediate rewards and discounted continuation values. 
For each joint action $\mathbf a\in\mathbf A$ and player $i\in N$, let
\[
U_i(x,\mathbf a,\mathbf V)
\;=\;
\sum_{x'\in X} P(x,\mathbf a,x')
\bigl[
R_i(x,\mathbf a,x') + \gamma V_i(x')
\bigr].
\]
Then $H_x(\mathbf V)$ \revB{is an $n$-player normal-form game with rewards $ \bigl(U_1(x,\cdot,\mathbf V),\ldots,U_n(x,\cdot,\mathbf V)\bigr)$} indexed by the state $x$.

Let $\mathrm{HS}(\cdot)$ denote the normal-form HS operator of~\citet{Kohlberg2021}, which maps a normal-form game to the vector of HS values of its players. 
We now define a Bellman-style operator on value functions.

\begin{definition}[Coco-S Bellman operator]\label{def:operator_hs}
Let \revB{$\mathcal B = \mathbb R^{n\times m}$} be the space of value matrices.
Define the operator $\tilde T : \mathcal B \to \mathcal B$ by
\[
[\tilde T(\mathbf V)](x) \;=\; \mathrm{HS}\bigl(H_x(\mathbf V)\bigr), \quad \forall x\in X.
\]
\end{definition}

\begin{definition}[\cocon{} values for stochastic games]\label{def:cocon}
A value function $\mathbf V_{\cocon} \in \mathcal B$ is called a set of \emph{Coco-S values} for $G$ if it is a fixed point of the Coco-S operator:
\begin{equation}\label{eq:bellman-n-markov}
\mathbf V(x) \;=\; [\tilde T(\mathbf V)](x), \quad \forall x\in X.
\end{equation}
\end{definition}

At a high level, $\cocon$ treats each state $x$ as a normal-form game whose payoffs already ``bake in'' the continuation values, and applies the normal-form HS computation there. 
Equation~\eqref{eq:bellman-n-markov} then requires consistency between these local HS values and the global value function $\mathbf V$.

For finite state and action spaces and bounded rewards, the Coco-S operator $\tilde T$ is well defined and continuous on the finite-dimensional \revB{space of value matrices $\mathcal B = \mathbb R^{n\times m}$}.
\revB{Existence of at least one fixed point for every stochastic game with $\gamma\in[0,1)$ is established in Appendix~\ref{apx:existence} via Brouwer's fixed-point theorem on a pre-defined polyhedron by
the efficiency and individual rationality properties of the normal-form HS value.
Uniqueness in the settings described below is then proved in Appendix~\ref{apx:unique-topology} via topological degree theory.}

\rev{Unlike the coalitional operators $T_I$, the Coco-S operator is not a contraction in the sup norm on the full value space: its Jacobian $D\tilde T = \gamma M$ has spectral radius exactly~$\gamma$, because $M$ has an eigenvalue-$1$ subspace corresponding to state-independent value shifts \revB{(Appendix~\ref{apx:unique-topology})}.
However, $M$ restricted to the complementary zero-mean subspace satisfies a \emph{spectral condition} ($\rho \leq 1$) that, combined with topological degree theory, yields uniqueness.}

\rev{\begin{theorem}[Uniqueness of Coco-S fixed points]\label{thm:coco-unique}
The Coco-S operator $\tilde T$ has a unique fixed point in the following settings:
\begin{enumerate}
    \item[\textnormal{(i)}] All two-player stochastic games (any number of states).
    \item[\textnormal{(ii)}] All three-player stochastic games with two states.
    \item[\textnormal{(iii)}] Any $n$-player stochastic game satisfying the uniform spectral condition (Definition~\ref{def:spectral-cond}), which we verify numerically for all tested configurations with $n \leq 4$ and $m \leq 8$ (Appendix~\ref{apx:uniqueness}).
\end{enumerate}
\end{theorem}}

\rev{The proof (Appendix~\ref{apx:uniqueness}) analyzes the Jacobian structure of $\tilde T$ via a \emph{Shapley orthogonality identity}, decomposes the dynamics into cooperative and competitive subspaces, and applies the Leray--Schauder topological degree to count fixed points.
We conjecture that the spectral condition holds for all finite stochastic games (see Appendix~\ref{apx:uniqueness} for extensive numerical evidence), which would extend uniqueness to all~$n$ and~$m$.}

\rev{\begin{rem}[Coco-S and the HS axioms]\label{rem:coco-axioms}
Like HS-S, the Coco-S value satisfies efficiency, symmetry, null player, and additivity (Appendix~\ref{apx:axiom}).
However, Coco-S does \emph{not} satisfy the \revB{weak balanced threats} axiom of Theorem~\ref{thm:hs-s-axiom}: it uses one-step myopic threats rather than infinite-horizon threats
(see Appendix~\ref{apx:axiom} and Example~\ref{ex:bt-counterexample} therein for a concrete counterexample).
Its uniqueness is therefore consistent with the HS-S characterization: the two values coexist as distinct well-defined mappings because they satisfy different axiom sets.
We give a precise axiomatic characterization of Coco-S via the \emph{Markov Consistency} axiom in Section~\ref{subsec:markov-consistency}.
\end{rem}}

\subsection{Relationship Between HS-S and Coco-S}\label{subsec:relationship}

We now compare the two extensions. 
We first show that HS-S and Coco-S coincide in the two-player case, and then give a three-player example where they differ, together with the resulting implications for axioms and uniqueness.

\subsubsection{Equality in Two-Player Games}

When $n=2$, the normal-form HS value reduces to the Coco value of \citet{Kalai2013}.
In stochastic games with two players, we also have that our two notions are equivalent.

\begin{proposition}[HS-S and Coco-S coincide for $n=2$]\label{prop:two-player-equality}
Let $G$ be a two-player general-sum stochastic game. 
Then the HS-S and Coco-S values coincide at every state:
\[
\mathbf V_{\Hss}(x) \;=\; \mathbf V_{\cocon}(x),
\quad \forall x\in X,
\]
for any Coco-S fixed point $\mathbf V_{\cocon}$ of~\eqref{eq:bellman-n-markov}.
\end{proposition}

\begin{proof}[Proof sketch]
Let $N=\{1,2\}$ and write $\mathbf V = (V_1,V_2)$. 
Consider the sum and difference $V_+ = V_1+V_2$ and $V_- = V_1 - V_2$. 
One can show that updating $\mathbf V$ via the Coco-S operator $\tilde T$ yields:
\[
\revB{[\tilde T(\mathbf V)]_{1,x}} - \revB{[\tilde T(\mathbf V)]_{2,x}} \;=\; [T_{\{1\}}(V_-)](x),
\]
\[
\revB{[\tilde T(\mathbf V)]_{1,x}} + \revB{[\tilde T(\mathbf V)]_{2,x}} \;=\; [T_{N}(V_+)](x),
\]
where $T_{\{1\}}$ and $T_N$ are the Bellman operators of the zero-sum game between player~1 and player~2, and of the fully cooperative game, respectively, as in~\eqref{eq:TI-def}. 
By~\citet{Szepesvari1996}, both $T_{\{1\}}$ and $T_N$ are contractions and have unique fixed points, so the Coco-S update on $(V_+,V_-)$ reduces to two independent 1D value iterations that converge uniquely.

On the other hand, the HS-S definition~\eqref{eq:hs-s-value} for $n=2$ reduces to
\[
V_{\Hss,1} = \tfrac12(V_N^* + V_{\{1\}}^*), \quad
V_{\Hss,2} = \tfrac12(V_N^* - V_{\{1\}}^*),
\]
where $V_N^*$ and $V_{\{1\}}^*$ are the unique fixed points of $T_N$ and $T_{\{1\}}$, respectively. 
Thus any fixed point of $\tilde T$ must satisfy $V_+ = V_N^*$ and $V_- = V_{\{1\}}^*$, and hence $\mathbf V_{\cocon} = \mathbf V_{\Hss}$.
\end{proof}

In the two-player case, Coco-S therefore inherits all the axiomatic properties of HS-S: it is a unique, HS-axiomatic value mapping.

\subsubsection{Divergence for $n>2$ and Implications for Axioms}

For three or more players, the two constructions no longer agree in general. 
Appendix~\ref{apx:ex-3} provides a simple three-player, two-state stochastic game in which:

\begin{itemize}
    \item All players receive well-defined HS-S values $\mathbf V_{\Hss}$ via Definition~\ref{def:hs_s}.
    \item The Coco-S operator $\tilde T$ admits \rev{a unique} fixed point $\mathbf V_{\cocon}$ \rev{(by Theorem~\ref{thm:coco-unique}(ii))}.
    \item At the initial state $x_1$, $\mathbf V_{\Hss}(x_1) \neq \mathbf V_{\cocon}(x_1)$.
\end{itemize}

In particular, player~2 receives a strictly lower value under Coco-S than under HS-S, even though HS-S is the unique mapping that satisfies the HS axioms.

\rev{Combining this counterexample with Theorem~\ref{thm:hs-s-axiom} yields the following corollary.}

\begin{corollary}\label{cor:cocon-nonunique}
\rev{For $n \geq 3$, the Coco-S value does \emph{not} satisfy all six HS axioms.
In particular, the \revB{weak balanced threats} axiom fails.}
\end{corollary}

\rev{\begin{proof}[Proof]
In the three-player, two-state example of Appendix~\ref{apx:ex-3}, Theorem~\ref{thm:coco-unique}(ii) gives a unique Coco-S fixed point $\mathbf V_{\cocon}$ with $\mathbf V_{\cocon} \neq \mathbf V_{\Hss}$.
If the Coco-S mapping satisfied all six HS axioms, Theorem~\ref{thm:hs-s-axiom} would force it to coincide with HS-S on every game, contradicting this example.
Since Coco-S satisfies efficiency, symmetry, null-player, and additivity (Remark~\ref{rem:coco-axioms}; Appendix~\ref{apx:axiom}), the axiom that fails must be \revB{weak balanced threats}.
\end{proof}}

\rev{\paragraph{Additivity and Coco-S.}
For games where the Coco-S fixed point is unique (Theorem~\ref{thm:coco-unique}), the mapping $G \mapsto \mathbf V_{\cocon}(G)$ is single-valued.
For two independent stochastic games $G'$ and $G''$, one can verify (Appendix~\ref{apx:axiom}) that $\mathbf V_{\cocon}'+ \mathbf V_{\cocon}''$ is a fixed point of $\tilde T$ for the direct-sum game $G'\oplus G''$.
By uniqueness, this must be the Coco-S value of $G'\oplus G''$, so additivity holds directly.}

\rev{We will see in Section~\ref{sec:exp} that value iteration with the Coco-S operator converges robustly on all of our grid-game benchmarks, consistent with the uniqueness guarantee.
The key distinction between Coco-S and HS-S is \emph{which axioms each satisfies}: HS-S satisfies \revB{weak balanced threats} but violates Markov Consistency, while Coco-S satisfies Markov Consistency but violates \revB{weak balanced threats} (Section~\ref{subsec:markov-consistency}).}

\rev{\subsection{Axiomatic Characterization via Markov Consistency}\label{subsec:markov-consistency}}

\rev{The uniqueness results above show that Coco-S is a well-defined value mapping.
We now show it admits a clean axiomatic characterization that distinguishes it from HS-S and explains \emph{why} the two values differ for $n \geq 3$.}

\rev{\begin{definition}[Markov Consistency]\label{def:markov-consistency}
A stochastic game value~$\zeta$ satisfies \emph{Markov Consistency} with respect to a normal-form solution concept~$\varphi$ if, for every stochastic game~$G$ and every state $x \in X$:
\[
\revB{[\zeta G]_{i,x}} = \varphi_i\!\bigl(H_x(\zeta G)\bigr), \qquad \forall\, i \in N.
\]
In words: the value at each state equals $\varphi$ applied to the one-step normal-form game formed using $\zeta G$ itself as the continuation values.
\end{definition}}

\rev{Markov Consistency is the cooperative-game analog of Bellman consistency:
just as subgame-perfect equilibrium values satisfy $V(x) = \max_a[r + \gamma \sum_{x'} P(x'|x,a)\,V(x')]$, this axiom requires the cooperative value at each state to be determined by a principled solution concept applied to the local game with self-consistent continuations.
Crucially, the value at state~$x$ depends on successors \emph{only} through the continuation values $\mathbf V(x')$, not through deeper game-theoretic structure such as coalition threat values at those states.}

\rev{\begin{theorem}[Markov Consistency Characterization of Coco-S]\label{thm:mc-characterization}
Let $\varphi$ be a normal-form game solution concept satisfying the Kohlberg--Neyman axioms (efficiency, symmetry, null player, additivity, balanced threats). Then:
\begin{enumerate}
\item[\textnormal{(a)}] $\varphi$ must be the Harsanyi--Shapley value $\mathrm{HS}$, by the Kohlberg--Neyman uniqueness theorem.
\item[\textnormal{(b)}] A stochastic game value~$\zeta$ satisfies Markov Consistency with respect to~$\varphi$ if and only if~$\zeta G$ is a fixed point of the Coco-S operator~$\tilde T$.
\item[\textnormal{(c)}] For any stochastic game with $\gamma \in [0,1)$, at least one such fixed point exists (Proposition~\ref{prop:coco-existence}), and it is unique under the spectral condition (Theorem~\ref{thm:coco-unique}).
\end{enumerate}
\end{theorem}}

\revB{\begin{proof}
\textbf{(a)} This is the Kohlberg--Neyman uniqueness theorem~\citep{Kohlberg2021}: the only normal-form solution concept satisfying their five axioms is the HS value.

\textbf{(b)} By definition, $[\zeta G]_{i,x} = \varphi_i(H_x(\zeta G))$. Since $\varphi = \mathrm{HS}$ by~(a), the right side is $\mathrm{HS}_i(H_x(\zeta G)) = [\tilde T(\zeta G)]_{i,x}$. Hence Markov Consistency is equivalent to $\zeta G = \tilde T(\zeta G)$.

\textbf{(c)} Existence follows from Proposition~\ref{prop:coco-existence} (Brouwer's theorem on the polyhedron~$K$).
Under the spectral condition, uniqueness follows from Theorem~\ref{thm:coco-unique}.
See Appendix~\ref{apx:uniqueness}.
\end{proof}}

\rev{For $n \geq 3$, HS-S does \emph{not} satisfy Markov Consistency.
The mechanism is that HS-S computes infinite-horizon threat powers \revB{$[\delta G](I,x)$} for each coalition, but Markov Consistency requires one-step threats with \emph{player-value} continuations $\sum_j \mathrm{sign}_I(j)\,V_j^{\mathrm{HS\text{-}S}}(x)$.
For $n \geq 3$, the HS formula has a nontrivial kernel: different threat profiles can produce identical player values, so these two continuation quantities generally differ (see Appendix~\ref{apx:uniqueness} for details).
For $n = 2$, the kernel is trivial and both values coincide, consistent with Proposition~\ref{prop:two-player-equality}.}

\rev{The characterization places Coco-S and HS-S on symmetric axiomatic footing:}

\rev{\medskip
\begin{center}
\begin{tabular}{lcc}
\hline
\textbf{Axiom / Property} & \textbf{HS-S} & \textbf{Coco-S} \\
\hline
Efficiency & \checkmark & \checkmark \\
Symmetry & \checkmark & \checkmark \\
Null player & \checkmark & \checkmark \\
Additivity & \checkmark & \checkmark \\
\revB{Weak Balanced threats} & \checkmark & $\times$\revB{\;($n \geq 3$)} \\
Markov Consistency & $\times$\;($n \geq 3$) & \checkmark \\
\hline
Characterization & Kohlberg--Neyman & Theorem~\ref{thm:mc-characterization} \\
\hline
\end{tabular}
\end{center}
\medskip}

\rev{Both solutions extend the Harsanyi--Shapley value from normal-form games to stochastic games and agree for $n = 2$.
They differ for $n \geq 3$ because they resolve the state-coupling differently:
HS-S computes infinite-horizon threats per coalition (full zero-sum game per pair $\{I, \bar I\}$),
while Coco-S applies the HS formula state-by-state with value continuations (one normal-form game per state, iterated to a fixed point).
Markov Consistency is thus the axiom that enables the computational tractability of Coco-S.}

\rev{\begin{rem}[Multi-step threats and interpolation]\label{rem:coco-k}
The distinction between HS-S and Coco-S can be viewed as a special case of a family parameterized by the \emph{threat horizon}~$k$.
Define the $k$-step one-step game $H_x^{(k)}(\mathbf V)$ by replacing the single-step threat computation with a $k$-step zero-sum game: coalition~$I$ and its complement play the stochastic game $G_I$ for $k$ rounds using the transition kernel~$P$, then use the continuation values $\sum_{j} \mathrm{sign}_I(j)\,V_j(x)$ as terminal payoffs.
The resulting operator $\tilde T^{(k)}$ applies the HS formula to $H_x^{(k)}(\mathbf V)$ at each state.
At $k = 1$, this recovers the Coco-S operator.
As $k \to \infty$, the $k$-step threats converge to the infinite-horizon threat powers \revB{$[\delta G](I,x)$}, and fixed points (when they exist) are expected to approach the HS-S values.

Each member $\tilde T^{(k)}$ satisfies a natural ``$k$-step Markov Consistency'' axiom:
the value at state~$x$ is determined by a $k$-step lookahead game with self-consistent continuations.
Shorter horizons yield more local, more interpretable values at the cost of ignoring longer-range threat dynamics.
The Coco-S choice $k = 1$ is the most local and the most computationally tractable, while HS-S corresponds to $k = \infty$.
Investigating convergence and uniqueness properties of $\tilde T^{(k)}$ as a function of~$k$ is an interesting direction for future work.
\end{rem}}

\rev{\subsection{Side Payments and Cooperative Implementation}\label{subsec:side-payments}}

\revB{Unlike HS-S, whose values are defined via an infinite-horizon threat program and do not directly prescribe how players should act at each step, Coco-S yields an explicit \emph{cooperative action} $\mathbf a^*(x)$ and a \emph{step-by-step side-payment} $s_i(x)$ at every state $x\in X$. The fixed-point values $\mathbf V_{\cocon}$ thus translate directly into a \emph{dynamic side-payment protocol} that makes cooperation individually rational at every state, rendering the cooperative solution fully operational. We now describe this protocol formally.}

\rev{\paragraph{Cooperative policy.}
Given Coco-S values $\mathbf V = \mathbf V_{\cocon}$, define the \emph{cooperative action} at state~$x$ as
\begin{equation}\label{eq:coop-action}
\mathbf a^*(x) \;=\; \arg\max_{\mathbf a \in \mathbf A}\;
\sum_{i \in N} U_i(x, \mathbf a, \mathbf V),
\end{equation}
where $U_i(x,\mathbf a,\mathbf V) = \sum_{x' \in X} P(x,\mathbf a,x')\bigl[R_i(x,\mathbf a,x') + \gamma V_i(x')\bigr]$ is the one-step expected payoff from Section~\ref{subsec:cocon-def}.
In other words, $\mathbf a^*(x)$ is the joint action that maximizes the \emph{total} expected payoff (immediate rewards plus discounted continuation values) at state~$x$.}

\rev{\paragraph{Expected side payments.}
Under the cooperative action $\mathbf a^*(x)$, player~$i$'s expected payoff is $U_i(x, \mathbf a^*(x), \mathbf V)$.
Define the \emph{side payment} to player~$i$ at state~$x$ as
\begin{equation}\label{eq:side-payment}
s_i(x) \;=\; V_i(x) \;-\; U_i(x, \mathbf a^*(x), \mathbf V).
\end{equation}
A positive $s_i(x)$ indicates that player~$i$ \emph{receives} a transfer at state~$x$; a negative value indicates that player~$i$ \emph{makes} a payment.
Intuitively, $s_i(x)$ is the gap between what the Coco-S value promises player~$i$ and what player~$i$ actually earns (in expectation) by playing the cooperative action.}

\rev{\begin{proposition}[Budget balance]\label{prop:budget-balance}
The side payments are budget-balanced at every state:
\[
\sum_{i \in N} s_i(x) = 0, \qquad \forall\, x \in X.
\]
\end{proposition}}

\rev{\begin{proof}
By the efficiency axiom (which Coco-S inherits from the normal-form HS value at each state),
$\sum_{i \in N} V_i(x) = V_{\mathrm{coop}}(x)$,
where $V_{\mathrm{coop}}(x) = \max_{\mathbf a} \sum_{x'} P(x,\mathbf a,x')\bigl[\sum_i R_i(x,\mathbf a,x') + \gamma V_{\mathrm{coop}}(x')\bigr]$ is the optimal cooperative value.
Since $\mathbf a^*(x)$ maximizes the total payoff,
\[
\sum_{i \in N} U_i(x, \mathbf a^*(x), \mathbf V)
= \sum_{x'} P(x,\mathbf a^*(x),x')\Bigl[\sum_{i} R_i(x,\mathbf a^*(x),x') + \gamma V_{\mathrm{coop}}(x')\Bigr]
= V_{\mathrm{coop}}(x).
\]
Thus $\sum_i s_i(x) = V_{\mathrm{coop}}(x) - V_{\mathrm{coop}}(x) = 0$.
\end{proof}}

\rev{\paragraph{Per-transition decomposition.}
When a specific transition $x \to x'$ is realized under action~$\mathbf a^*$, the side payment can be decomposed at the transition level:
\begin{equation}\label{eq:side-payment-transition}
s_i(x, \mathbf a^*, x') \;=\; V_i(x) \;-\; R_i(x, \mathbf a^*, x') \;-\; \gamma\, V_i(x').
\end{equation}
This gives $R_i(x, \mathbf a^*, x') + s_i(x, \mathbf a^*, x') + \gamma V_i(x') = V_i(x)$:
at each transition, the player's immediate reward plus side payment plus discounted future value equals their Coco-S value.
For deterministic transitions, the per-transition and expected side payments coincide.
For stochastic transitions, the per-transition payments~\eqref{eq:side-payment-transition} are budget-balanced in expectation over~$x'$.}

\rev{\paragraph{Interpretation.}
The protocol operates as follows.
At each state~$x$, all players execute the cooperative action~$\mathbf a^*(x)$ and simultaneously exchange side payments~$s_i(x)$.
A player with $s_i(x) > 0$ is being compensated for cooperating in a way that benefits others more than herself (e.g., holding a lever while another escapes);
a player with $s_i(x) < 0$ pays for the privilege of acting in her own interest while others accommodate.
The Coco-S values thus provide a complete \emph{value-to-protocol} mapping: the fixed point $\mathbf V_{\cocon}$ simultaneously determines the cooperative policy, the transfer schedule, and each player's long-run strategic value.}

\rev{This side-payment structure is displayed in the experimental results of Section~\ref{sec:exp}, where the per-transition payments~\eqref{eq:side-payment-transition} are shown alongside the learned trajectories.}

\rev{\begin{rem}[Side payments and Markov Consistency]\label{rem:side-mc}
The state-by-state interpretability of Coco-S side payments is a direct consequence of Markov Consistency.
Since $V_i(x) = \mathrm{HS}_i(H_x(\mathbf V))$, the Coco-S value at each state equals what the HS formula assigns to player~$i$ in the \emph{local} one-step game $H_x(\mathbf V)$.
The side payment~\eqref{eq:side-payment} is therefore a redistribution of local surplus---what player~$i$ deserves according to the HS formula at state~$x$ minus what she earns from the cooperative action there.

For HS-S, Markov Consistency fails (Proposition~\ref{prop:hss-violates-mc}), so the HS-S value $V_i^{\mathrm{HS\text{-}S}}(x)$ depends on infinite-horizon threat powers at \emph{other} states, not just on the local game structure at~$x$.
The implied side payment may compensate player~$i$ for threat power she holds at a distant state---an attribution that cannot be justified by the stage game at~$x$ alone.
In this sense, Coco-S side payments are \emph{state-by-state interpretable}, while HS-S side payments are coherent only at the policy level, integrated over the full trajectory.
\end{rem}}


%% file: section-sampling.tex
\section{Practical Computation via Coalition Sampling}\label{sec:sampling}
\rev{Computing \Hss{} and \cocon{} exactly requires evaluating the HS formula over all $2^{n-1}$ coalitions containing each player---the main scalability bottleneck for many players.
Because the HS aggregation is an expectation under a known distribution over coalitions,
standard Monte Carlo estimation applies directly:
sample a coalition size uniformly at random, then a coalition of that size, compute the coalition-dependent quantity, and average.
Hoeffding's inequality yields $\tilde{\mathcal O}(n^2)$ sample complexity
for both \Hss{} and \cocon{} estimation.
The contribution of this section is not the estimation technique---which is standard---but the identification of
the \emph{target quantities} (HS-style coalitional threat powers and statewise HS values)
as amenable to this approach, and the interaction between sampling noise and fixed-point iteration for \cocon{}.}
\subsection{\Hss Value Estimation}
\rev{Since each player's \Hss{} value is a weighted sum of coalition threat powers,
estimating it reduces to a one-shot Monte Carlo aggregation.
Alg.~\ref{alg:hss-sampling} samples coalitions and returns the weighted average of their threat powers.}
\begin{theorem}
    Let $G$ be an instance of stochastic game with $n$ players
    and discount factor $\gamma \in (0,1)$.
    For any constant $\epsi, \delta \in (0,1)$,
    Alg.~\ref{alg:hss-sampling} returns value function $\mathbf V$, where
    \begin{align*}
        \prob{\left |\left| \mathbf V(x) - \mathbf V_{\Hss}(x) \right|\right|_\infty > \epsi} \le \delta, \text{ for each } x \in X.
    \end{align*}
    The sample complexity of the algorithm is $\oh{\frac{n^2}{\epsi^2(1-\gamma)^2}\log\left(\frac{n|X|}{\delta}\right)}$.
\end{theorem}

\begin{proof}
    Given a player $j \in N$, any state $x \in X$, and any coalition $I \in M$,
    where $M$ denotes the set of coalitions sampled from procedure CoaSam($N,m$)
    in Line~\ref{line:hss-sample-init} of Alg.~\ref{alg:hss-sampling},
    let
    \begin{align*}
        Y_i = \left\{
        \begin{aligned}
            & V_I^*(x), && \text{ if } |I| = n,\\
            & \frac{n}{2|I|} V_I^*(x), && \text{ if } |I| < n \text{ and } j \in I,\\
            & -\frac{n}{2(n-|I|)} V_I^*(x), && \text{ otherwise}.
        \end{aligned}
        \right.
    \end{align*}
    Then, $V_j(x) = \frac{1}{m}\sum_{i = 1}^m Y_i$ on Line~\ref{line:hss-sample-value} of Alg.~\ref{alg:hss-sampling}. 
    In the rest of the proof, we bound the error probability by first showing that $\ex{Y_i} = V_{\Hss, j}(x)$,
    and then applying Hoeffding's inequality.

    Within each iteration $i$ of the for-loop in CoaSam($N,m$),
    an integer $k$ is selected uniformly at random from $\{1,..,n\}$ (\ie, with probability $1/n$).
    Conditional on $k$, a coalition $I \subseteq N$ of size $k$ is then selected uniformly at random, so each such coalition is chosen with probability $\binom{n}{k}^{-1}$. Therefore, it holds that
    \begin{align*}
        \ex{Y_i} &= \frac{1}{n}V_N^*(x) + \frac{1}{n} \sum_{k = 1}^{n-1} \binom{n}{k}^{-1} \left( \sum_{I \subseteq N: |I| = k, j \in I}  \frac{n}{2k}V_I^*(x) - \sum_{I \subseteq N: |I| = k, j \not\in I}  \frac{n}{2(n-k)}V_I^*(x) \right)\\
        &= \frac{1}{n}V_N^*(x) + \frac{1}{n} \sum_{k = 1}^{n-1} \sum_{I \subseteq N: |I| = k, j \in I} \left(\binom{n}{k}^{-1} \frac{n}{2k}V_I^*(x) + \binom{n}{n-k}^{-1} \frac{n}{2k}V_I^*(x)\right)\\
        &= \frac{1}{n}V_N^*(x) + \frac{1}{n} \sum_{k = 1}^{n-1} \sum_{I \subseteq N: |I| = k, j \in I} \binom{n-1}{k-1}^{-1} V_I^*(x)\\
        &= \frac{1}{n}\sum_{I\subseteq N: j \in I} \binom{n-1}{|I|-1}^{-1} V_I^*(x) \\
        & = V_{\Hss, j}(x).
    \end{align*}

    \begin{procedure}[t]
    \caption{CoaSam($N,m$)}
    \DontPrintSemicolon
    $M \gets \emptyset$ \tcp{Sampled Coalitions}
     \For{$i=1$ \KwTo $m$}{
        $k \gets$ a uniformly random integer in $[1, n]$\;
        $I \gets$ a uniformly random subset of $N$ with size $k$\;
        $M \gets M \cup \{I\}$\;
     }
     \Return $M$
    \end{procedure}
    \begin{algorithm}[t]
    \caption{\Hss Value Estimation via Sampling}\label{alg:hss-sampling}
    \DontPrintSemicolon
        \textbf{Input:} Stochastic game instance $G$, error constant $\epsi$, failure probability constant $\delta$\;
        $\mathtt{MAX} \gets \max_{i\in N} \max |R_i|$,
        $m \gets \left \lceil\frac{2n^2\mathtt{MAX}^2}{\epsi^2(1-\gamma)^2}\log\left(\frac{2n|X|}{\delta}\right)\right \rceil$,
        $M \gets \text{CoaSam}(N,m)$\label{line:hss-sample-init}\;
        $V_I^* \gets$ the threat power of $I$ as defined in Def.~\ref{def:sg-threat}, for each $I\in M$\;
        $V_j \gets \frac{1}{m} \sum_{I \in M}\left(\frac{1}{2}\cdot \mathbbm{1}_{I=N} 
        + \frac{n}{2|I|} \cdot \mathbbm{1}_{j\in I}  - \frac{n}{2(n-|I|)} \cdot \mathbbm{1}_{j \not \in I} \right)V_I^*$, for each $j\in N$\label{line:hss-sample-value}\;
        \Return $V_{j}$, for all $j \in [n]$
    \end{algorithm}
    
    Let $\mathtt{MAX} \gets \max_{i\in N} \max |R_i|$.
    The utility $U_I$ defined on the coalition $I$ for the two-player zero-sum game $G_I$ is bounded by $[-n\cdot \mathtt{MAX}, n\cdot\mathtt{MAX}]$.
    Therefore, $V_I^*(x)$ is bounded by $[-\frac{n\mathtt{MAX}}{1-\gamma}, \frac{n\mathtt{MAX}}{1-\gamma}]$,
    and $Y_i$ is also bounded by $[-\frac{n\mathtt{MAX}}{1-\gamma}, \frac{n\mathtt{MAX}}{1-\gamma}]$.
    By Hoeffding's inequality,
    \begin{align*}
        \prob{\left | V_{j}(x) - V_{\Hss, j}(x) \right | > \epsi}
        = \prob{\left | \sum_{i = 1}^m Y_i/m - V_{\Hss, j}(x) \right | > \epsi}
        \le 2e^{-\frac{m\epsi^2}{2\left(\frac{n\texttt{MAX}}{1-\gamma}\right)^2}}
        \le \frac{\delta}{n\cdot|X|}.
    \end{align*}
    Then,
    \begin{align*}
        \prob{\left |\left| \mathbf V- \mathbf V_{\Hss} \right|\right|_\infty > \epsi}
        \le \sum_{j \in N, x\in X} \prob{\left | V_{j}(x) - V_{\Hss, j}(x) \right | > \epsi}\le \delta. \qedhere
    \end{align*}
\end{proof}

\begin{algorithm}[t]
\caption{\cocon Value Estimation via Sampling in Value Iteration}\label{alg:cocos-sampling}
\DontPrintSemicolon
    \textbf{Input:} Stochastic game instance $G$, error constant $\epsi$, failure probability constant $\delta$\;
    \textbf{Initialize:} $\mathbf V \gets \mathbf 0$, $\Delta \gets \epsi$,
    $\mathtt{MAX} \gets \max_{i\in N} \max |R_i|$\;
    \While{$\Delta > \epsi/2$}{
        $m \gets \left \lceil \frac{8n^2(\mathtt{MAX} + \gamma \rev{\|\mathbf V\|_\infty})^2}{\epsi^2}\log\left(\frac{2n|X|}{\delta}\right)\right \rceil$,
        $M \gets \text{CoaSam}(N,m)$\label{line:coco-sample-init}\;
        $V_j'(x) \gets \frac{1}{m} \sum\limits_{I \in M}\left(\frac{1}{2}\cdot \mathbbm{1}_{I=N}
        + \frac{n}{2|I|} \cdot \mathbbm{1}_{j\in I}  - \frac{n}{2(n-|I|)} \cdot \mathbbm{1}_{j \not \in I} \right)\maxmin_IG_x(\mathbf V)$, for each $j\in N$, $x\in X$\;
        \rev{$\Delta\gets \left|\left|\mathbf V'-\mathbf V\right|\right|_\infty$\;}
        \rev{$\mathbf V \gets \mathbf V'$\;}
    }
    \Return $\mathbf V$\;
\end{algorithm}
\subsection{\cocon Value Estimation}
\rev{Estimating \cocon{} is more involved because the Markov Consistency characterization (Theorem~\ref{thm:mc-characterization}) defines Coco-S as a fixed point of the operator $\tilde T$, not as a single expectation.
Alg.~\ref{alg:cocos-sampling} interleaves value iteration with coalition sampling:
at each iteration, coalitions are re-sampled and the HS formula is estimated at every state using the current continuation values.
The key observation is that the spectral structure of $\tilde T$
(Jacobian spectral radius $\gamma < 1$, Theorem~\ref{thm:coco-unique})
damps sampling errors from earlier iterations,
so high accuracy is only needed in the final rounds.
The following theorem provides the formal guarantee.}
\begin{theorem}
    Let $G$ be an instance of stochastic game with $n$ players.
    For any constant $\epsi, \delta\in (0,1)$,
    If Alg.~\ref{alg:cocos-sampling} successfully returns value function $\mathbf V$, then,
    it holds that
    \begin{align*}
        \prob{\left |\left| \mathbf V - \hs(G(\mathbf V)) \right|\right|_\infty > \epsi} \le \delta.
    \end{align*}
    The sample complexity at every iteration is $\oh{\frac{n^2}{\epsi^2}\log\left(\frac{n|X|}{\delta}\right)}$,
    where $\mathtt{MAX} = \max_{i\in N} \max |R_i|$.
\end{theorem}
\rev{\begin{rem}
The guarantee is stated as a residual bound ($\|\mathbf V - \tilde T(\mathbf V)\|_\infty \leq \epsi$).
By Theorem~\ref{thm:coco-unique}, for games where the spectral condition holds (proved for $n=2$ and $n=3, m=2$; conjectured generally), $\tilde T$ has a unique fixed point $\mathbf V^*$, and
$\|\mathbf V - \mathbf V^*\|_\infty \leq \epsi/(1-\gamma)$.
Thus the guarantee becomes \emph{unconditional}: the algorithm converges to the true Coco-S value.
\end{rem}}
\begin{proof}
    Consider the last iteration of the while loop, where \rev{$\Delta \leq \epsi/2$}.
    Given a player $j \in N$ and any state $x\in X$,
    in Line~\ref{line:coco-sample-init} of Alg.~\ref{alg:cocos-sampling},
    let \[Y_i = \left(\frac{1}{2}\cdot \mathbbm{1}_{I=N} 
        + \frac{n}{2|I|} \cdot \mathbbm{1}_{j\in I}  - \frac{n}{2(n-|I|)} \cdot \mathbbm{1}_{j \not \in I} \right)\maxmin_I G_x(\mathbf V),\]
    where $I$ is the $i$-th coalition in $M$,
    where $M$ denotes the set of coalitions sampled from procedure CoaSam($N,m$). 
    Then,
    \begin{align*}
        \ex{Y_i}&= \frac{1}{n} \maxmin_N G_x(\mathbf V) + \frac{1}{n} \sum_{S \subsetneq N} \binom{n}{k}^{-1}\left(\frac{n}{2|S|} \cdot \mathbbm{1}_{j\in S}  - \frac{n}{2(n-|S|)} \cdot \mathbbm{1}_{j \not \in S} \right)\maxmin_S G_x(\mathbf V)\\
        &= \frac{1}{n}\sum_{S\subseteq N: j \in S} \binom{n-1}{k-1}^{-1} \maxmin_S G_x(\mathbf V)\\
        &= \hs_j(G_x(\mathbf V)).
    \end{align*}

    For each player $j \in N$ in $G_x(\mathbf V)$, it holds that
    $|U_j(x, \mathbf a, \mathbf V)| \le \mathtt{MAX} + \gamma \rev{\|\mathbf V\|_\infty}$.
    Therefore, $\maxmin_I G_x(\mathbf V) \in [-n\left(\mathtt{MAX} + \gamma \rev{\|\mathbf V\|_\infty}\right), n\left(\mathtt{MAX} + \gamma \rev{\|\mathbf V\|_\infty}\right)]$ for each $I\subseteq N$ and $x \in X$,
    then, $Y_i \in [-n\left(\mathtt{MAX} + \gamma \rev{\|\mathbf V\|_\infty}\right), n\left(\mathtt{MAX} + \gamma \rev{\|\mathbf V\|_\infty}\right)]$.
    By Hoeffding's inequality,
    \begin{align*}
        \prob{\left|V_j'(x) - \hs_j(G_x(\mathbf V))\right| > \frac{\epsi}{2}}
        = \prob{\left|\sum_{i=1}^m Y_i - m\hs_j(G_x(\mathbf V))\right| > \frac{m\epsi}{2}}
        \le 2e^{-\frac{m\epsi^2}{8\mathtt{MAX}^2}}\le \frac{\delta}{n\cdot |X|}.
    \end{align*}
    Furthermore,
    \begin{align*}
        \prob{\left|V_j'(x) - \hs_j(G_x(\mathbf V))\right| > \frac{\epsi}{2}}
        &\ge \prob{\left|V_j(x) - \hs_j(G_x(\mathbf V))\right|-\left|V_j'(x) - V_j(x)\right| > \frac{\epsi}{2}}\\
        &\ge \prob{\left|V_j(x) - \hs_j(G_x(\mathbf V))\right| > \epsi}
    \end{align*}
    where the last inequality follows from $\left|V_j'(x) - V_j(x)\right| \le \left|\left|\mathbf V'-\mathbf V\right|\right|_\infty \le \frac{\epsi}{2}$.
    Therefore, it holds that $\prob{\left|V_j(x) - \hs_j(G_x(\mathbf V))\right| > \epsi} \le \frac{\delta}{n\cdot |X|}$.
    Then,
    \begin{align*}
    \prob{\left |\left| \mathbf V - \hs(G(\mathbf V)) \right|\right|_\infty > \epsi}
    \le \sum_{j \in N, x\in X}\prob{\left|V_j(x) - \hs_j(G_x(\mathbf V))\right| > \epsi}
    \le \delta. \qedhere
    \end{align*}
\end{proof}

%% file: experiments.tex

In this section, we learn $\mathbf V_{\Hss}, \mathbf V_{\cocon}$ and extract the resulting policies and side payments
on a standard test suite of games generalized
to more than two players.

\textbf{Baselines.}
As a security-level baseline, we implement MMBE, the generalized Bellman equations with the $\maxmin$ operator
corresponding to the per-player security level in our axioms. This baseline isolates the benefit of cooperative side
payments beyond worst-case play. As an equilibrium-based learning baseline, we use CE, Correlated-$Q$~\citep{Greenwald2003},
which is the canonical joint-action Q-learning algorithm for computing correlated equilibria in general-sum Markov games.
Because the game model is known, we implement all methods, including Correlated-$Q$, via value iteration.
As is typical for such equilibrium-based value-iteration methods, Correlated-$Q$ does not enjoy general convergence
guarantees in multi-player general-sum Markov games, and in our experiments it indeed fails to converge in most of the games we study.

In addition to the exact evaluations above, we also evaluate sampling-based variants of $\Hss$ and $\cocon$ in Section~\ref{sec:sampling}.
The exact computation of $\Hss$ and $\cocon$ values requires evaluating all possible coalitions of players at each state,
which requires exponential time in the number of players.
We replace this step with Monte Carlo sampling of coalitions.
Concretely, in the experiments, we run these algorithms as heuristics by sampling a fixed number of coalitions.
\footnote{Source code to reproduce the results is available in the supplementary material. }

  \begin{figure}[t]
    \centering
        \includegraphics[width=\columnwidth]{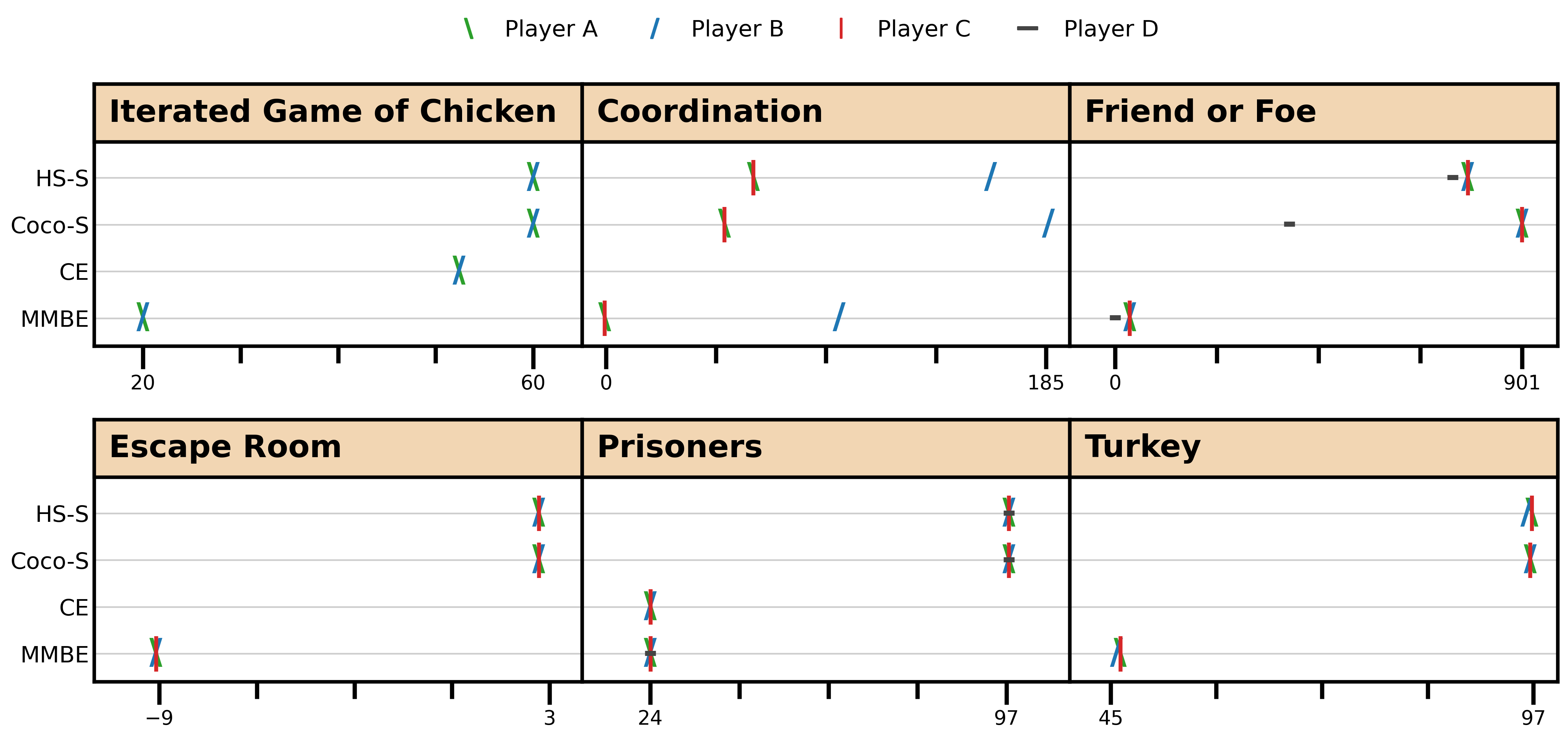}
    \caption{The strategic value for each player in each game, as computed by the corresponding algorithm.
    CE denotes Correlated-$Q$~\citep{Greenwald2003},
    and MMBE denotes the generalized Bellman equations with the maxmin operator. 
    The two rightmost columns report results on the grid games.
    }
    \label{fig:overall_results}
\end{figure}
\textbf{Summary of results.}
We find that both $\Hss$ and $\cocon$ learn sensible
assessments of the strategic strength of each player and enable
maximizing the overall combined score while preserving the competitive nature of the game.
This is achieved through side payments that encourage other players to act in ways that may
not be immediately advantageous.
Usually, $\Hss$ and $\cocon$ agree on at least the \textit{relative strength} of the players, although
this is not always the case (see Fig. \ref{fig:overall_results}). 
Frequently, they disagree on the nominal strength of the players.
Additionally, compared to MMBE, $\Hss$ and $\cocon$ typically achieve higher strategic values and lead to better overall outcomes.
This trend aligns with the axiom of individual rationality satisfied by $\Hss$, 
whereas we do not establish the same guarantee for $\cocon$.
Consistent with these limitations, Correlated-$Q$ did not converge on most of our games,
with the exception of Prisoners grid game and \rev{Iterated Game of Chicken}.

We find that the side payments at each state transition learned
by $\cocon$ agree better with our intuition (see the discussion below).
\rev{This is a direct consequence of the Markov Consistency axiom (Section~\ref{subsec:markov-consistency}):
because $\cocon$ applies the HS formula state-by-state with self-consistent continuation values,
each side payment reflects the local surplus redistribution at that state.
For $\Hss$, which uses infinite-horizon threat powers, the side payments are coherent
at the policy level but may attribute value from distant states to the current one.}

In symmetric games like escape room game and Prisoners grid game, both $\Hss$ and $\cocon$
find a series of side payments that ultimately result in symmetric values for all players.
In games like Coordination grid game, where one player has a significant advantage from the start,
both algorithms learn a final value that is proportionate to the players' starting strengths.
Additionally, in Friend-or-Foe grid game, we see a large nominal disagreement between $\Hss$ and $\cocon$ about the strength
of the weaker player. 

\subsection{Game Environments}\label{subsec:game-rules}
\begin{wraptable}{r}{0.4\textwidth}
  \centering
  \caption{Payoff matrix for iterated game of chicken.}
  \label{table:ipd_payoff_matrix}
  \vspace{0.5em} 
  \begin{tabular}{c|c|c}
    \textbf{A / B} & \textbf{C} & \textbf{D} \\ \hline
    \textbf{C} & (6, 6) & (2, 7) \\
    \textbf{D}    & (7, 2) & (0, 0) \\
  \end{tabular}
\end{wraptable}
\textbf{Iterated Game of Chicken.}
Chicken is a classic social dilemma in which two players choose to either cooperate or defect.
The payoff structure captures a tension between individual incentives and collective risk: 
unilateral defection yields the highest payoff to the defector while the cooperator receives a lower payoff; mutual cooperation leads to a moderate outcome for both players; and mutual defection results in the worst outcome due to a costly \rev{``collision.''}
We provide the payoff matrix used in our experiments in Table~\ref{table:ipd_payoff_matrix}.
This environment evaluates how different solution concepts and learning objectives balance immediate gains against long-term outcomes, and how well they sustain cooperation under strategic adaptation.
The discount factor $\gamma$ is set to $0.9$.

\textbf{Escape Room Game.}
An escape room game, ER($n$, $m$), involves $n$ players who must cooperate to escape a room with at least $m$ players pressing a button that opens the exit door.
Each player occupies one of three discrete positions—start, lever, or door.
At each step, every player simultaneously chooses an action that updates their position. Rewards are determined by the number of players who choose lever in that step. 
If fewer than $m$ players choose lever, then players who stay receive reward $0$, and others receive reward $-1$.
If at least $m$ players choose lever, the door is opened: any player who chooses door in that step receives reward $10$, while other players staying receive reward $0$, moving receive reward $-1$, and the game terminates.
In our experiments, we set $n=3$ and $m=2$.
The discount factor $\gamma$ is set to $1$ for ease of interpretability.

\textbf{Grid Games.}
In Grid games, players compete on a grid of $m \times n$ squares.
Each square can be occupied by at most one player.
Each player has a designated starting square and a set of individual and shared goal squares
where rewards are received.
The players can observe the positions of themselves and other players on the grid.
Additionally, there are walls and semi-walls that impede movement.
During each round, all players simultaneously choose an action from the options of moving up, down, left, right, or sticking in place.
Each move incurs a step cost of $-1.0$, even when the player is unable to move as intended. Sticking
incurs a step cost of $-0.1$.

When a player selects a move without obstacles, they move in the chosen direction.
If a player tries to move through a wall or to a square already occupied by a sticking player, they stay in their current square.
If a player attempts to move through a semi-wall, they have a 50\% chance of doing so, otherwise they stay in their current location.
If two players try to move into the same square, one is randomly selected to move and the other stays in their current location.

The game concludes when one of the players reaches a goal square, which has a positive reward assigned to it. 
If multiple players reach their goal squares simultaneously, they all receive rewards.
In our experiments, unless otherwise stated, the rewards for reaching the goal are set to 100, the cost of taking a step is -1, and the reward for staying in the same place is -0.1.
The discount factor $\gamma$ is set to 1 for ease of interpretability.

Agents are represented on the grid as A, B, C, and D\@.
The goal squares for each agent are drawn with unique directional lines.
In the case where agents have a shared goal, all sets of lines will be displayed.
The path taken by the agent is shown as a sequence of arrows pointing from the agent's current square to its next.
Each time an agent moves to a new square, the corresponding arrow of the path is labeled with the time step.
A \rev{``stick''} and failed actions are illustrated as another time label in the same square.
The side payments and total trajectory values for each agent are displayed below the game, with positive values indicating an agent received a payment and negative values indicating that the agent made a payment.

\textbf{Two State Game.}
To verify the effectiveness of our sample-based algorithms, we consider an
$n$-player extension of the two state game from Appendix~\ref{apx:ex-3-unequal}.
The game has two states, $x_1$ and $x_2$,
where $x_1$ is the decision state and $x_2$ is the absorbing state.

At state $x_1$, only a subset of players are active and may choose between $s$ (stay) and $m$ (move).
All other players are inactive at $x_1$.
The transition to $x_2$ requires unanimity among the active players: 
if every active player chooses $m$, the game transitions to $x_2$;
otherwise, it remains at $x_1$.
Rewards at $x_1$ are always 0 for all players. 
Upon reaching $x_2$, the game stays there forever regardless of actions, 
and each player receives a fixed payoff initialized by the environment.

This construction highlights a simple bargaining dynamic: 
because any active player can unilaterally prevent the transition, 
the active players can credibly threaten to keep the game at 
$x_1$ unless players who benefit from reaching $x_2$ offer sufficient compensation (e.g., via side payments).

In our experiments, we use $n=12$ players and initialize the fixed payoffs
at state $x_2$ so that 8 players receive $1$
and 4 players receive $-2$.
3 players with positive reward at $x_2$ are active players.
All results are averaged over 10 independent runs.

\subsection{Results}\label{subsec:example-games}
In this section, we compare the learned policies and side payments on specific examples of iterated game of chicken, escape room game and grid games:
Prisoners; Friend-or-Foe; Coordination; and Turkey. All of the grid games are generalizations
of commonly used $2$-player grid games \citep{Hu2003,Greenwald2003,Sodomka2013} to more than $2$ players.
We additionally evaluate computational efficiency by comparing our sampling-based 
estimators against exact computation for both \Hss and \cocon.

\begin{wraptable}{r}{0.6\textwidth}\label{table:chicken_results}
    \centering
    \caption{A learned CE, $\Hss$ and $\cocon$ policy in Iterated Game of Chicken. The values shown are probabilities of choosing each pair of actions.}
    \label{tab:ipd_prob}
    \begin{tabular}{l|ll}
\shortstack{CE/\Hss/\\\cocon} &  C  & D    \\
\hline
C     & 0.5/1/1    & 0.25/0/0     \\ 
D     & 0.25/0/0    & 0/0/0     \\ 
\end{tabular}
\end{wraptable}
\textbf{Iterated Game of Chicken.}
In this game (Table~\ref{table:chicken_results}), both $\Hss$ and $\cocon$ 
learn to cooperate consistently,
resulting in high strategic values for both players.
In contrast, Correlated Equilibrium (CE) learns a mixed strategy that includes defection with a non-negligible probability.

\textbf{Escape Room Game.}
In the escape room game, both $\Hss$ and $\cocon$ successfully learn to cooperate to escape the room in a single step (Fig.~\ref{fig:overall_results}).
Each player attains a strategic value of $2.67$ (total value $8$), 
corresponding to the coordinated outcome in which one agent 
reaches the door while two agents hold the lever.
In contrast, CE fails to converge in this game even
when we use $\gamma < 1$.

\begin{figure}[ht]
    \centering
    \begin{minipage}[t]{0.4\textwidth}
        \centering
        \vspace*{0pt}
        \includegraphics[width=\linewidth]{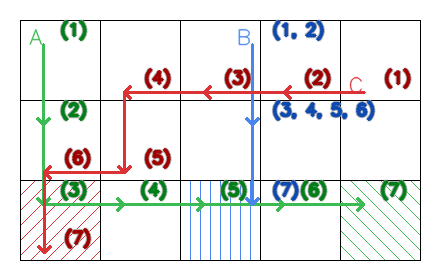}
    \end{minipage}
    \begin{minipage}[t]{0.48\textwidth}
        \centering
        \vspace*{0pt}
        \small
        \begin{tabular}{lll}
        State & $\Hss$        & $\cocon$       \\ \hline
        1     & (-0.1, -0.2, 0.3)    & (-7.6, 15.2, -7.6)    \\ 
        2     & (0.2, -0.3, 0.1)    & (-5.0, -7.9, 12.9)    \\ 
        3     & (-16.3, 32.1, -15.8)    & (-30.4, 52.2, -21.8)    \\ 
        4     & (-16.2, 32.7, -16.6)    & (-13.7, 54.0, -40.3)    \\ 
        5     & (0.5, -0.4, -0.0)    & (12.4, -25.1, 12.7)    \\ 
        6     & (0.0, -0.0, -0.0)    & (0.0, 0.0, 0.0)    \\ 
        \hline
        V     & (62.0, 161.6, 62.0)    & (49.8, 185.9, 49.8)    \\ 
        SP     & (-32.0, 64.0, -32.0)    & (-44.2, 88.3, -44.2)    \\ 
        \end{tabular}
    \end{minipage}
    \caption{A learned $\Hss$ and $\cocon$ trajectory in Coordination, with the side payments as computed by each algorithm. The value of the trajectory is indicated by (V) and the total side payments by (SP). }
    \label{fig:coordination_coco_figure}
\end{figure}

\textbf{Coordination Grid Game.}
In the Coordination game (as illustrated in Fig. \ref{fig:coordination_coco_figure}),
Players A, B, and C each
have their own goals they need to reach without colliding by coordinating their moves across the grid. Notice that Players
A and C are symmetric, but player B is closer to her goal than the other players. Therefore, one would expect that the strategic value of Player B is stronger than the others; and that therefore, Players A and C should pay Player B to allow them to make it to their respective goals. 

%
In Fig. \ref{fig:coordination_coco_figure},
we show the trajectory learned by the
algorithms.
Also, in the table of
Fig. \ref{fig:coordination_coco_figure}, we show
the values for the side payments made at each step along the trajectory,
as well as the total value (V) and the total side payments (SP).
Both $\mathbf V_{\Hss}$
and $\mathbf V_{\cocon}$ agree with the above intuition,
while Correlated-$Q$ did not converge.
Player B is the strongest, and Players A and C have to
pay Player B to stick while they coordinate
their passing.
The $\cocon$ side payments shown in the table
agree better with our intuition.
For example, in State 1, why should Player B pay the other players
to stick, when it is against his immediate self-interest?
$\cocon$ agrees with our intuition by having the other players
pay Player B to stick.

Additionally, $\Hss$ and $\cocon$ disagree on just how
strong Player B is. 
While $\Hss$ and $\cocon$ often agree on the relative
strength of the players, they do not always.
For example,
at State 5, there is a strong disagreement about the
strength of Player B: in this state, Player A is
occupying the goal of Player B. The $\Hss$ value
considers all players to be roughly equal, since
Player A cannot proceed to his own goal without
moving off of the goal of Player B. However,
the $\cocon$ value takes the threat of Players A
and C working together much more seriously.
\begin{figure}[ht]
    \centering
    \includegraphics[width=\columnwidth]{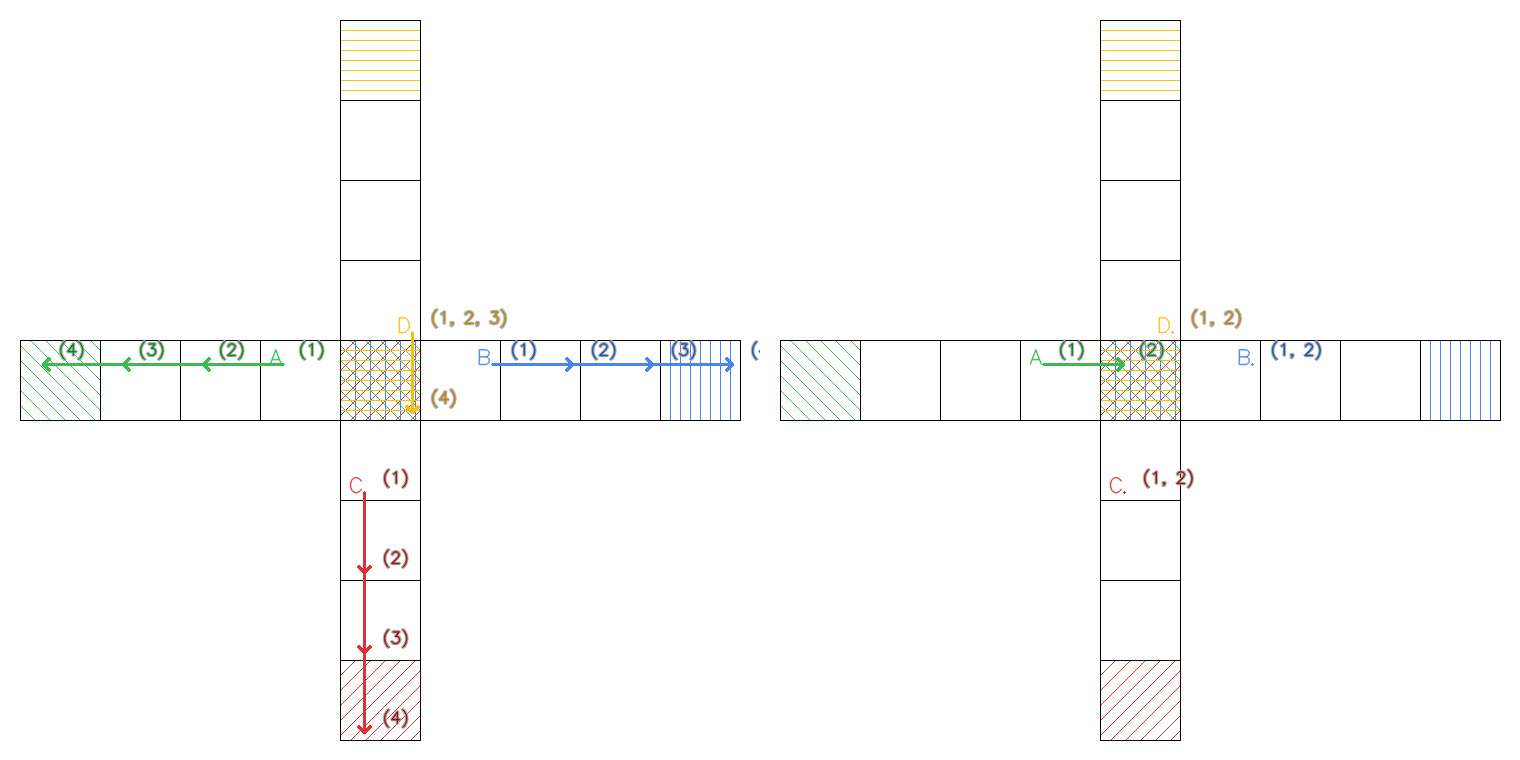}
    {\small \begin{tabular}{lll}
State & $\Hss$        & $\cocon$       \\ \hline
1     & (33.2, 33.2, 33.2, -99.7)    & (33.2, 33.2, 33.2, -99.7)    \\ 
2     & (-32.8, -32.8, -32.8, 98.4)    & (-32.8, -32.8, -32.8, 98.4)    \\ 
3     & (0.0, 0.0, 0.0, -0.0)    & (0.0, 0.0, 0.0, 0.0)    \\ 
\hline
V     & (97.5, 97.5, 97.5, 97.5)    & (97.5, 97.5, 97.5, 97.5)    \\ 
SP     & (0.5, 0.5, 0.5, -1.4)    & (0.5, 0.5, 0.5, -1.4)    \\ 
\end{tabular}}
    \caption{Prisoners. \textbf{Top:} A trajectory of $\Hss, \cocon$ (Left) and Correlated-$Q$ (Right). \textbf{Bottom:} Side payments for the trajectory in top left.}\label{fig:prisoners_coco_figure}
\end{figure}
\textbf{Prisoners Grid Game.}
The game depicted in Fig. \ref{fig:prisoners_coco_figure} is based on the classic normal-form Prisoners' Dilemma game,
with each agent having her own goal located at the end of her respective hallway and a shared goal in the center.
In this grid game, moving towards the shared goal (defecting) is the rational strategy for each agent 
If any agent chooses to move towards the shared goal, the others also prefer to move towards it to potentially win the tiebreaker.
However, if agents cooperate and move towards their own goals, they all can receive a higher expected value.

The side payments table in Fig. \ref{fig:prisoners_coco_figure} illustrates the payments exchanged during the players' progression.
Player D strategically pays Players A, B, and C to move away from the shared objective, gaining a significant advantage over them.
As a result, A, B, and C become vulnerable and are forced to pay Player D to stay in place temporarily, in order to position themselves to reach their individual goals.
Once each player is in a position to score, no further side payments are made among them. Notably, $\Hss$ and $\cocon$ agree exactly on
the values of the players at each state in this game, as well as the side payments. 
The Correlated-$Q$ policy has each player choose their rational strategy of attempting to move into the shared goal, resulting in a win with a probability of 0.25 and an expected value of 24.0, which is significantly lower than the $\Hss$ value.

\begin{figure}[ht]
    \centering
    \includegraphics[width=0.5\columnwidth]{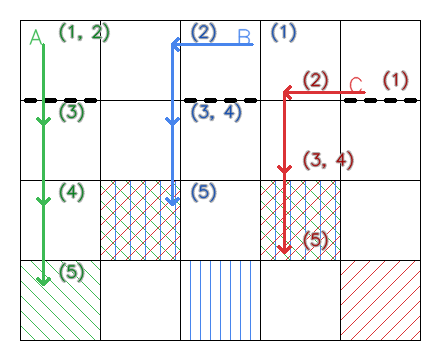} \\
    \begin{tabular}{lll}
State & $\Hss$        & $\cocon$       \\ \hline
1     & (75.4, -37.4, -36.8)    & (65.9, -32.3, -32.3)    \\ 
2     & (-9.1, 3.9, 3.9)    & (0.2, -0.7, -0.7)    \\ 
3     & (-65.5, 32.8, 32.8)    & (-65.5, 32.7, 32.7)    \\ 
4     & (0.0, 0.0, 0.0)    & (0.0, -0.0, -0.0)    \\ 
\hline
V     & (96.8, 96.2, 96.8)    & (96.6, 96.6, 96.6)    \\ 
SP     & (0.8, -0.7, -0.1)    & (0.6, -0.3, -0.3)    \\ 
\end{tabular}

    \includegraphics[width=0.4\columnwidth]{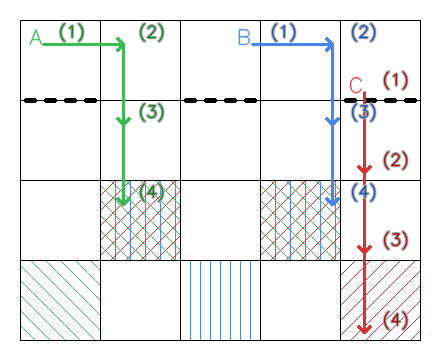}
    \includegraphics[width=0.4\columnwidth]{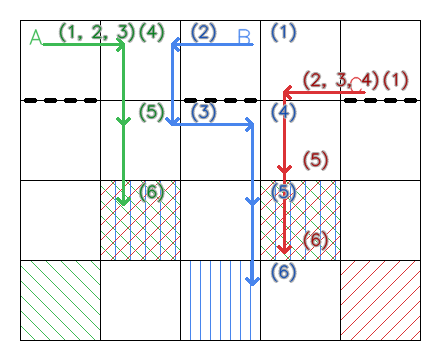} \\

\caption{Multiple trajectories generated by the $\Hss$ and $\cocon$ policy in the Turkey Game.
  The first (Top) shows Player A attempting the semi-wall, the second (Bottom-Left) 
  shows Player C moving through the semi-wall. The final trajectory (Bottom-Right),
  shows Player B moving out of the way of Player A for a guaranteed movement.}\label{fig:turkey_coco_figure}

\end{figure}
\textbf{Turkey Grid Game.}
The game shown in Figure \ref{fig:turkey_coco_figure} involves agents with individual goals located three steps below their starting positions.
Semi-walls, represented by thick dashed lines, are placed between the agent and its goal, with a probability of 0.5 for success if an agent attempts to pass through it.
Additionally, there are two shared goals placed three spaces from each pair of agents.



For this game, the trajectory corresponding to the
$\Hss, \cocon$ policy is not deterministic,
but depends on what happens when a player
attempts to pass through a semi-wall.
In the depicted trajectory in Fig. \ref{fig:turkey_coco_figure},
Player A attempts to pass through the semi-wall and was unsuccessful.
They took this risky action because they were paid by both Players B and C to do so, which allowed them to move around their own semi-wall with guaranteed success.
Once Player A has passed through the semi-wall, they coerced  the cooperation of Players B and C via a payment to stick while A gets into position for a score.

\begin{figure}[ht]
    \centering
    \includegraphics[width=0.8\columnwidth]{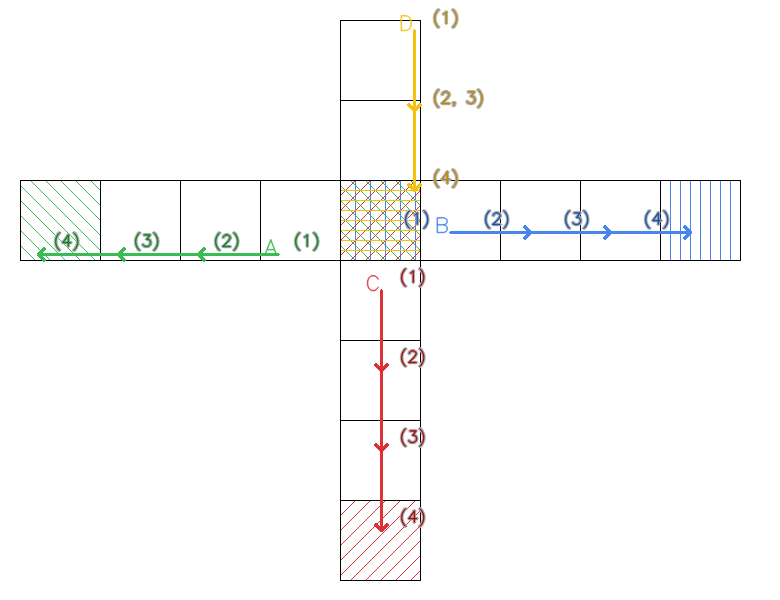}
     \begin{tabular}{lll}
State & $\Hss$           \\ \hline
1     & (116.3, 116.3, 116.3, -349.0)   \\
2     & (-332.8, -332.8, -332.8, 998.4) \\
3     & (0.0, 0.0, 0.0, 0.0)  \\
\hline
V     & (780.6, 780.6, 780.6, 747.2) \\
SP     & (-216.4, -216.4, -216.4, 649.3) \\
     \end{tabular}
     \begin{tabular}{lll}
State & $\cocon$       \\ \hline
1     & (236.8, 236.8, 236.8, -710.5)    \\ 
2     & (-332.8, -332.8, -332.8, 998.4)    \\ 
3     & (0.0, 0.0, 0.0, 0.0)    \\ 
\hline
V     & (901.0, 901.0, 901.0, 385.8)    \\ 
SP     & (-96.0, -96.0, -96.0, 287.9)    \\ 
     \end{tabular}
     
    \caption{A Friend-or-Foe trajectory with side payments.}
    \label{fig:friend_or_foe_coco_figure}
  \end{figure}
\textbf{Friend-or-Foe Grid Game.}
The game of Friend-or-Foe, depicted in Figure \ref{fig:friend_or_foe_coco_figure}, has
a weak player (Player D) who starts two steps away from a shared goal with reward $100$.
The other players start one step from the shared goal, but have their own goals worth
$1000$ three steps away. If the other players try to move to their high-value
goals, the weaker player can act as a spoiler (which is also in her self interest)
by moving to the shared goal and ending the game. 



Both $\Hss$ and $\cocon$ determine that Player D should pay the others to move away from the shared goal, but there is
a large disagreement about the amount of the payment. $\cocon$ assigns a larger payment since Player D is powerless
in the normal-form game at the first state. After the first state, Player D is in a stronger position than the
others and $\Hss$ and $\cocon$ agree on the strategic value on the rest of the trajectory. 


\begin{table}[ht]
    \centering
    \caption{Runtime and maximum absolute deviation between exact and sampled solutions in Two State Game with 12 players. All results are averaged over 10 independent runs.}
    \label{table:sampling_results}
    \begin{tabular}{c|cccc}
\shortstack{CE/\Hss/\\\cocon} &  \Hss  & \Hss Sampling & \cocon & \cocon Sampling    \\
\hline
Runtime ($s$)     & 160   & 101 ($\sigma = 2$)  & 386  & 240 ($\sigma = 13$) \\ 
Max Abs Deviation     & - & 0.33 ($\sigma = 0.08$) & -  & 0.13 ($\sigma = 0.036$)
\end{tabular}
\end{table}

\textbf{Sample-Based Estimation for \Hss and \cocon.}
To evaluate the performance of our sampling-based algorithms (Sec.~\ref{sec:sampling}), 
we apply them to the Two State Game described earlier.
Table~\ref{table:sampling_results} reports both runtime and the maximum absolute deviation in strategic values between the exact computation and the sampled estimates.
With 12 players, exact computation requires enumerating $2^{12}=4,096$ coalitions,
whereas our sampling methods use $3,000$ sampled coalitions.
Sampling substantially reduces computation time for both solution concepts 
while keeping estimation error small. 
Notably, \cocon yields a smaller maximum deviation than \Hss ($0.13$ vs.\ $0.33$), 
indicating more accurate value approximation in this setting, albeit at a higher overall runtime.


%% file: appendix.tex
\section{Computation of 3-Player Banana Example}\label{apx:ex-3}

\begin{example}

    \textbf{Case: $I = \{1\}$, $\overline{I} = \{2,3\}$.}
    Then the payoff matrix for player $I$ is 
      \[R_1 = 
    \begin{array}{cc|cccc}
          &  \multicolumn{5}{c}{ \overline{I}= \{2,3\} }\\
          &  & (B,B) & (B, NB) & (NB, B) & (NB, NB) \\ \hline
          & R & 2 & 2 & 2 & 2 \\
        \smash{\raisebox{.3\normalbaselineskip}{$I=\{ 1 \}$}}  & C & 4 & 4 & 4 & 0
    \end{array} \]
    
    The payoff matrix for $\overline I$ is $-R_1$.
    Then $\maxmin( G_I ) = \maxmin (R_1) = 2$, and hence $\maxmin( G_{\overline{I}} ) = -2$.
    
    \textbf{Case: $I = \{2\}$, $\overline{I} = \{1,3\}$.}
    Then the payoff matrix for player $I$ is 
      \[R_2 = 
    \begin{array}{cc | cccc}
      &   \multicolumn{4}{c}{ \overline{I}=\{ 1,3\} }\\
      &  & (R,B) & (R, NB) & (C, B) & (C, NB) \\ \hline
      & B & -2 & -2 & -4 & -4 \\
    \smash{\raisebox{.3\normalbaselineskip}{$I=\{ 2\}$}}  & NB & -2 & -2 & -4 & 0
    \end{array} \]
    
    The payoff matrix for $\overline I$ is $-R_2$.
    Then $\maxmin( G_I ) = \maxmin (R_2) = -4$, and
    $\maxmin( G_{\overline{I}} ) = 4$.
    
    \textbf{Case: $I = \{3\}$, $\overline{I} = \{1,2\}$.}
    This computation is the same as the preceding
    case, giving
    $\maxmin( G_I ) = \maxmin (R_3) = -4$, and
    $\maxmin( G_{\overline{I}} ) = 4$.
    
    \textbf{Case: $I = \{1,2,3\}$,  $\overline{I} = \varnothing$.}
    Then 
    \[
    \maxmax( G_I ) = \max_{\mathbf{a}  \in A_1 \times A_2 \times A_3} (R_1(\mathbf{a} ) + R_2(\mathbf{a} ) + R_3(\mathbf{a} )) = 4.
    \]
    
    Having computed the $\maxmin$ for each coalitional game $G_I$, let $\phi_I = \maxmin(G_I)$
    
    We now compute the $\hs$ values:
    \begin{align*}
      \hs_1 &= \frac{1}{3}( \phi_{ \{ 1 \} } + \frac{1}{2}( \phi_{ \{1, 2 \} } + \phi_{ \{1,3 \} } ) + \phi_{ \{ 1, 2, 3 \} } ) = \frac{10}{3},  \\
      \hs_2 &= \frac{1}{3}( \phi_{ \{ 2 \} } + \frac{1}{2}( \phi_{ \{2, 3 \} } + \phi_{ \{1 , 2 \} } ) + \phi_{ \{ 1, 2, 3 \} } ) = \frac{1}{3},  \\
      \hs_3 &= \hs_2 = \frac{1}{3}.
    \end{align*}

\end{example}

\section{3-Player Example of $\mathbf V_{\Hss} \neq \mathbf V_{\cocon}$}
\label{apx:ex-3-unequal}
\begin{example}
    \revB{The stochastic game $G(N, X, \mathbf A, P, \mathbf R, \gamma)$} is defined as follows
    \begin{itemize}
        \item \textbf{Players}. $N = \{1,2,3\}$.
        \item \textbf{States}. $X = \{x_1, x_2\}$, where $x_1$ is the decision state
            and $x_2$ is the absorbing state.
        \item \textbf{Actions and Transitions}. At $x_1$, only player 1 acts 
            with $A_1(x_1) = \{s\text{ (stay)}, m\text{ (move)}\}$, where 
            $P(x_1, s, x_1) = 1$ and $P(x_1, m, x_2) = 1$. Player 2 and 3 are passive.
            At $x_2$, all players' actions are irrelevant, and they always stay at $x_2$.
        \item \textbf{Rewards.} $\mathbf R(x_1, \cdot, \cdot) = (0, 0, 0)$
            and $\mathbf R(x_2, \cdot, \cdot) = (1, -2, 1)$.
        \item \textbf{Discount Factor.} $\gamma = 0.5$.
    \end{itemize}
\end{example}
\textbf{\Hss Value Calculation.} 
Since all players remain in $x_2$ once it is reached,
it is clear that 
\begin{align*}
    & \revB{[\delta G](N, x_2)} = 0, \\
    & \revB{[\delta G](\{1\}, x_2)} = \revB{[\delta G](\{3\}, x_2)} = -\revB{[\delta G](\{2,3\}, x_2)} = -\revB{[\delta G](\{1,2\}, x_2)} = 4,\\
    & \revB{[\delta G](\{2\}, x_2)} = -\revB{[\delta G](\{1,3\}, x_2)} = -8.
\end{align*}
The \Hss value at state $x_2$ is 
\[\mathbf V_{\Hss}(x_2) = \left( \frac{1}{3}(0 + \frac{1}{2}(-4+8) + 4),
        \frac{1}{3}(0 + \frac{1}{2}(-4-4) - 8),
        \frac{1}{3}(0 + \frac{1}{2}(-4+8) + 4) \right) = (2, -4, 2).\]
At state $x_1$, 
\begin{align*}
    &\left\{
    \begin{aligned}
    & \revB{[\delta G](N, x_1)} = \max\left\{\frac{1}{2} \revB{[\delta G](N, x_1)}, \frac{1}{2}\revB{[\delta G](N, x_2)}\right \}\\
    & \revB{[\delta G](\{1\}, x_1)} = -\revB{[\delta G](\{2,3\}, x_1)} = \max\left\{ \frac{1}{2} \revB{[\delta G](\{1\}, x_1)}, \frac{1}{2}\revB{[\delta G](\{1\}, x_2)} \right\}\\
    & \revB{[\delta G](\{2\}, x_1)} = -\revB{[\delta G](\{1,3\}, x_1)} = \min\left\{ \frac{1}{2} \revB{[\delta G](\{2\}, x_1)}, \frac{1}{2}\revB{[\delta G](\{2\}, x_2)} \right\}\\
    & \revB{[\delta G](\{3\}, x_1)} = -\revB{[\delta G](\{1,2\}, x_1)} = \min\left\{ \frac{1}{2} \revB{[\delta G](\{3\}, x_1)}, \frac{1}{2}\revB{[\delta G](\{3\}, x_2)} \right\}
\end{aligned}
\right .\\
&\implies\,\left\{
\begin{aligned}
    & \revB{[\delta G](N, x_1)} = 0\\
    & \revB{[\delta G](\{1\}, x_1)} = -\revB{[\delta G](\{2,3\}, x_1)} = 2\\
    & \revB{[\delta G](\{2\}, x_1)} = -\revB{[\delta G](\{1,3\}, x_1)} = -4\\
    & \revB{[\delta G](\{3\}, x_1)} = -\revB{[\delta G](\{1,2\}, x_1)} = 0
\end{aligned}\right ..
\end{align*}
Then, the \Hss value at state $x_1$ is
\[\mathbf V_{\Hss}(x_1) = \left( \frac{1}{3}(0 + \frac{1}{2}(0+4) + 2),
        \frac{1}{3}(0 + \frac{1}{2}(0-2) - 4),
        \frac{1}{3}(0 + \frac{1}{2}(4-2) + 0) \right) 
        = (\frac{4}{3}, -\frac{5}{3}, \frac{1}{3}).\]

\textbf{\cocon Value Calculation.}
The rewards of game $G(\mathbf V)$ is following
\begin{align*}
    U_i(x_1, s) = \frac{1}{2}V_i(x_1), \,
    U_i(x_1, m) = \frac{1}{2}V_i(x_2),\,
    U_i(x_2) = R_i(x_2, \cdot, \cdot) + \frac{1}{2}V_i(x_2)
\end{align*}
Then, for state $x_2$,
\begin{align*}
    & \maxmin_N(G_{x_2}(\mathbf V)) = \frac{1}{2}\left( V_1(x_2) + V_2(x_2)+ \rev{V_3(x_2)} \right),\\
    & \maxmin_{1}(G_{x_2}(\mathbf V)) = -\maxmin_{2,3}(G_{x_2}(\mathbf V))
         = 2 + \frac{1}{2}\left( V_1(x_2) - V_2(x_2)- V_3(x_2) \right),\\
    & \maxmin_{2}(G_{x_2}(\mathbf V)) = -\maxmin_{1,3}(G_{x_2}(\mathbf V))
         = -4 + \frac{1}{2}\left(- V_1(x_2) + V_2(x_2)- V_3(x_2) \right),\\
    & \maxmin_{3}(G_{x_2}(\mathbf V)) = -\maxmin_{1,2}(G_{x_2}(\mathbf V))
         = 2 + \frac{1}{2}\left(-V_1(x_2) - V_2(x_2)+ V_3(x_2) \right)
\end{align*}
\begin{align*}
    \left\{
    \begin{aligned}
        & V_1(x_2) = 1 +\frac{1}{2}V_1(x_2)\\
        & V_2(x_2) = -2 +\frac{1}{2}V_2(x_2)\\
        & V_3(x_2) = 1 +\frac{1}{2}V_3(x_2)
    \end{aligned}\right . \implies \left\{
    \begin{aligned}
        & V_1(x_2) = 2\\
        & V_2(x_2) = -4\\
        & V_3(x_2) = 2
    \end{aligned}\right . .
\end{align*}
Therefore, $\mathbf V_{\cocon}(x_2) = (2, -4, 2)$.

For state $x_1$,
\begin{align*}
    & \maxmin_N(G_{x_1}(\mathbf V)) =
         \max\left\{ \frac{1}{2}
        (V_1(x_1) + V_2(x_1) + V_3(x_1)),
        0 \right\}\\
    & \maxmin_{1}(G_{x_1}(\mathbf V)) =
        -\maxmin_{2,3}(G_{x_1}(\mathbf V)) =
        \max\left\{ \frac{1}{2}
        (V_1(x_1) - V_2(x_1) - V_3(x_1)),
        2 \right\}\\
    & \maxmin_{2}(G_{x_1}(\mathbf V)) =
        -\maxmin_{1,3}(G_{x_1}(\mathbf V)) =
        \min\left\{ \frac{1}{2} 
        (-V_1(x_1) + V_2(x_1) - V_3(x_1)),
        -4 \right\}\\
    & \maxmin_{3}(G_{x_1}(\mathbf V)) =
        -\maxmin_{1,2}(G_{x_1}(\mathbf V)) =
        \min\left\{ \frac{1}{2} 
        (-V_1(x_1) - V_2(x_1) + V_3(x_1)),
        2 \right\}
\end{align*}
\begin{align*}
    &\left\{
    \begin{aligned}
        & V_1(x_1) = \frac{1}{3}\left( \maxmin_N(G_{x_1}(\mathbf V)) + \frac{1}{2}(\maxmin_{1,2}(G_{x_1}(\mathbf V))+\maxmin_{1,3}(G_{x_1}(\mathbf V)))
        +\maxmin_1(G_{x_1}(\mathbf V)) \right)\\
        & V_2(x_1) = \frac{1}{3}\left( \maxmin_N(G_{x_1}(\mathbf V)) + \frac{1}{2}(\maxmin_{2,3}(G_{x_1}(\mathbf V))+\maxmin_{1,2}(G_{x_1}(\mathbf V)))
        +\maxmin_2(G_{x_1}(\mathbf V)) \right)\\
        & V_3(x_1) = \frac{1}{3}\left( \maxmin_N(G_{x_1}(\mathbf V)) + \frac{1}{2}(\maxmin_{1,3}(G_{x_1}(\mathbf V))+\maxmin_{2,3}(G_{x_1}(\mathbf V)))
        +\maxmin_3(G_{x_1}(\mathbf V)) \right)
    \end{aligned}\right .\\
    &\implies \, \left\{
    \begin{aligned}
        & V_1(x_1) = \frac{5}{4}\\
        & V_2(x_1) = -\frac{7}{4}\\
        & V_3(x_1) = \frac{1}{2}
    \end{aligned}\right . .
\end{align*}
Therefore, $\mathbf V_{\cocon}(x_1) = (\frac{5}{4}, -\frac{7}{4}, \frac{1}{2})$.

\input{apx_axiom}

\input{apx_uniqueness}

%% file: apx_axiom.tex
\section{Axioms for Stochastic Games and Uniqueness of \Hss}\label{apx:axiom}
This appendix formalizes the axioms we extend from normal-form games to 
stochastic games, proves that the \Hss value satisfies them, 
and shows that \Hss is the unique mapping consistent with these axioms.

\subsection{Extending Axioms from Normal-Form Games~\cite{Kohlberg2021} to Stochastic Games}\label{sec:axiom}


\begin{definition}[Sum of Games]
  Let \revB{$G_1(N, X_1, \mathbf A_1, P_1, \mathbf R_1, \gamma)$}
  and \revB{$G_2(N, X_2, \mathbf A_2, P_2, \mathbf R_2, \gamma)$}
  be two independent stochastic games.
  Then $G:= G_1 \oplus G_2$ is the stochastic game
  \revB{$(N, X, \mathbf A, P, \mathbf R)$,}
  where $X = X_1\times X_2$,
  $\mathbf A = \mathbf A_1 \times \mathbf A_2$,
  and for any $x' = (x_1', x_2'), x'' = (x_1'', x_2'') \in X$ and $\mathbf a = (\mathbf a_1, \mathbf a_2)$,
  \begin{align*}
    &P(x', \mathbf a, x'') =  P_1(x_1', \mathbf a_1, x_1'') \cdot P_2(x_2', \mathbf a_2, x_2'')\\
    &\mathbf R(x', \mathbf a, x'') = \mathbf R_1(x_1', \mathbf a_1, x_1'') + \mathbf R_2(x_2', \mathbf a_2, x_2'').
  \end{align*}
\end{definition}
\begin{definition}[Null Player]
  We say the player $i$ is a null player in stochastic game
  \revB{$G(N, X, \mathbf A, P, \mathbf R)$,} if
  \begin{itemize}
    \item $R_i(x', \mathbf a, x'') = 0$ for all $\mathbf a \in \mathbf A$
    and $x', x'' \in X$;
    \item $\mathbf R(x', \mathbf a, x'') = \mathbf R(x', \mathbf b, x'')$
    and $P(x', \mathbf a, x'') = (x', \mathbf b, x'')$,
    if $\mathbf a, \mathbf b \in \mathbf A$ and $a_j = b_j$ for all $j\neq i$.
  \end{itemize}
\end{definition}
\begin{definition}[Interchangeable Players]
  We say the players $i$ and $j$ are interchangeable in stochastic game
  \revB{$G(N, X, \mathbf A, P, \mathbf R)$,} if
  \begin{itemize}
    \item $A_i = A_j$ and $R_i = R_j$;
    \item $\mathbf R(x', \mathbf a, x'') = \mathbf R(x', \mathbf b, x'')$ 
    and $P(x', \mathbf a, x'') = P(x', \mathbf b, x'')$, 
    for all $\mathbf a, \mathbf b \in \mathbf A$, where
    $a_i = b_j$, $a_j = b_i$, and $a_k = b_k$ for all $k \neq i, j$.
  \end{itemize}
\end{definition}
In the following, we establish the additivity property of the threat power of coalition $\delta$ as follows.
\begin{proposition}\label{prop:add-threat}
    \revB{The threat power $\delta: \mathbb G(N) \to \left(2^N \times X \to \mathbb R\right)$ satisfies additivity:
    \[[\delta(G' \oplus G'')](I, x) = [\delta G'](I, x_1) + [\delta G''](I, x_2), \quad \forall I\subseteq N,\, x=(x_1,x_2)\in X,\]
    where $G', G'' \in \mathbb G(N)$ are two independent stochastic games.}
\end{proposition}
\begin{proof}
    \revB{Let $G' = (N, X_1, \mathbf A_1, P_1, \mathbf R_1, \gamma)$ and $G'' = (N, X_2, \mathbf A_2, P_2, \mathbf R_2, \gamma)$ be two independent stochastic games, and let $G = G'\oplus G'' = (N, X, \mathbf A, P, \mathbf R, \gamma)$ with $X = X_1\times X_2$.}
    By the definition of threat power $\delta$, we know that,
    for each coalition $I\subset N$,
    \begin{align*}
        &\revB{[\delta G'](I, x_1)} = \max_{\mathbf a_{1, I}}\min_{\mathbf a_{2, \bar I}} \sum_{x_1'\in X_1} P_1(x_1, \mathbf a_1, x_1')\left[U_{1,S}(x_1, \mathbf a_1, x_1') + \gamma\revB{[\delta G'](I, x_1')} \right], \forall x_1 \in X_1\\
        &\revB{[\delta G''](I, x_2)} = \max_{\mathbf a_{2, I}}\min_{\mathbf a_{2, \bar I}} \sum_{x_2'\in X_2} P_2(x_2, \mathbf a_2, x_2')\left[U_{2,S}(x_2, \mathbf a_2, x_2') + \gamma\revB{[\delta G''](I, x_2')} \right], \forall x_2 \in X_2
    \end{align*}
    By summing the above equations, it holds that, for each coalition $I\subset N$
    and state $x = (x_1, x_2)\in X$,
    \begin{align*}
        &\revB{[\delta G'](I, x_1)} + \revB{[\delta G''](I, x_2)} = \max_{\mathbf a_{1, I}}\min_{\mathbf a_{2, \bar I}} \sum_{x_1'\in X_1} P_1(x_1, \mathbf a_1, x_1')\left[U_{1,S}(x_1, \mathbf a_1, x_1') + \gamma\revB{[\delta G'](I, x_1')} \right]\\
        &\hspace{3em} +\max_{\mathbf a_{2, I}}\min_{\mathbf a_{2, \bar I}} \sum_{x_2'\in X_2} P_2(x_2, \mathbf a_2, x_2')\left[U_{2,S}(x_2, \mathbf a_2, x_2') + \gamma\revB{[\delta G''](I, x_2')} \right]\\
        &\hspace{1em}=\max_{(\mathbf a_{1, I}, \mathbf a_{2, I})}\min_{(\mathbf a_{1, \bar I}, \mathbf a_{2, \bar I})} \Biggl(\sum_{x_1'\in X_1} P_1(x_1, \mathbf a_1, x_1')\left[U_{1,S}(x_1, \mathbf a_1, x_1') + \gamma\revB{[\delta G'](I, x_1')} \right]\\
        &\hspace{3em} + \sum_{x_2'\in X_2} P_2(x_2, \mathbf a_2, x_2')\left[U_{2,S}(x_2, \mathbf a_2, x_2') + \gamma\revB{[\delta G''](I, x_2')} \right] \Biggr)\\
        %
        %
        &\hspace{1em}= \max_{\mathbf a_{I}}\min_{\mathbf a_{\bar I}} \sum_{(x_1', x_2') \in X} P(x, \mathbf a, x') \left[U_{1,S}(x_1, \mathbf a_1, x_1') +U_{2,S}(x_2, \mathbf a_2, x_2') + \gamma\left(\revB{[\delta G'](I, x_1')}+\revB{[\delta G''](I, x_2')}\right) \right]\\
        &\hspace{1em}= \max_{\mathbf a_{I}}\min_{\mathbf a_{\bar I}} \sum_{(x_1', x_2') \in X} P(x, \mathbf a, x') \left[U_{S}(x, \mathbf a, x') + \gamma\left(\revB{[\delta G'](I, x_1')}+\revB{[\delta G''](I, x_2')}\right) \right].
    \end{align*}
    Due to the uniqueness of threat power, it holds that
    \[\revB{[\delta G'](I, x_1) + [\delta G''](I, x_2) = [\delta G](I, x)}, \forall\, x_1 \in X_1, x_2 \in X_2, x = (x_1, x_2),\]
    which completes the proof.
\end{proof}
In the following, we adapt the axioms defined in~\cite{Kohlberg2021} for
normal form games to the stochastic games.
\revB{One notable departure from the original formulation concerns the balanced threats axiom.
Kohlberg and Neyman require that \emph{all} coalition threat powers being zero implies every player's value is zero.
We relax this to \emph{Weak Balanced Threats}: only proper-coalition threats $[\delta G](S, x) = 0$ for $S \subsetneq N$ are required to vanish; the grand-coalition threat $[\delta G](N, x)$ is left unconstrained.
Crucially, this relaxation does not weaken the uniqueness result: the proof of Theorem~\ref{theo:normal_axiom} relies only on the weak version, so the characterization of \Hss\ as the unique value mapping satisfying the stochastic axioms continues to hold under this relaxed condition.}
\begin{restatable}{axiom}{SGaxiom}\label{ax:SG}
    For any stochastic game \revB{$G(N, X, \mathbf A, P, \mathbf R, \gamma)\in \mathbb G(N)$,}
    let \revB{$\delta G: 2^N \times X \to \reals$} be the threat power defined in Def.~\ref{def:sg-threat},
    and $\{(\mathcal Q^*_i, \mathcal V^*_i): i\in N\}$ be the solution of the generalized
    Bellman equations (Equation~\eqref{eq:bellman-q}) on game $G$ with operator $\otimes = \maxmin$.
    We then consider the following axioms on a mapping \revB{$\zeta: \mathbb G(N) \to \mathbb R^{n \times m}$:}
    \begin{itemize}
        \item \textit{Efficiency}: \revB{$\sum_{i\in N} [\zeta G]_{i,x} = [\delta G](N, x)$ for all $x\in X$.}
        \item \textit{Symmetry}: If $i$ and $j$ are interchangeable in $G$, then \revB{$[\zeta G]_{i,x} = [\zeta G]_{j,x}$ for all $x\in X$.}
        \item \textit{Null player}: If $i$ is a null player in $G$, then \revB{$[\zeta G]_{i,x} = 0$ for all $x\in X$.}
        \item \textit{Additivity}: For any two independent stochastic games $G'\in \mathbb{G}(N)$ and $ G''\in \mathbb{G}(N)$, it satisfies that \revB{$[\zeta (G' \oplus G'')]_{i,x} = [\zeta G']_{i,x_1} + [\zeta G'']_{i,x_2}$, for all $i\in N$, $x_1\in X_1$, $x_2\in X_2$ and $x = (x_1, x_2)$.}
        \item \revB{\textit{Weak Balanced threats}: If $[\delta G](S, x) = 0$ for all $S\subsetneq N$ and $x\in X$, then $[\zeta G]_{i,x} = [\zeta G]_{j,x}$ for all $i, j\in N$ and $x\in X$.}
        \item \textit{Individual rationality}: \revB{$[\zeta G]_{i,x} \ge \mathcal V^*_i(x)$ for all $x\in X$.}
    \end{itemize}
\end{restatable}








\subsection{Axiomatic Compliance of \Hss}\label{apx:hss-satisf}
In the following, we verify that \Hss\ value ($V_{\Hss}$) satisfies all axioms stated
in Axiom~\ref{ax:SG}.

\textbf{Efficiency.} For any game \revB{$G(N, X, \mathbf A, P, \mathbf R, \gamma)\in \mathbb G(N)$,} 
by \Hss calculation defined in Def.~\ref{def:hs_s}, it holds that
\begin{align*}
    &\sum_{i\in N}V_{\Hss, i}(x) = \frac{1}{n}\sum_{i\in N} \sum_{I\subseteq N: i\in I} \binom{n-1}{|I|-1}^{-1} \revB{[\delta G](I, x)}\\
    & = \frac{1}{n} \sum_{I\subseteq N: I \neq \emptyset} |I|\cdot \binom{n-1}{|I|-1}^{-1} \revB{[\delta G](I, x)}\\
    & = \revB{[\delta G](N, x)} + \frac{1}{2n} \sum_{I\subsetneq N: I \neq \emptyset} \left(|I|\cdot \binom{n-1}{|I - 1|}^{-1} \revB{[\delta G](I, x)} + (n-|I|)\cdot \binom{n-1}{n-|I|-1}^{-1} \revB{[\delta G](\bar I, x)}\right)\\
    & = \revB{[\delta G](N, x)},
\end{align*}
where the last equation follows from
$|I|\cdot \binom{n-1}{|I - 1|}^{-1} = (n-|I|)\cdot \binom{n-1}{n-|I|-1}^{-1}$
and \revB{$[\delta G](I, x) = - [\delta G](\bar I, x)$}.

Efficiency axiom is satisfied.

\textbf{Symmetry.}
\revB{Let $G$ be a stochastic game.}
Suppose that players $i$ and $j$ are interchangeable.
Let $S \subseteq N$, $x, x'\in X$,
and $\mathbf a, \mathbf b \in \mathbf A$, where
$i, j \not \in S$,
$a_i = b_j$, $a_j = b_i$ and $a_k = b_k$ for all $k\neq i, j$.
From interchangeability,
it holds that, $P(x, \mathbf a, x') = P(x, \mathbf b, x')$, 
$R_i = R_j$, and $\mathbb R(x, \mathbf a, x') = \mathbb R(x, \mathbf b, x')$.
Furthermore,
\begin{align*}
    U_{S+i}(x, \mathbf a, x') &= \sum_{k\in S+i} R_k(x, \mathbf a, x') - \sum_{k\in \bar S - i} R_k(x, \mathbf a, x') \\
    &= \sum_{k\in S} R_k(x, \mathbf b, x') - \sum_{k\in \bar S - i - j} R_k(x, \mathbf b, x') \\
    &= U_{S+j}(x, \mathbf b, x').
\end{align*}
Therefore, for any state $x \in X$,
\begin{align*}
    \revB{[\delta G](S+i, x)} &= \max_{\mathbf a_{S+i}}\min_{\mathbf a_{\bar{S}-i}} \sum_{x'\in X}P(x, \mathbf a, x')
    \left[U_{S+i}(x, \mathbf a, x') + \gamma \revB{[\delta G](S+i, x')}\right] \\
    & = \max_{\mathbf a_{S+i}}\min_{\mathbf a_{\bar{S}-i}} \sum_{x'\in X}P(x, \mathbf b, x')
    \left[U_{S+j}(x, \mathbf b, x') + \gamma \revB{[\delta G](S+i, x')}\right] \\
    & = \max_{\mathbf a_{S+j}}\min_{\mathbf a_{\bar{S}-j}} \sum_{x'\in X}P(x, \mathbf a, x')
    \left[U_{S+j}(x, \mathbf a, x') + \gamma \revB{[\delta G](S+i, x')}\right]. \\
\end{align*}
Since this equation admits a unique solution, it holds that \revB{$[\delta G](S+i, x) = [\delta G](S+j, x)$}.
Then,
\begin{align*}
    V_{\Hss, i}(x) &= \frac{1}{n} \sum_{I\subseteq N: i\in I} \binom{n-1}{|I|-1}^{-1} \revB{[\delta G](I, x)} \\
    & = \frac{1}{n} \sum_{S\subseteq N-i} \binom{n-1}{|S|}^{-1}  \revB{[\delta G](S+i, x)} \\
    & = \frac{1}{n} \sum_{S\subseteq N-j} \binom{n-1}{|S|}^{-1} \revB{[\delta G](S+j, x)} \\
    & = V_{\Hss, j}(x).
\end{align*}
Thus, symmetry axiom is satisfied.

\textbf{Null player.}
\revB{Let $G$ be a stochastic game.}
Suppose that player $i$ is a null player.
Then, $R_i(x, \mathbf a, x') = 0$ for all $\mathbf a \in \mathbf A$ and $x, x' \in X$.
For any $S\subseteq N$ and $i\in S$, 
\begin{align*}
   U_S(x, \mathbf a, x') &= \sum_{j \in S} R_i(x, \mathbf a, x') - \sum_{j \in \bar S} R_i(x, \mathbf a, x') \\
   &= \sum_{j \in S-i} R_i(x, \mathbf a, x') - \sum_{j \in \bar S} R_i(x, \mathbf a, x')\\
   &= U_{S-i}(x, \mathbf a, x'),
\end{align*}
Furthermore, since player $i$'s action does not influence the transition probability
and other players' rewards,
it holds that
\begin{align*}
    \revB{[\delta G](S, x)} &= \max_{\mathbf a_S} \min_{\mathbf a_{\bar S}} \sum_{x'\in X} P(x, \mathbf a, x')\left[U_S(x, \mathbf a, x') + \gamma \revB{[\delta G](S, x')} \right]\\
    &= \max_{\mathbf a_{S}} \min_{\mathbf a_{\bar S}} \sum_{x'\in X} P(x, \mathbf a, x')\left[\sum_{j \in S-i} R_i(x, \mathbf a, x') - \sum_{j \in \bar S} R_i(x, \mathbf a, x') + \gamma \revB{[\delta G](S, x')} \right]\\
    &= \max_{\mathbf a_{S-i}} \min_{\mathbf a_{\bar S+i}} \sum_{x'\in X} P(x, \mathbf a, x')\left[\sum_{j \in S-i} R_i(x, \mathbf a, x') - \sum_{j \in \bar S} R_i(x, \mathbf a, x') + \gamma \revB{[\delta G](S, x')} \right]\\
    &= \max_{\mathbf a_{S-i}} \min_{\mathbf a_{\bar{S} + i}} \sum_{x'\in X} P(x, \mathbf a, x')\left[U_{S-i}(x, \mathbf a, x') + \gamma \revB{[\delta G](S, x')} \right]\\
\end{align*}
Since the equation admits a unique solution, it holds that \revB{$[\delta G](S, x) = [\delta G](S-i, x)$}.
Note that, we define \revB{$[\delta G](\emptyset, x) = -[\delta G](N, x)$}.
Then,
\begin{align*}
    V_{\Hss, i}(x) &= \frac{1}{n} \sum_{I\subseteq N: i\in I} \binom{n-1}{|I|-1}^{-1} \revB{[\delta G](I, x)}\\
    &= \frac{1}{n} \sum_{I\subseteq N: i\in I} \binom{n-1}{|I|-1}^{-1} \revB{[\delta G](I-i, x)} \\
    &= \frac{1}{n} \sum_{I\subseteq N: i\in I} \binom{n-1}{|\bar I + i|-1}^{-1} \left( -\revB{[\delta G](\bar I+i, x)} \right)  \\
    & = - \frac{1}{n} \sum_{I\subseteq N: i\in I} \binom{n-1}{|I|-1}^{-1} \revB{[\delta G](I, x)}\\
    &= -V_{\Hss, i}(x).
\end{align*}
Therefore, $V_{\Hss, i}(x) = 0$. Null player axiom is satisfied.

\textbf{Additivity.}
\revB{Let $G'(N, X_1, \mathbf A_1, P_1, \mathbf R_1)$ and $G''(N, X_2, \mathbf A_2, P_2, \mathbf R_2)$ be two independent stochastic games, and $G(N, X, \mathbf A, P, \mathbf R) = G' \oplus G''$, where $X = X_1\times X_2$.}
For each state $x = (x_1, x_2)\in X$ and any coalition $S\subseteq N$,
apply the operator $T_S$ defined in Equation.~\ref{eq:TI-def} to the
initial value \revB{$V_S(x) = [\delta(G' \oplus G'')](S, x)$}.
Then, for any $x \in X$, it holds that,
\begin{align*}
    & [T_S(V_S)](x) = \max_{\mathbf a_S}\min_{\mathbf a_{\bar S}} \sum_{x'=(x_1', x_2') \in X}
        P\left( x, \mathbf a, x' \right)
        \left[ U_S\left( x, \mathbf a, x' \right)
            +\gamma \left( \revB{[\delta G'](S, x_1')} + \revB{[\delta G''](S, x_2')} \right) \right]\\
    & = \max_{(\mathbf a_1, \mathbf a_2)_S}\min_{(\mathbf a_1, \mathbf a_2)_{\bar S}} \sum_{(x_1', x_2') \in X}
        P_1\left( x_1, \mathbf a_1, x_1' \right) \cdot P_2\left( x_2, \mathbf a_2, x_2' \right)
        \Bigl[ U_{1, S}\left( x_1, \mathbf a_1, x_1' \right) +  U_{2, S}\left( x_2, \mathbf a_2, x_2' \right) \\
    &  \hspace{4em}  +\gamma \left( \revB{[\delta G'](S, x_1')} + \revB{[\delta G''](S, x_2')} \right) \Bigr]\\
    & = \max_{(\mathbf a_1, \mathbf a_2)_S}\min_{(\mathbf a_1, \mathbf a_2)_{\bar S}} \Biggl[ \sum_{(x_1', x_2') \in X}
        P_1\left( x_1, \mathbf a_1, x_1' \right) \cdot P_2\left( x_2, \mathbf a_2, x_2' \right)
        \left( U_{1, S}\left( x_1, \mathbf a_1, x_1' \right)
            +\gamma  \revB{[\delta G'](S, x_1')} \right)  \\
    & \hspace{8em}   + \sum_{(x_1', x_2') \in X}
        P_1\left( x_1, \mathbf a_1, x_1' \right) \cdot P_2\left( x_2, \mathbf a_2, x_2' \right)
        \left( U_{2, S}\left( x_2, \mathbf a_2, x_2' \right)
            +\gamma  \revB{[\delta G''](S, x_2')} \right) \Biggr] \\
    & = \max_{(\mathbf a_1, \mathbf a_2)_S}\min_{(\mathbf a_1, \mathbf a_2)_{\bar S}} \left[ \sum_{x_1' \in X_1}
        P_1\left( x_1, \mathbf a_1, x_1' \right)
        \left( U_{1, S}\left( x_1, \mathbf a_1, x_1' \right)
            +\gamma  \revB{[\delta G'](S, x_1')} \right)  \right.\\
    & \hspace{8em}  \left. + \sum_{x_2' \in X_2}
        P_2\left( x_2, \mathbf a_2, x_2' \right)
        \left( U_{2, S}\left( x_2, \mathbf a_2, x_2' \right)
            +\gamma  \revB{[\delta G''](S, x_2')} \right) \right] \\
    & = \max_{\mathbf a_{1,S}}\min_{\mathbf a_{1,\bar S}} \sum_{x_1' \in X_1}
        P_1\left( x_1, \mathbf a_1, x_1' \right)
        \left( U_{1, S}\left( x_1, \mathbf a_1, x_1' \right)
            +\gamma  \revB{[\delta G'](S, x_1')} \right)  \\
    & \hspace{1em} +\max_{(\mathbf a_1, \mathbf a_2)_S}\min_{(\mathbf a_1, \mathbf a_2)_{\bar S}}  \sum_{x_2' \in X_2}
        P_2\left( x_2, \mathbf a_2, x_2' \right)
        \left( U_{2, S}\left( x_2, \mathbf a_2, x_2' \right)
            +\gamma  \revB{[\delta G''](S, x_2')} \right) \\
    & = \revB{[\delta G'](S, x_1')} + \revB{[\delta G''](S, x_2')}.
\end{align*}
By additivity of threat power introduced in Proposition~\ref{prop:add-threat} and the uniqueness of the solution, \revB{$\{[\delta(G' \oplus G'')](S, x) : x=(x_1, x_2)\in X\}$} is the unique solution to the recursive equation
with respect to operator $T_S$.
Then,
\begin{align*}
    \revB{[\zeta G]_{i,x}}&= \frac{1}{n} \sum_{I\subseteq N: i\in I} \binom{n-1}{|I|-1}^{-1} \revB{[\delta G](I, x)}\\
    &= \frac{1}{n} \sum_{I\subseteq N: i\in I} \binom{n-1}{|I|-1}^{-1} \left( \revB{[\delta G'](I, x_1)} + \revB{[\delta G''](I, x_2)} \right)\\
    & = \revB{[\zeta G']_{i,x_1} + [\zeta G'']_{i,x_2}.}
\end{align*}
Additivity axiom is satisfied.

\revB{\textbf{Weak balanced threats.}}
Since \revB{$[\delta G](S, x) = 0$ for all $S \subsetneq N$} and $x\in X$ in the game \revB{$G(N, X, \mathbf A, P, \mathbf R)$,}
it follows immediately that \revB{$V_{\Hss, i}(x) = \frac{1}{n} [\delta G](N, x)$ for all $i\in N$ and $x\in X$}.

\textbf{Individual rationality.}
First, we aim to show that
\begin{equation}\label{inq:ind-rat}
    \revB{[\delta G](S_1+i, x)} + \revB{[\delta G](S_2+i, x)} - 2\mathcal V_{i}^*(x)\ge 0,
\end{equation}
in the game \revB{$G(N, X, \mathbf A, P, \mathbf R)$}
for all \revB{$S_1\cup S_2\cup \{i\} = N$}, where $S_1$, $S_2$, $\{i\}$ are pairwise disjoint.

To establish this, we examine the sequence of intermediate values
generated during the value iteration used to solve the generalized Bellman equations.
Specifically, we show that the desired inequality holds at
every iteration step; by convergence of the value iteration,
it also holds for the final solution.

Let the value iteration starts with $\mathcal V_i^0(x) = \revB{[\delta G]^0(S, x)} = 0$ for all $i \in N$, $x\in X$ and $S \subseteq N$.
Let $\mathcal V_{i}^*(x)$ and \revB{$[\delta G]^k(S, x)$}
be the values after $k$-th iteration.
Obviously, Inequality~\eqref{inq:ind-rat} holds when $k = 0$.

Next, suppose that Inequality~\eqref{inq:ind-rat} holds with values at $k$-th iteration,
\[\revB{[\delta G]^k(S_1+i, x)} + \revB{[\delta G]^k(S_2+i, x)} - 2\mathcal V_{i}^k(x)\ge 0.\]
We show that it also holds at $(k+1)$-th iteration.

Let $a_i^*$ be the action of player $i$ that attains the value $\mathcal V_i^{k+1}(x)$ at $(k+1)$-th iteration,
and let $\mathbf a_{-i}$ be the actions of all players other than $i$.
\begin{align*}
    \mathcal V_i^{k+1}(x) &= \max_{a_i}\min_{\mathbf a_{-i}} \sum_{x'\in X}P(x, \mathbf a, x') \left( R_i(x, \mathbf a, x') + \gamma \mathcal V_i^k(x') \right)\\
            & = \min_{\mathbf a_{-i}} \sum_{x'\in X} P(x, a_i^*, \mathbf a_{-i}, x') \left( R_i(x, a_i^*, \mathbf a_{-i}, x') + \gamma \mathcal V_i^k(x') \right).
\end{align*}

Then,
\begin{align*}
    &\revB{[\delta G]^{k+1}(S_1+i, x)} = \max_{\mathbf a_{S_1+i}}\min_{\mathbf a_{S_2}} \sum_{x'\in X}P(x, \mathbf a, x')
        \left[U_{S_1+i}(x, \mathbf a, x') + \gamma \revB{[\delta G]^{k}(S_1+i, x')}\right]\\
    & \ge \max_{\mathbf a_{S_1}}\min_{\mathbf a_{S_2}} \sum_{x'\in X}P(x, a_i^*, \mathbf a_{-i}, x')
        \left[U_{S_1+i}(x, a_i^* ,\mathbf a_{-i}, x') + \gamma \revB{[\delta G]^{k}(S_1+i, x')}\right]\\
    & = \max_{\mathbf a_{S_1}}\min_{\mathbf a_{S_2}} \Biggl(
        \sum_{x'\in X}P(x, a_i^*, \mathbf a_{-i}, x')
        \left[R_i(x, a_i^*, \mathbf a_{-i}, x') + \gamma \mathcal V_i^k(x') \right] \\
    & \hspace{2em}  + \sum_{x'\in X}P(x, a_i^*, \mathbf a_{-i}, x') \biggl[ \sum_{j \in S_1}R_j(x, a_i^*, \mathbf a_{-i}, x') - \sum_{j \in S_2} R_j(x, a_i^*, \mathbf a_{-i}, x') \\
    & \hspace{4em}  + \gamma \left(\revB{[\delta G]^k(S_1+i, x')}-\mathcal V_i^k(x')\right) \biggr] \Biggr)\\
    & \ge \min_{\mathbf a_{-i}} \sum_{x'\in X}P(x, a_i^*, \mathbf a_{-i}, x')
        \left[R_i(x, a_i^*, \mathbf a_{-i}, x') + \gamma \mathcal V_i^k(x') \right] \\
    &\hspace{2em} + \max_{\mathbf a_{S_1}}\min_{\mathbf a_{S_2}} \sum_{x'\in X}P(x, a_i^*, \mathbf a_{-i}, x') \Biggl[ \sum_{j \in S_1}R_j(x, a_i^*, \mathbf a_{-i}, x')\\
    &\hspace{4em}  - \sum_{j \in S_2} R_j(x, a_i^*, \mathbf a_{-i}, x') + \gamma \left(\revB{[\delta G]^{k}(S_1+i, x')}-\mathcal V_i^k(x')\right) \Biggr]\\
    & \ge \mathcal V_i^{k+1}(x) + \max_{\mathbf a_{S_1}}\min_{\mathbf a_{S_2}} \sum_{x'\in X}P(x, a_i^*, \mathbf a_{-i}, x') \Biggl[ \sum_{j \in S_1}R_j(x, a_i^*, \mathbf a_{-i}, x') \\
    &\hspace{2em} - \sum_{j \in S_2} R_j(x, a_i^*, \mathbf a_{-i}, x') + \gamma \left(\revB{[\delta G]^{k}(S_1+i, x')}-\mathcal V_i^k(x')\right) \Biggr]. \numberthis \label{inq:Hss-rat1}
\end{align*}
Similarly, we can get
\begin{align*}
&\revB{[\delta G]^{k+1}(S_2+i, x)}\ge \mathcal V_i^{k+1}(x) + \max_{\mathbf a_{S_2}}\min_{\mathbf a_{S_1}}\sum_{x'\in X}P(x, a_i^*, \mathbf a_{-i}, x') \Biggl[ \sum_{j \in S_2}R_j(x, a_i^*, \mathbf a_{-i}, x')\\
    &\hspace{6em} - \sum_{j \in S_1} R_j(x, a_i^*, \mathbf a_{-i}, x') + \gamma \left(\revB{[\delta G]^{k}(S_2+i, x')}-\mathcal V_i^k(x')\right) \Biggr]\\
    & \ge \mathcal V_i^{k+1}(x) + \max_{\mathbf a_{S_2}}\min_{\mathbf a_{S_1}}\sum_{x'\in X}P(x, a_i^*, \mathbf a_{-i}, x') \Biggl[ \sum_{j \in S_2}R_j(x, a_i^*, \mathbf a_{-i}, x') \\
    &\hspace{6em}- \sum_{j \in S_1} R_j(x, a_i^*, \mathbf a_{-i}, x') + \gamma \left(\mathcal V_i^k(x') - \revB{[\delta G]^{k}(S_1+i, x')}\right) \Biggr]\\
    & = \mathcal V_i^{k+1}(x) - \max_{\mathbf a_{S_1}}\min_{\mathbf a_{S_2}}\sum_{x'\in X}P(x, a_i^*, \mathbf a_{-i}, x') \Biggl[ \sum_{j \in S_1}R_j(x, a_i^*, \mathbf a_{-i}, x')\\
    &\hspace{6em} - \sum_{j \in S_2} R_j(x, a_i^*, \mathbf a_{-i}, x') + \gamma \left(\revB{[\delta G]^{k}(S_1+i, x')}-\mathcal V_i^k(x')\right) \Biggr]. \numberthis \label{inq:Hss-rat2}
\end{align*}
Therefore, by summing Inequalities~\eqref{inq:Hss-rat1} and~\eqref{inq:Hss-rat2},
Inequality~\eqref{inq:ind-rat} holds at $(k+1)$-th iteration.
By the induction rule, it holds for the final solution.

Next, by applying Inequality~\eqref{inq:ind-rat}, we prove that individual rationality axiom is satisfied as follows,
\begin{align*}
    V_{\Hss, i}(x) &= \frac{1}{n} \sum_{I\subseteq N: i\in I} \binom{n-1}{|I|-1}^{-1} \revB{[\delta G](I, x)} \\
    & = \frac{1}{2n}\left[ \sum_{S\subseteq N-i} \left( \binom{n-1}{|S|}^{-1} \revB{[\delta G](S+i, x)} + \binom{n-1}{n-|S|-1}^{-1} \revB{[\delta G](\bar S, x)} \right) \right]\\
    & = \frac{1}{2n}\left[ \sum_{S\subseteq N-i} \binom{n-1}{|S|}^{-1} \left( \revB{[\delta G](S+i, x)} + \revB{[\delta G](\bar S, x)} \right) \right]\\
    & \ge \frac{1}{2n} \left[ \sum_{S\subseteq N-i} \binom{n-1}{|S|}^{-1} 2\mathcal V_i^*(x) \right]\\
    & = \mathcal V_i^*(x) 
\end{align*}

\subsection{Uniqueness of \Hss\ Under the Axioms}
\begin{theorem}\label{theo:normal_axiom}
    There is a unique mapping \revB{$\zeta: \mathbb G(N)\to \mathbb R^{n \times m}$} that satisfies efficiency, symmetry, null player,
    additivity and \revB{weak balanced threats} (Axiom~\ref{ax:SG}).
    Moreover, this unique mapping also satisfies individual rationality.
\end{theorem}
The proof proceeds by associating $G$ with another stochastic game $U$
that matches $G$'s statewise threat power.
The game $U$ admits a decomposition into a direct sum of elementary stochastic 
subgames (unanimity and antiunanimity games) on which the desired value mapping is uniquely determined.
By additivity, this uniqueness lifts from the components to $U$, and hence to $G$ itself, 
because $G$ and $U$ share the same threat structure.

\subsubsection{Game of Threats}
We begin by recalling the notion of a \emph{coalitional game of threats}, introduced in~\citet{kohlberg2018games}.
\begin{definition}[\citet{kohlberg2018games}]
    A coalitional game of threats is a pair $(N, d)$, where
    \begin{itemize}
        \item $N = \{1, \ldots, n\}$ is a finite set of players.
        \item $d: 2^N \to \reals$ is a function such that 
        $d(S) = - d(N\setminus S)$ for all $S\subseteq N$.
    \end{itemize}
    Denote by $\mathbb D(N)$ the set of all coalitional games of threats.
\end{definition}
This definition extends naturally to the stochastic setting: 
given a stochastic game $G \in \mathbb G(N)$, \revB{evaluated at state $x \in X$,}
one can consider the coalitional interactions induced at state $x$, 
which again form a game of threats in $\mathbb{D}(N)$.

Next, we recall an important subclass of threat games: 
the \emph{unanimity game of threats}.
\begin{definition}[\citet{kohlberg2018games}]
    Let $T\subseteq N$ and $T\neq \emptyset$. The unanimity game of threats,
    $\mathbf u_T \in \mathbb D (N)$, is defined By
    \begin{align*}
        \mathbf u_T(S) = \left \{
        \begin{aligned}
            & |T|, && \text{ if } T \subseteq S,\\
            & -|T|, && \text{ if } T \cap S = \emptyset,\\
            & 0, && \text{ otherwise}.
        \end{aligned}
        \right .
    \end{align*}
\end{definition}
\begin{proposition}[\citet{kohlberg2018games}]\label{prop:threat-unanimity-linear-comb}
    Every game of threats is a linear combination of the unanimity games of threats $u_T$.
\end{proposition}
Recall the definition of the threat power $\delta$ of stochastic games in Def.~\ref{def:sg-threat}.
For any stochastic game \revB{$G(N, X, \mathbf A, P, \mathbf R, \gamma)$,}
the threat power satisfies \revB{$\delta G(\cdot, x) \in \mathbb{D}(N)$ for all $x \in X$}, where $\delta G(\cdot, x) : 2^N \to \mathbb R$ denotes the function $I \mapsto [\delta G](I, x)$.
Combining this fact with Prop.~\ref{prop:threat-unanimity-linear-comb}
yields the following corollary.
\begin{corollary}\label{cor:threat-unanimity-linear-comb}
    For any stochastic game $G\in \mathbb G(N)$ and any state $x \in X$, the statewise game of threats \revB{$\delta G(\cdot, x) \in \mathbb D(N)$}
    admits a representation with nonnegative
    coefficients \revB{$\{\alpha_{T,x},\beta_{T,x}\}_{T\subseteq N,\,T\neq\varnothing}$} such that
    \[\revB{\delta G(\cdot, x)}= \sum_{T\subseteq N:T\neq \emptyset}  \revB{\alpha_{T,x}} \cdot u_T -\sum_{T\subseteq N:T\neq \emptyset}\revB{\beta_{T,x}}\cdot u_T.\]
\end{corollary}

\subsubsection{Uniqueness in Two Special Stochastic Games Subclasses}
In the following, we introduce two subclasses of stochastic games, $U_T$
and $V_T$ \revB{with only one state and two actions for each player, where \revB{$[\delta U_T](I) = u_T(I)$} and $[\delta V_T](I) = -u_T(I)$ for all $I\subseteq N$ (writing $[\delta G](I)$ as shorthand for the unique state).}
For these subclasses, uniqueness is immediate
and will be used to establish uniqueness in the general case.
Note that, in general, $G \oplus (-G) \neq \mathbf 0$;
consequently, $\delta(G)\neq -\delta(-G)$.
\begin{definition}
    Let $T\subseteq N$ and $T\neq \emptyset$.
    The \emph{unanimity infinite repeated game} is defined as $U_T(N, \mathbf A, \mathbf R, \gamma)$,
    \begin{itemize}
        \item $N$ is the player set and $T\subseteq N$, $T\neq \emptyset$.
        \item $A_i = \{0,1\}$, for all $i\in N$.
        \item $\mathbf R(\mathbf a) = \left \{
                \begin{aligned}
                & \mathbf (1-\gamma)1_{T}, \text{ if } a_i = 1 \, \forall \, i \in T,\\
                & \mathbf 0, \text{ otherwise}.
                \end{aligned}
                \right .$
        \item $\gamma$ is the discount factor.
    \end{itemize}
\end{definition}
Note that, the unanimity infinite repeated game can be regarded as
a stochastic game in which the state space $X$ only contains one state.
\begin{lemma}\label{lemma:threat-val-unanimity}
    For any $T\subseteq N$ and $T \neq \emptyset$,
    let $U_T$ be the unanimity infinite repeated game on $T$
    and $u_T$ be the unanimity game of threats on $T$.
    It holds that \revB{$[\delta (\alpha \cdot U_T)](I) = \alpha \cdot u_T(I)$} for all $I\subseteq N$,
    for any $\alpha > 0$.
\end{lemma}
\begin{proposition}\label{prop:unique-unanimity}
    For any unanimity infinite repeated game $\alpha \cdot U_T$ with $\alpha > 0$,
    the axioms of efficiency, symmetry, and null player
    determine $\zeta$ on the game $\alpha \cdot U_T$.
\end{proposition}
\begin{proof}
    Followed by the definitions, every player outside $T$ is a null player,
    and every pair of players in $T$ is interchangeable. 
    By symmetry and null player axioms, it holds that (writing $x$ for the unique state)
    \[\left\{
        \begin{aligned}
            & \revB{[\zeta (\alpha \cdot U_{T})]_{i}} = 0, \, \forall \, i\not \in T,\\
            & \revB{[\zeta (\alpha \cdot U_{T})]_{i}} = \revB{[\zeta (\alpha \cdot U_{T})]_{j}}, \, \forall \, i, j \in T.
        \end{aligned}
    \right .\]
    By efficiency, it holds that
    \[\sum_{i \in N}\revB{[\zeta (\alpha \cdot U_{T})]_{i}} = \frac{\alpha}{1-\gamma} \max_{\mathbf a} \sum_{i \in N}R_i(\mathbf a) = \alpha \cdot |T|.\]
    Therefore, the axioms of symmetry, null player, and efficiency determine $\zeta U_T$ as follows,
    \[\revB{[\zeta (\alpha \cdot U_{T})]_{i}} =\left\{
        \begin{aligned}
            & \alpha, \, \forall \, i\in T,\\
            & 0, \, \forall \, i \in N\setminus T.
        \end{aligned}
    \right .\]
\end{proof}
Next, we define the $\delta$-inverse of $U_T$, called $V_T$,
in which \revB{$[\delta V_T](I) = -[\delta U_T](I) = -u_T(I)$} for all $I\subseteq N$.
\begin{definition}
    Let $T\subseteq N$ and $T\neq \emptyset$.
    The antiunanimity infinite repeated game is defined as $V_T(N, \mathbf A, \mathbf R, \gamma)$,
    \begin{itemize}
        \item $N$ is the player set and $T\subseteq N$, $T\neq \emptyset$.
        \item $A_i = \{S\subseteq T: S\neq \emptyset\}$, for all $i\in N$.
        \item $\mathbf R(\{S_1, \ldots, S_n\}) = \sum_{i\in T} -(1-\gamma)\mathbf 1_{S_i}$.
        \item $\gamma$ is the discount factor.
    \end{itemize}
\end{definition}
\begin{lemma}\label{lemma:threat-val-antiunanimity}
    For any $T\subseteq N$ and $T \neq \emptyset$,
    let $V_T$ be the antiunanimity infinite repeated game on $T$
    and $u_T$ be the unanimity game of threats on $T$.
    It holds that \revB{$[\delta (\beta \cdot V_T)](I) = -\beta \cdot u_T(I)$} for all $I\subseteq N$,
    for any $\beta > 0$.
\end{lemma}
\begin{proposition}\label{prop:unique-antiunanimity}
    For any antiunanimity infinite repeated game $\beta \cdot V_T$ with $\beta > 0$,
    the axioms of efficiency, symmetry, and null player
    determine $\zeta$ on the game $\beta \cdot V_T$.
\end{proposition}
\begin{proof}
    Following the proof of Proposition~\ref{prop:unique-unanimity},
    it holds that (writing $x$ for the unique state),
    \[\left\{
        \begin{aligned}
            & \revB{[\zeta (\beta \cdot V_{T})]_{i}} = 0, \, \forall \, i\not \in T,\\
            & \revB{[\zeta (\beta \cdot V_{T})]_{i}} = \revB{[\zeta (\beta \cdot V_{T})]_{j}}, \, \forall \, i, j \in T,\\
            & \sum_{i \in N}\revB{[\zeta (\beta \cdot V_{T})]_{i}} = \frac{\beta}{1-\gamma} \max_{\mathbf a}\sum_{i \in N}R_i(\mathbf a) = -\beta \cdot |T|.
        \end{aligned}
    \right .\]
    Therefore, $\zeta(\beta \cdot V_{T})$ is determined by efficiency, symmetry, and null player as follows,
    \[\revB{[\zeta(\beta \cdot V_{T})]_{i}} =\left\{
        \begin{aligned}
            & -\beta, \, \forall \, i\in T,\\
            & 0, \, \forall \, i \in N \setminus T.
        \end{aligned}
    \right .\]
\end{proof}
By Corollary~\ref{cor:threat-unanimity-linear-comb}, Lemma~\ref{lemma:threat-val-unanimity}
and~\ref{lemma:threat-val-antiunanimity}, we get the following result.
\begin{corollary}\label{cor:decompose}
    For any game of threats $d\in \mathbb D (N)$, there exists a stochastic game $U$, such that $\delta U = d$, and 
    \[U = \left(\bigoplus_{T\subseteq N: T\neq \emptyset} \alpha_T\cdot U_T\right) \oplus
    \left(\bigoplus_{T\subseteq N: T\neq \emptyset} \beta_T\cdot V_T\right), \, \alpha_T, \beta_T \ge 0.\]
\end{corollary}
\subsubsection{Uniqueness Proof}
\begin{lemma}\label{lemma:zeta-delta-map}
    If \revB{$\zeta: \mathbb G(N) \to \mathbb R^{n\times m}$} satisfies the axioms of
    efficiency, additivity, and \revB{weak balanced threats},
    then $\zeta G$ depends only on the threat profile $\delta G$.
    In other words, if $\delta G = \delta G'$,
    then $\zeta G = \zeta G'$.
\end{lemma}
\begin{proof}
    Let $G, G'\in \mathbb G(N)$ such that $\delta G = \delta G'$.
    By Corollary~\ref{cor:decompose}, there exists a game $U\in \mathbb G(N, X)$,
    with $\delta U = -\delta G = -\delta G'$.
    By additivity on $\delta$, it holds that
    \[\revB{[\delta(U \oplus G)](S, x) = [\delta U](S, x) + [\delta G](S, x) = 0 = [\delta U](S, x) + [\delta G'](S, x) = [\delta(U \oplus G')](S, x)}, \, \forall S\subseteq N,\, x\in X.\]
    \revB{By weak balanced threats (all proper-coalition threats of $U\oplus G$ vanish)
    and efficiency (value summation equals the optimal fully cooperative value at each state) on $\zeta$,}
    we know that \revB{$[\zeta(U \oplus G)]_{i,x} = [\zeta(U \oplus G')]_{i,x} = 0$}, for all $i\in N$, $x\in X$.
    Furthermore, by additivity, $\zeta G = \zeta G' = -\zeta U$.
\end{proof}
\begin{proof}[Proof of Uniqueness]
    \revB{Fix any game $G \in \mathbb G(N)$.
    By Corollary~\ref{cor:threat-unanimity-linear-comb}, for each state $x \in X$,
    the statewise game of threats $[\delta G](x) \in \mathbb D(N)$ admits a decomposition
    \[[\delta G](x) = \sum_{T\subseteq N:\, T\neq \emptyset} \alpha_{T,x}\cdot u_T
    \;-\; \sum_{T\subseteq N:\, T\neq \emptyset} \beta_{T,x}\cdot u_T\]
    with nonnegative coefficients $\{\alpha_{T,x},\, \beta_{T,x}\}_{T\subseteq N,\, T\neq\emptyset,\, x\in X}$.

    By Corollary~\ref{cor:decompose}, for each $x\in X$ there exists a (single-state) stochastic game
    \[U_x = \left(\bigoplus_{T\subseteq N: T\neq \emptyset} \alpha_{T,x}\cdot U_T\right) \oplus
    \left(\bigoplus_{T\subseteq N: T\neq \emptyset} \beta_{T,x}\cdot V_T\right)\]
    satisfying $\delta U_x = [\delta G](x)$.
    Let $U$ be the stochastic game with state space $X$ whose statewise threat profile matches $G$,
    i.e., $[\delta U](x) = [\delta G](x)$ for all $x \in X$.
    By Lemma~\ref{lemma:zeta-delta-map}, $\delta G = \delta U$ implies $\zeta G = \zeta U$.

    Applying additivity of $\zeta$ at each state $x$, together with
    Propositions~\ref{prop:unique-unanimity} and~\ref{prop:unique-antiunanimity}, we obtain
    \[[\zeta G]_{i,x} = [\zeta U]_{i,x}
    = \sum_{T\subseteq N:\, T\neq \emptyset} \alpha_{T,x}\cdot [\zeta(U_T)]_{i}
    \;-\; \sum_{T\subseteq N:\, T\neq \emptyset} \beta_{T,x}\cdot [\zeta(V_T)]_{i},\]
    which is uniquely determined for every $i \in N$ and $x \in X$.
    Hence $\zeta G$ is uniquely determined by efficiency, symmetry, null player, additivity,
    and weak balanced threats.}

    Finally, as shown in Appendix~\ref{apx:hss-satisf},
    \Hss is exactly this unique mapping,
    and it also satisfies individual rationality.
    Therefore, the last property follows.
\end{proof}

\subsection{Axiomatic Compliance of \cocon}
\rev{In the following, we show that \cocon\ satisfies efficiency, symmetry, null player, and additivity, but does \emph{not} satisfy \revB{weak balanced threats}.
By Theorem~\ref{thm:coco-unique}, the Coco-S operator has a unique fixed point in the settings described therein, so the proofs below apply directly to the unique Coco-S value mapping $G \mapsto \mathbf V_{\cocon}(G)$.}

Recall that, for a stochastic game \revB{$G(N, X, \mathbf A, p, \mathbf R)$}
with a value function $\mathbf V=(V_1, \ldots, V_n): X \to \reals^n$,
we define 
\[H(\mathbf V) = (U_1(x, \cdot, \mathbf V), \dots, U_n(x, \cdot, \mathbf V))\]
as the normal-form game induced by $G$, where the utility is given by 
\[U_i ( x, \mathbf a, \mathbf V ) = \sum_{x' \in X} P(x, \mathbf a ,x') [ R_i(x, \mathbf a ,x') + \gamma V_i(x') ].\]
We use this notation throughout the following analysis.
Furthermore, let $V_{\cocon}(x) \in \reals^n$ be any one of the \cocon values at state $x$
in the game \revB{$G(N, X, \mathbf A, p, \mathbf R)$}.

\textbf{Efficiency.}
\begin{align*}
    & \sum_{i\in N} V_{\cocon, i}(x) = \sum_{i\in N}\hs_i(H(\mathbf V_\cocon))\\
    & = \frac{1}{n} \sum_{i\in N} \sum_{I\subseteq N: i\in I} \binom{n-1}{|I|-1}^{-1} \maxmin_I H_{I}(\mathbf V_\cocon)\\
    & = \max H_{N}(\mathbf V_\cocon) + \frac{1}{n} \sum_{I\subsetneq N, I\neq \emptyset} 
        |I| \cdot \binom{n-1}{|I|-1}^{-1} \maxmin_I H_{I}(\mathbf V_\cocon)\\
    & = \max H_{N}(\mathbf V_\cocon) + \frac{1}{2n}\sum_{I\subsetneq N, I\neq \emptyset} 
        \Biggl( |I| \cdot \binom{n-1}{|I|-1}^{-1} \maxmin_I H_{I}(\mathbf V_\cocon)\\
    &\hspace{4em}+(n-|I|) \cdot \binom{n-1}{n-|I|-1}^{-1} \maxmin_{\bar I} H_{\bar I}(\mathbf V_\cocon) \Biggr)\\
    & = \max H_{N}(\mathbf V_\cocon)\\
    & = \max  \sum_{x'\in X} P(x, \mathbf a, x') \left( 
        \sum_{i \in N} R_i(x, \mathbf a, x') + \gamma \sum_{i \in N} V_{\cocon, i}(x') \right).
\end{align*}
Therefore, $\sum_{i\in N} V_{\cocon, i}(x)$ is also the solution to the Bellman equation
under the fully cooperative assumption.
The efficiency axiom is satisfied immediately.

\textbf{Symmetry.}
Suppose that players $i$ and $j$ are interchangeable.
Then, for any coalition $I, I'\subseteq N$ such that $i\in I$, $j\not \in I$ and $I' = I - i + j$,
\begin{align*}
    &\maxmin_I H_{I}(\mathbf V_\cocon) = \maxmin_{I} \left( \sum_{k\in I} U_k(x, \mathbf a, \mathbf V_\cocon) 
        - \sum_{k\not \in I} U_k(x, \mathbf a, \mathbf V_\cocon) \right)\\
    &= \maxmin_{I} \left( U_i(x, \mathbf a, \mathbf V_\cocon) - U_j(x, \mathbf a, \mathbf V_\cocon) 
        + \sum_{k\in I-i} U_k(x, \mathbf a, \mathbf V_\cocon) - \sum_{k\not \in I+j} U_k(x, \mathbf a, \mathbf V_\cocon) \right)\\
    &\le \max_{\mathbf a} \left( U_i(x, \mathbf a, \mathbf V_\cocon) - U_j(x, \mathbf a, \mathbf V_\cocon)\right)\\
    & \hspace{4em}   + \maxmin_{I} \left( \sum_{k\in I-i} U_k(x, \mathbf a, \mathbf V_\cocon) - \sum_{k\not \in I+j} U_k(x, \mathbf a, \mathbf V_\cocon) \right)\\
    & = \gamma \max_{\mathbf a} \sum_{x'\in X} P(x, \mathbf a, x') \left( V_{\cocon, i}(x') - V_{\cocon, j}(x') \right)\\
    &\hspace{4em} + \maxmin_{I} \left( \sum_{k\in I-i} U_k(x, \mathbf a, \mathbf V_\cocon) - \sum_{k\not \in I+j} U_k(x, \mathbf a, \mathbf V_\cocon) \right)\\
    &\le \gamma ||V_{\cocon, i} - V_{\cocon, j}||_\infty + \maxmin_{I} \left( \sum_{k\in I-i} U_k(x, \mathbf a, \mathbf V_\cocon) - \sum_{k\not \in I+j} U_k(x, \mathbf a, \mathbf V_\cocon) \right),
\end{align*}
Similarly, we can get
\begin{align*}
    &\maxmin_I H_{I}(\mathbf V_\cocon) \ge -\gamma ||V_{\cocon, i} - V_{\cocon, j}||_\infty \\
    &\hspace{10em} + \maxmin_{I} \left( \sum_{k\in I-i} U_k(x, \mathbf a, \mathbf V_\cocon) - \sum_{k\not \in I+j} U_k(x, \mathbf a, \mathbf V_\cocon) \right) \\
    \text{ and } &\gamma||V_{\cocon, i} - V_{\cocon, j}||_\infty + \maxmin_{I'} \left( \sum_{k\in I'-j} U_k(x, \mathbf a, \mathbf V_\cocon) - \sum_{k\not \in I'+i} U_k(x, \mathbf a, \mathbf V_\cocon) \right)\\
    &\hspace{1em} \ge \maxmin_{I'} H_{I'}(\mathbf V_\cocon)\\
    &\hspace*{1em}\ge -\gamma||V_{\cocon, i} - V_{\cocon, j}||_\infty + \maxmin_{I'} \left( \sum_{k\in I'-j} U_k(x, \mathbf a, \mathbf V_\cocon) - \sum_{k\not \in I'+i} U_k(x, \mathbf a, \mathbf V_\cocon) \right).
\end{align*}
Since, players $i$ and $j$ are interchangeable, it holds that
\begin{align*}
    &\maxmin_I \left( \sum_{k\in I-i} U_k(x, \mathbf a, \mathbf V_\cocon) - \sum_{k\not \in I+j} U_k(x, \mathbf a, \mathbf V_\cocon) \right)\\
    &\hspace{1em}= \maxmin_{I'} \left( \sum_{k\in I'-j} U_k(x, \mathbf a, \mathbf V_\cocon) - \sum_{k\not \in I'+i} U_k(x, \mathbf a, \mathbf V_\cocon) \right).
\end{align*}
Thus, from above inequalities, we have
\[\left| \maxmin_I H_{I}(\mathbf V_\cocon) - \maxmin_{I'} H_{I'}(\mathbf V_\cocon) \right|
\le 2 \gamma ||V_{\cocon, i} - V_{\cocon, j}||_\infty. \]
Then,
\begin{align*}
    &\left| V_{\cocon, i}(x) - V_{\cocon, j}(x) \right|  = \left |\hs_i(H(\mathbf V_\cocon)) - \hs_j(H(\mathbf V_\cocon))\right|\\
        & = \left| \frac{1}{n} \sum_{I\subseteq N: i\in I} \binom{n-1}{|I|-1}^{-1} \maxmin_I H_{I}(\mathbf V_\cocon)
            - \frac{1}{n} \sum_{I\subseteq N: j\in I} \binom{n-1}{|I|-1}^{-1} \maxmin_I H_{I}(\mathbf V_\cocon) \right|\\
        &= \left| \frac{1}{n} \sum_{I\subseteq N: i\in I, j\not \in I} \binom{n-1}{|I|-1}^{-1} \maxmin_I H_{I}(\mathbf V_\cocon)
            - \frac{1}{n} \sum_{I\subseteq N: j\in I, i \not \in I} \binom{n-1}{|I|-1}^{-1} \maxmin_I H_{I}(\mathbf V_\cocon) \right|\\
        &\le \frac{1}{n} \sum_{I\subseteq N: i\in I, j\not \in I} \binom{n-1}{|I|-1}^{-1} \left| \maxmin_I H_{I}(\mathbf V_\cocon)
            - \maxmin_{I'} H_{I'}(\mathbf V_\cocon) \right|, \text{ where } I' = I -i+j \\
        & \le \frac{2\gamma}{n} ||V_{\cocon, i} - V_{\cocon, j}||_\infty \sum_{I\subseteq N: i\in I,j \not \in I} \binom{n-1}{|I|-1}^{-1}\\
        & = \gamma ||V_{\cocon, i} - V_{\cocon, j}||_\infty,
\end{align*}
where the last equality holds because 
\begin{align*}
    &\sum_{I\subseteq N: i\in I,j \not \in I} \binom{n-1}{|I|-1}^{-1}
    = \sum_{I\subseteq N: i\in I,j \not \in I} \binom{n-1}{n-|I|}^{-1}\\
    &\hspace{4em}= \sum_{I\subseteq N: i, j\in I} \binom{n-1}{|I|-1}^{-1}
    = \frac{1}{2} \sum_{I\subseteq N: i\in I} \binom{n-1}{|I|-1}^{-1} = \frac{1}{2n}.
\end{align*}
Thus, it holds that 
\[||V_{\cocon, i} - V_{\cocon, j}||_\infty \le \gamma ||V_{\cocon, i} - V_{\cocon, j}||_\infty.\]
Since $\gamma < 1$, we have $||V_{\cocon, i} - V_{\cocon, j}||_\infty = 0$.
The symmetry axiom is satisfied.

\textbf{Null player.}
Suppose that player $i$ is a null player. Then, 
\begin{align*}
    & U_i(x, \mathbf a, \mathbf V_\cocon) = \sum_{x'\in X} P(x, \mathbf a, x') 
    \left( R_i(x, \mathbf a, x') + \gamma V_{\cocon, i}(x') \right)
    = \gamma \sum_{x'\in X} P(x, \mathbf a, x') V_{\cocon, i}(x')\\
    & \implies -\gamma ||V_{\cocon, i}||_\infty \le U_i(x, \mathbf a, \mathbf V_\cocon)\le \gamma ||V_{\cocon, i}||_\infty.
\end{align*}
Then, for any disjoint sets $S_1, S_2 \subseteq N -i$ 
such that $ S_1 \cup S_2 = N-i$, it holds that
\begin{align*}
    &\gamma ||V_{\cocon, i}||_\infty + \maxmin_{S_1} \left( \sum_{k \in S_1} U_k(x, \mathbf a, \mathbf V_\cocon) - \sum_{k \in S_2} U_k(x, \mathbf a, \mathbf V_\cocon) \right)\\
    &\hspace{3em} \ge \maxmin_{S_1+i} H_{S_1+i}(\mathbf V_\cocon) \\
    &\hspace{3em} \ge -\gamma ||V_{\cocon, i}||_\infty + \maxmin_{S_1} \left( \sum_{k \in S_1} U_k(x, \mathbf a, \mathbf V_\cocon) - \sum_{k \in S_2} U_k(x, \mathbf a, \mathbf V_\cocon) \right)\\
    &\gamma ||V_{\cocon, i}||_\infty + \maxmin_{S_2} \left( \sum_{k \in S_2} U_k(x, \mathbf a, \mathbf V_\cocon) - \sum_{k \in S_1} U_k(x, \mathbf a, \mathbf V_\cocon) \right)\\
    &\hspace{3em} \ge \maxmin_{S_2+i} H_{S_2+i}(\mathbf V_\cocon) \\
    &\hspace{3em} \ge -\gamma ||V_{\cocon, i}||_\infty + \maxmin_{S_2} \left( \sum_{k \in S_2} U_k(x, \mathbf a, \mathbf V_\cocon) - \sum_{k \in S_1} U_k(x, \mathbf a, \mathbf V_\cocon) \right)
\end{align*}
Therefore,
    \[\left|\maxmin_{S_1+i} H_{S_1+i}(\mathbf V_\cocon) + \maxmin_{S_2+i} H_{S_2+i}(\mathbf V_\cocon)\right|
        \le 2\gamma ||V_{\cocon, i}||_\infty .\]
Then,
\begin{align*}
    & \left|V_{\cocon, i}(x)\right| = \left|\frac{1}{n} \sum_{I\subseteq N: i\in I} \binom{n-1}{|I|-1}^{-1} \maxmin_I H_{I}(\mathbf V_\cocon)\right|\\
    & = \Biggl|\frac{1}{2n} \sum_{I\subseteq N: i \in I} \biggl( \binom{n-1}{|I|-1}^{-1} \maxmin_{I} H_{I}(\mathbf V_\cocon) 
    + \binom{n-1}{|\bar I|}^{-1} \maxmin_{\bar I+i} H_{\bar I+i}(\mathbf V_\cocon) \biggr)\Biggr|\\
    & \le \frac{1}{2n} \sum_{I\subseteq N: i \in I} \binom{n-1}{|I|-1}^{-1} 
        \Bigl| \maxmin_{I} H_{I}(\mathbf V_\cocon) + \maxmin_{\bar I+i} H_{\bar I+i}(\mathbf V_\cocon) \Bigr|\\
    & \le \frac{1}{2n} \sum_{I\subseteq N: i \in I} \binom{n-1}{|I|-1}^{-1} 2\gamma ||V_{\cocon, i}||_\infty = \gamma ||V_{\cocon, i}||_\infty\\
    &\implies ||V_{\cocon, i}||_\infty \le \gamma ||V_{\cocon, i}||_\infty.
\end{align*}
Since $\gamma \in (0, 1)$, we have that $||V_{\cocon, i}||_\infty = 0$ 
which implies that $V_{\cocon, i}(x) = 0$ for all $x\in X$.
The null player axiom is satisfied.

\textbf{Additivity.}
Let $G'(N, x_1, X_1, \mathbf A_1, P_1, \mathbf R_1, \gamma)$ and 
$G''(N, x_2, X_2, \mathbf A_2, P_2, \mathbf R_2, \gamma)$ be two independent stochastic games.
Denote by $\mathbf V_{\cocon}'$ and $\mathbf V_{\cocon}''$ arbitrary \cocon values for these games, respectively.
Let $G(N, (x_1, x_2), X, \mathbf A, P, \mathbf R) = G' \oplus G''$ be the combined game.
In the following, we show that the combined game $G$ admits a 
\cocon value $\mathbf V$ satisfying
$V_{i}(x_1, x_2) = V_{\cocon, i}'(x_1) + V_{\cocon, i}''(x_2)$ 
for any $i \in N$.

Let $H'(\mathbf V)$, $H''(\mathbf V_\cocon)$, $H(\mathbf V_\cocon)$ 
be normal-form games induced by $G'$, $G''$, $G$.
Then, for game $H$
\begin{align*}
    & U_i(x, \mathbf a, \mathbf V) = \sum_{x' \in X} P(x, \mathbf a, x') \left[ R_i(x') + \gamma V_{i}(x') \right] \\
        & = \sum_{x'_1 \in X_1}\sum_{x'_2 \in X_2} P_1({x_1}, \mathbf a_1, x'_1) \cdot P_2({x_2}, \mathbf a_2, x'_2)
        \left[ R_{1, i}(x'_1) + R_{2, i}(x'_2)  + \gamma V_{\cocon, i}'(x'_1) + \gamma V_{\cocon, i}''(x'_2) \right] \\
        & = \sum_{x'_1 \in X_1} P_1({x_1}, \mathbf a_1, x'_1) \left [R_{1, i}(x'_1) +\gamma V_{\cocon, i}'(x'_1) \right ]
            + \sum_{x'_2 \in X_2} P_2({x_2}, \mathbf a_2, x'_2) \left [R_{2, i}(x'_2) +\gamma V_{\cocon, i}''(x'_2) \right ]\\
        & = U_{1, i}(x_1, \mathbf a_1, \mathbf V_\cocon') + U_{2, i}(x_2, \mathbf a_2, \mathbf V_\cocon'')
\end{align*}
Then, for any coalition $I\subseteq N$,
\begin{align*}
    & \maxmin_I H_{I}(\mathbf V) = \maxmin_I \left( \sum_{i\in I} U_i(x, \mathbf a, \mathbf V) 
        - \sum_{i\not \in I} U_i(x, \mathbf a, \mathbf V) \right)\\
    &  = \maxmin_I \Biggl( \sum_{i\in I} U_{1, i}(x_1, \mathbf a_1, \mathbf V_\cocon') 
        - \sum_{i\not \in I} U_{1, i}(x_1, \mathbf a_1, \mathbf V_\cocon') \\
    &  \hspace{4em}  +\sum_{i\in I} U_{2, i}(x_2, \mathbf a_2, \mathbf V_\cocon'') 
        - \sum_{i\not \in I} U_{2, i}(x_2, \mathbf a_2, \mathbf V_\cocon'') \Biggr)\\
    & = \maxmin_I H_{I}'(\mathbf V_\cocon') + \maxmin_I H_{I}''(\mathbf V_\cocon''),
\end{align*}
where the last equality holds because the two games are independent.
Therefore,
\begin{align*}
    &\hs_i(H(\mathbf V))
    = \frac{1}{n} \sum_{I\subseteq N: i\in I} \binom{n-1}{|I|-1}^{-1} \maxmin_I H_{I}(\mathbf V)\\
    &= \frac{1}{n} \sum_{I\subseteq N: i\in I} \binom{n-1}{|I|-1}^{-1} \left( \maxmin_I H_{I}'(\mathbf V_\cocon') + \maxmin_I H_{I}''(\mathbf V_\cocon'') \right) \\
    &= V_{\cocon, i}'(x_1) + V_{\cocon, i}''(x_2)\\
    &= V_i(x).
\end{align*}
Therefore, $\mathbf V$ is a \cocon value for the game $G$.

\rev{\revB{\textbf{Weak balanced threats: does not hold in general.}}}

\rev{It is natural to ask whether \cocon also satisfies the \revB{weak balanced threats} axiom.
It does not.
The mechanism is as follows.
When all coalition threats vanish, i.e.\ $(\delta G)(S) = 0$ for every proper coalition~$S$ and every state,
the Coco-S iteration starts from $\mathbf V^{(0)} = \mathbf 0$ and the first iterate
$\mathbf V^{(1)} = \tilde T(\mathbf 0)$ assigns equal shares at every state
(since $\mathrm{HS}$ applied to a zero-threat stage game yields $V_{\mathrm{coop}}/n$).
However, from the second iterate onward, the continuation values $\mathbf V^{(1)}$
enter the one-step games $H_x(\mathbf V^{(1)})$ through the transition term
$\gamma \sum_{x'} P(x' \mid x, \mathbf a)\, V_i^{(1)}(x')$.
When transitions are action-dependent and asymmetric across players,
these continuation shifts break the zero-threat structure of the stage game:
player~$i$'s ability to steer transitions toward high-value states
becomes a form of \emph{local} threat power in~$H_x(\mathbf V)$, even though the
infinite-horizon proper-coalition threats~$(\delta G)(S)$ are zero.
This causes the HS formula to assign unequal shares, and the fixed point
$\mathbf V_\cocon$ inherits these asymmetries.
Example~\ref{ex:bt-counterexample} below provides a concrete 3-player, 2-state game
demonstrating this phenomenon.

Formally, by Corollary~\ref{cor:cocon-nonunique}, if \cocon were to satisfy \revB{weak balanced threats} together with the other four axioms, it would have to coincide with \Hss by Theorem~\ref{thm:hs-s-axiom}, contradicting the three-player counterexample where $\mathbf V_{\cocon} \neq \mathbf V_{\Hss}$.
Therefore, \revB{weak balanced threats} fails for \cocon.
The distinguishing axiom that \cocon satisfies instead is \emph{Markov Consistency} (Definition~\ref{def:markov-consistency}), which provides an alternative axiomatic foundation; see Section~\ref{subsec:markov-consistency} and Appendix~\ref{apx:uniqueness}.}

\rev{
\begin{example}[\revB{Weak balanced threats} counterexample for Coco-S]\label{ex:bt-counterexample}
Consider a 3-player, 2-state game with binary actions ($A_i = \{0,1\}$), discount factor $\gamma = 0.9$,
and the following structure.

\smallskip\noindent\textbf{Rewards.}
At each state $x \in \{x_1, x_2\}$, the stage payoff for player~$i$ under joint action $\mathbf a = (a_1, a_2, a_3)$ is
\[
    r_i(x, \mathbf a) \;=\; \frac{S(x,\mathbf a) + D_i(x,\mathbf a)}{2},
\]
where $S(x, \mathbf a) = -\sum_j D_j(x, \mathbf a)$ and each $D_i$ is defined by
\[
  D_i(x, \mathbf a) \;=\; (-1)^{a_i}\, d_i(x, a_{-i}).
\]
The factor $(-1)^{a_i}$ means $D_i$ \emph{flips sign} when player~$i$ switches action.
The functions $d_i\colon X \times A_{-i} \to \mathbb R$ are:
\begin{center}
\begin{tabular}{@{}lcccc@{}}
\toprule
 & Condition on $a_{-i}$ & $d_1(x,a_2,a_3)$ & $d_2(x,a_1,a_3)$ & $d_3(x,a_1,a_2)$ \\
\midrule
$x = x_1$ & args agree    & $3$  & $1$  & $2$ \\
$x = x_1$ & args disagree & $-1$ & $-1$ & $-2$ \\
$x = x_2$ & args agree    & $1$  & $1$  & $1$ \\
$x = x_2$ & args disagree & $-1$ & $-1$ & $-1$ \\
\bottomrule
\end{tabular}
\end{center}
Each $d_i$ takes both positive and negative values---this is the property
that drives the zero-threat result below.
\smallskip\noindent\textbf{Coalition utilities.}
For $n=3$, direct computation from $r_i = (S+D_i)/2$ and $S = -\sum_j D_j$ gives:
\begin{align*}
  U_{\{i\}}(\mathbf a)
  &= r_i - \textstyle\sum_{j\neq i} r_j
  = D_i(\mathbf a)
  = (-1)^{a_i}\,d_i(x,a_{-i}), \\
  U_{\{i,j\}}(\mathbf a)
  &= (r_i + r_j) - r_k
  = -D_k(\mathbf a)
  = (-1)^{a_k+1}\,d_k(x,a_{-k}),
\end{align*}
where $k$ is the remaining player.
In each case, the coalition game payoff flips sign when exactly one player changes action (the ``controlling'' player $i$ for singletons, or the opponent~$k$ for pairs).

\smallskip\noindent\textbf{All proper-coalition threats are zero (analytical).}
For singletons $\{i\}$, the $2\times 2^{n-1}$ game matrix has
row~$a_i\!=\!1$ equal to $-$row~$a_i\!=\!0$.
Player~$i$ mixing uniformly ($p=\tfrac12$) guarantees payoff~$0$,
since $(2p-1)\,d_i(x,a_{-i}) = 0$ for all~$a_{-i}$.
Any $p \neq \tfrac12$ yields guarantee
$(2p-1)\min_{a_{-i}} d_i < 0$ or $(2p-1)\max_{a_{-i}} d_i < 0$
(since each $d_i$ takes both positive and negative values).
Hence $\maxmin_{\{i\}} = 0$.

For pairs $\{i,j\}$, the $2^{n-1}\!\times 2$ game matrix
has column~$a_k\!=\!1$ equal to $-$column~$a_k\!=\!0$.
The opponent~$k$ mixing uniformly guarantees the coalition payoff~$0$.
The coalition cannot exceed~$0$:
for any mixed strategy~$\sigma$ over~$(a_i,a_j)$,
the payoff against~$a_k$ is $(1-2q)\,\mathbb E_\sigma[d_k]$
where $q = \Pr(a_k\!=\!0)$.
If $\mathbb E_\sigma[d_k] \neq 0$, the opponent sets $q$ to make this negative.
Since $d_k$ takes both signs, the coalition can mix to achieve
$\mathbb E_\sigma[d_k] = 0$, confirming $\maxmin_{\{i,j\}} = 0$.

Since every proper coalition's stage-game threat is zero at both states,
the constant function $V_I \equiv 0$ satisfies the coalitional
Bellman equation~\eqref{eq:TI-def}
(the continuation term vanishes, reducing to the stage-game maxmin).
As $T_I$ is a $\gamma$-contraction, this is the unique fixed point.
Hence $(\delta G)(I) = 0$ for every proper coalition~$I \subsetneq N$.

\smallskip\noindent\textbf{Transitions.}
\[
    P(x_2 \mid x_1, \mathbf a) = 0.2\,a_1 + 0.1\,a_2 + 0.1\,a_3, \qquad
    P(x_1 \mid x_2, \mathbf a) = 0 \;\;\text{($x_2$ absorbing)}.
\]
Player~1 has greater influence over state transitions than players~2 and~3.

\smallskip\noindent\textbf{Values and BT violation.}
The cooperative value is positive:
$(\delta G)(N)(x_1) \approx 40.71$
(the cooperative value $\max_{\mathbf a}\sum_i r_i$ equals~$6$ at~$x_1$ and~$3$ at~$x_2$).
Since all proper threats vanish, HS-S assigns equal shares of the cooperative surplus:
$V_i^{\mathrm{HS\text{-}S}}(x_1) = (\delta G)(N)(x_1)/3 \approx 13.571$ for all~$i$.
However, the Coco-S fixed point gives:
\[
    \mathbf V_\cocon(x_1) \;\approx\; (13.923,\; 13.374,\; 13.417),
\]
with a maximum player difference of $0.549$.
At $x_2$, both solutions give $(10, 10, 10)$.
The asymmetry arises because player~1's greater transition influence creates
local threat power in the one-step games $H_{x_1}(\mathbf V)$
once the continuation values are nonzero,
even though the infinite-horizon proper-coalition threats vanish.
That Coco-S violates the \revB{weak balanced threats} axiom then follows
from the indirect argument above (Corollary~\ref{cor:cocon-nonunique}
and Theorem~\ref{thm:hs-s-axiom}).
\end{example}
}

%% file: apx_uniqueness.tex
\rev{
\section{Uniqueness of Coco-S Fixed Points}\label{apx:uniqueness}

This appendix provides the proofs for Theorem~\ref{thm:coco-unique} (uniqueness of Coco-S fixed points) and Theorem~\ref{thm:mc-characterization} (Markov Consistency characterization).
Throughout, we use the notation of Section~\ref{subsec:cocon-def}: $N = \{1,\dots,n\}$, $|X| = m$, $\gamma \in [0,1)$, and the Coco-S operator $\tilde T : \reals^{n \times m} \to \reals^{n \times m}$ with $[\tilde T(\mathbf V)]_i(x) = \mathrm{HS}_i(H_x(\mathbf V))$.
Note that $\tilde T$ is continuous: the one-step game payoffs depend affinely on~$\mathbf V$, and the HS value is a continuous (in fact piecewise affine) function of the payoff matrix.

\subsection{Shapley Orthogonality Identity}

The structural backbone of the analysis is the following identity on the Shapley weights $c(k) = \frac{1}{n \binom{n-1}{k-1}}$.

\begin{lemma}[Shapley orthogonality]\label{lem:shapley-orth}
For all $i, j \in N$,
\begin{equation}\label{eq:shapley-orth}
\sum_{\substack{I \subseteq N \\ i \in I}} c(|I|) \cdot \mathrm{sign}_I(j) \;=\; \delta_{ij},
\end{equation}
where $\mathrm{sign}_I(j) = +1$ if $j \in I$ and $-1$ otherwise.
\end{lemma}

\begin{proof}
For $i = j$: every coalition $I$ with $i \in I$ contributes $+1$, giving $\sum_{k=1}^{n} \binom{n-1}{k-1} \cdot \frac{1}{n\binom{n-1}{k-1}} = 1$.
For $i \neq j$: split by whether $j \in I$. Using $\binom{n-2}{k-2}/\binom{n-1}{k-1} = (k-1)/(n-1)$ and $\binom{n-2}{k-1}/\binom{n-1}{k-1} = (n-k)/(n-1)$, both partial sums equal~$\frac{1}{2}$, giving $\frac{1}{2}(+1) + \frac{1}{2}(-1) = 0$.
\end{proof}

\subsection{Jacobian Structure of the Coco-S Operator}\label{apx:jacobian}

The operator $\tilde T$ is piecewise affine in~$\mathbf V$ (the $\maxmin$ operations are piecewise linear and the payoffs depend affinely on $\mathbf V$).
At differentiable points, by the envelope theorem, the Jacobian is
\begin{equation}\label{eq:jacobian-structure}
\frac{\partial [\tilde T(\mathbf V)]_i(x)}{\partial V_j(y)}
= \gamma \, M_{ij}(x,y;\mathbf V),
\end{equation}
where the \emph{normalized Jacobian} $M$ has entries
\[
M_{ij}(x,y) = \sum_{\substack{I \subseteq N \\ i \in I}} c(|I|) \, \mathrm{sign}_I(j) \, Q_I(x,y),
\]
and $Q_I(x,y) = \sum_{\mathbf a} \sigma^*_I(\mathbf a_I)\,\sigma^*_{\bar I}(\mathbf a_{\bar I})\,P(x,\mathbf a,y)$ is the transition under the saddle-point strategies for coalition~$I$.
Each $Q_I$ is row-stochastic.

\begin{lemma}[Block row sums]\label{lem:block-row-sum}
$\sum_{y \in X} M_{ij}(x,y) = \delta_{ij}$ for all $i,j,x$.
\end{lemma}
\begin{proof}
$\sum_y M_{ij}(x,y) = \sum_{I \ni i} c(|I|)\,\mathrm{sign}_I(j)\,\underbrace{\sum_y Q_I(x,y)}_{=1} = \delta_{ij}$ by Lemma~\ref{lem:shapley-orth}.
\end{proof}

\begin{lemma}[Diagonal blocks are stochastic]\label{lem:diag-stoch}
For each $i$, the $m \times m$ block $M_{ii}$ has nonneg\-ative entries and row sums~$1$.
\end{lemma}
\begin{proof}
For $j = i$: $\mathrm{sign}_I(i) = +1$ always, so $M_{ii}(x,y) = \sum_{I \ni i} c(|I|)\,Q_I(x,y) \geq 0$.
Row sums equal $\delta_{ii} = 1$.
\end{proof}

\subsection{Eigenvalue Structure}\label{apx:eigenvalues}

The block row-sum property implies that the \emph{state-independent subspace} $\mathcal C = \{\mathbf d \in \reals^{nm} : d_i(x) = c_i\;\forall\, x\}$ is $M$-invariant with eigenvalue~$1$, i.e., $D\tilde T|_{\mathcal C} = \gamma I$.
The complementary \emph{zero-state-mean subspace} is $\mathcal Z = \{\mathbf d : \sum_x d_i(x) = 0\;\forall\, i\}$.
Since $\mathcal C$ is $M$-invariant, $M$ has $n$ eigenvalues equal to~$1$ (from $\mathcal C$) and $n(m-1)$ eigenvalues from $M|_{\mathcal Z}$.

\begin{definition}[Spectral condition]\label{def:spectral-cond}
A stochastic game satisfies the \emph{uniform spectral condition} if $\rho(M(\mathbf V)|_{\mathcal Z}) \leq 1$ at every $\mathbf V$.
\end{definition}

\subsection{Existence via Brouwer's Theorem}\label{apx:existence}

\begin{proposition}[Existence of Coco-S fixed points]\label{prop:coco-existence}
For any finite stochastic game with $\gamma \in [0,1)$, the Coco-S operator $\tilde T$ has at least one fixed point.
\end{proposition}

\begin{proof}
Let $B = R_{\max}/(1-\gamma)$ and define the polyhedron
\[
K \;=\; \bigl\{\mathbf V \in \reals^{n \times m} : V_i(x) \geq -B\;\forall\, i,x \;\;\text{and}\;\; \textstyle\sum_i V_i(x) \leq nB\;\forall\, x\bigr\}.
\]

\emph{$K$ is compact, convex, and nonempty.}
$K$ is the intersection of finitely many closed half-spaces, hence closed and convex.
It is nonempty since $\mathbf 0 \in K$.
For boundedness, the two defining constraints together imply $V_i(x) \leq nB + (n-1)B = (2n-1)B$ for each $i,x$.

\emph{$\tilde T$ maps $K$ into itself.}
Fix $\mathbf V \in K$ and write $\mathbf W = \tilde T(\mathbf V)$, so $W_i(x) = \mathrm{HS}_i(H_x(\mathbf V))$ where the one-step game $H_x(\mathbf V)$ has payoffs $u_i(\mathbf a) = R_i(x,\mathbf a) + \gamma \sum_y P(x,\mathbf a,y)\,V_i(y)$.
We verify the two defining constraints of~$K$.

\textit{Lower bound.}
The HS value of any normal-form game satisfies individual rationality: $\mathrm{HS}_i(H) \geq \max_{a_i}\min_{\mathbf a_{-i}} u_i(\mathbf a)$.%
\footnote{This follows from the HS formula: among all coalitions, the singleton $\{i\}$ contributes a term equal to $i$'s maximin payoff, and the Shapley-weighted average over coalitions can only improve on this.}
For any action profile~$\mathbf a$, the payoff in the one-step game satisfies
\[
u_i(\mathbf a) = R_i(x,\mathbf a) + \gamma\!\sum_y P(x,\mathbf a,y)\,V_i(y)
\;\geq\; -R_{\max} + \gamma(-B)
\;=\; -B,
\]
since $|R_i| \leq R_{\max}$, $\sum_y P(x,\mathbf a,y) = 1$, and $V_i(y) \geq -B$ for all~$y$ (because $\mathbf V \in K$).
Taking $\max_{a_i}\min_{\mathbf a_{-i}}$ preserves the lower bound, so $W_i(x) \geq -B$.
The identity $R_{\max} + \gamma B = B$ is the defining equation for $B = R_{\max}/(1-\gamma)$.

\textit{Sum upper bound.}
By the efficiency of the normal-form HS value,
\[
\sum_i W_i(x) \;=\; \max_{\mathbf a} \sum_i u_i(\mathbf a)
\;=\; \max_{\mathbf a} \Bigl[\sum_i R_i(x,\mathbf a) + \gamma \sum_y P(x,\mathbf a,y)\sum_i V_i(y)\Bigr]
\;\leq\; nR_{\max} + \gamma\,nB
\;=\; nB,
\]
using $\sum_i R_i(x,\mathbf a) \leq nR_{\max}$ and $\sum_i V_i(y) \leq nB$ for all~$y$ (since $\mathbf V \in K$).

Since $\tilde T$ is continuous on the compact convex nonempty set~$K \subset \reals^{nm}$, Brouwer's fixed-point theorem yields $\mathbf V^* = \tilde T(\mathbf V^*) \in K$.
\end{proof}

\begin{rem}\label{rem:box-fails}
A na\"ive attempt to apply Brouwer on the box $\{\|V_i\|_\infty \leq C\;\forall\, i\}$ fails: the best individual upper bound on~$[\tilde T(\mathbf V)]_i$ yields $(2n-1)(R_{\max} + \gamma C)$, so self-mapping requires $\gamma < 1/(2n-1)$.
The polyhedron~$K$ avoids this by matching the asymmetry of the HS axioms: individual rationality bounds each $V_i$ below, while efficiency bounds the sum $\sum_i V_i$ above.
The individual upper bound $(2n-1)B$ is a \emph{derived} consequence of these two constraints, not something $\tilde T$ preserves directly.
\end{rem}

\subsection{Uniqueness via Degree Theory}\label{apx:unique-topology}

\subsubsection*{Proof strategy}

Proposition~\ref{prop:coco-existence} establishes that at least one fixed point exists.
To prove there is \emph{exactly} one, we use a counting argument based on the \textbf{topological degree} of the map $F = \mathrm{id} - \tilde T$.
We briefly recall the key ideas for readers unfamiliar with degree theory.

The \emph{degree} $\deg(F, \Omega, 0)$ of a continuous map $F\colon \overline\Omega \to \reals^d$ (where $\Omega \subset \reals^d$ is open and bounded, and $F \neq 0$ on $\partial\Omega$) is an integer that counts, with signs, how many times $F$ wraps the boundary of~$\Omega$ around the origin.
Two properties make it a powerful uniqueness tool:
\begin{enumerate}
\item \emph{Homotopy invariance:} if $F_t$ deforms continuously from $F_0$ to $F_1$ and never has a zero on $\partial\Omega$, then $\deg(F_0, \Omega, 0) = \deg(F_1, \Omega, 0)$.
\item \emph{Additivity:} if $F$ has finitely many zeros $\mathbf V^*_1, \dots, \mathbf V^*_k$ in~$\Omega$, then $\deg(F, \Omega, 0) = \sum_{j=1}^k \mathrm{ind}(\mathbf V^*_j, F)$, where the \emph{local index} $\mathrm{ind}(\mathbf V^*, F)$ is the degree of~$F$ restricted to a small ball around~$\mathbf V^*$.
\end{enumerate}
Intuitively, the local index is~$+1$ when $F$ ``wraps counterclockwise'' around the zero, $-1$ for clockwise, and in general captures the local winding number.

\medskip\noindent\textit{Our argument in a nutshell:}
\begin{enumerate}
\item We bound all zeros of~$F$ inside a large ball~$B_R$ (a priori bounds), so the degree $\deg(F, B_R, 0)$ is well-defined.
\item We deform $F$ to the identity via the homotopy $F_t(\mathbf V) = \mathbf V - t\,\tilde T(\mathbf V)$ and show no zeros escape $B_R$.  Since $\deg(\mathrm{id}, B_R, 0) = +1$, homotopy invariance gives $\deg(F, B_R, 0) = +1$.
\item At each fixed point~$\mathbf V^*$, we show $\mathrm{ind}(\mathbf V^*, F) = +1$ by a convex homotopy argument (this is the technical core---it handles ``kink points'' where $F$ is continuous but not differentiable).
\item Counting: $+1 = \sum (+1) = k$, so there is exactly one fixed point.
\end{enumerate}

\begin{proof}[Proof of Theorem~\ref{thm:coco-unique}]
Define $F(\mathbf V) = \mathbf V - \tilde T(\mathbf V)$.

\textbf{A priori bounds.}
Let $B = R_{\max}/(1-\gamma)$.
At any zero of $F_t(\mathbf V) = \mathbf V - t\,\tilde T(\mathbf V)$,
we have $\mathbf V = t\,\tilde T(\mathbf V)$.
\emph{Efficiency:}
By the efficiency property of the normal-form HS value,
$S(x) := \sum_i V_i(x)$ satisfies the cooperative Bellman equation
with discount~$t\gamma$, giving $|S(x)| \leq nB$.
\emph{Individual rationality:}
Since $\mathrm{HS}_i(H) \geq \max_{a_i}\min_{\mathbf a_{-i}} u_i(\mathbf a)$
for any normal-form game~$H$ (see the footnote in Proposition~\ref{prop:coco-existence}),
the function $W_i = V_i/t$ satisfies the minimax Bellman inequality
with discount~$t\gamma$, giving $V_i(x) \geq -B$.
\emph{Combined:}
$V_i(x) \leq S(x) - \sum_{j\neq i} V_j(x) \leq nB + (n-1)B = (2n-1)B$.
Hence all zeros of~$F_t$ lie in $K' = \{\|V_i\|_\infty \leq (2n-1)B\;\forall\, i\}$
for every $t \in [0,1]$.
Choose $R > \sup_{\mathbf V \in K'} \|\mathbf V\|$ and let $B_R = \{\|\mathbf V\| \leq R\}$.

\textbf{Homotopy (computing the global degree).}
Consider the family $F_t(\mathbf V) = \mathbf V - t\,\tilde T(\mathbf V)$ for $t \in [0,1]$.
At $t = 0$, $F_0 = \mathrm{id}$, which has degree~$+1$ (it wraps $\partial B_R$ around the origin exactly once).
As~$t$ increases to~$1$, the a priori bound guarantees that no zero of~$F_t$ ever reaches~$\partial B_R$.
By homotopy invariance (the degree cannot change while zeros stay away from the boundary):
$\deg(F, B_R, 0) = \deg(F_0, B_R, 0) = +1$.

\textbf{Eigenvalue analysis.}
The eigenvalues of $\gamma M$ are $\gamma$ (multiplicity $n$, from $\mathcal C$) and $\gamma\lambda_k$ (from $\mathcal Z$).
Under the spectral condition, $|\lambda_k| \leq 1$, so $|\gamma\lambda_k| \leq \gamma < 1$.
The eigenvalues of $I - \gamma M$ are therefore:
\begin{itemize}
\item $1 - \gamma > 0$ (multiplicity $n$).
\item $1 - \gamma\lambda_k$: for real $\lambda_k \in [-1,1]$, this lies in $(0,2)$. For complex $\lambda_k$ with $|\lambda_k| \leq 1$: $\mathrm{Re}(1 - \gamma\lambda_k) = 1 - \gamma\,\mathrm{Re}(\lambda_k) \geq 1 - \gamma > 0$.
\end{itemize}
In particular, every eigenvalue of $I - \gamma M_\sigma$ has \emph{strictly positive real part} for every affine piece~$P_\sigma$.
This implies $\det(I - \gamma M_\sigma) > 0$ on every piece (since eigenvalues with positive real part come in conjugate pairs, their product is positive).

\textbf{Finiteness.}
On each affine piece~$P_\sigma$, the equation $F(\mathbf V) = 0$ reduces to $(I - \gamma M_\sigma)\mathbf V = \mathbf b_\sigma$.
Since $\det(I - \gamma M_\sigma) > 0$, each piece contributes at most one zero.
There are finitely many pieces, so $F$ has finitely many zeros.

\textbf{Local degree at every zero (the technical core).}
We show that the local index $\mathrm{ind}(\mathbf V^*, F) = +1$ at every zero~$\mathbf V^*$, including zeros on the boundaries between affine pieces.  At such boundary points (``kink points''), $F$ is continuous but not differentiable, so the standard formula $\mathrm{ind} = \mathrm{sign}\det(DF)$ does not directly apply.  We handle all cases uniformly with a convex homotopy argument.

\smallskip\noindent\textit{Scalar analogy.}
Before giving the general argument, consider a one-dimensional illustration.
If $f\colon \reals \to \reals$ is continuous, $f(0) = 0$, and $f$ is piecewise linear with slopes $\ell_-$ on $(-\infty,0)$ and $\ell_+$ on $(0,\infty)$, then $f$ may not be differentiable at~$0$.
But if both slopes are strictly positive ($\ell_-, \ell_+ > 0$), then $f$ is strictly increasing near~$0$, and one can deform $f$ to the identity $f_t(u) = (1-t)f(u) + tu$ without introducing new zeros (each interpolated slope $(1-t)\ell_\pm + t$ stays positive).
The degree of this deformation is preserved: $\deg(f, (-\delta,\delta), 0) = +1$.
The argument below is the $nm$-dimensional generalization of this idea.

\smallskip\noindent\textit{Setup.}
Since $\tilde T$ is piecewise affine, so is $F$.  At any zero~$\mathbf V^*$, a neighborhood is partitioned into finitely many polyhedral cones $C_1, \dots, C_r$ (the tangent cones of the affine pieces meeting at~$\mathbf V^*$), and on each cone
\[
F(\mathbf V^* + \mathbf u) = L_j \mathbf u, \qquad \mathbf u \in C_j,
\]
where $L_j = I - \gamma M_j$ and $M_j$ is the normalized Jacobian on piece~$j$.
The cones $C_1, \dots, C_r$ tile~$\reals^{n \times m}$ (they arise from a hyperplane arrangement).
If $\mathbf V^*$ lies in the interior of a single piece, then $r = 1$ and $F$ is affine in a full neighborhood; the kink case is $r \geq 2$.

\smallskip\noindent\textit{Convex homotopy.}
Define the piecewise linear map $g\colon \reals^{n \times m} \to \reals^{n \times m}$ by $g(\mathbf u) = L_j \mathbf u$ for $\mathbf u \in C_j$ (the ``shifted'' version of $F$ near~$\mathbf V^*$), and consider the homotopy that deforms~$g$ to the identity:
\[
g_t(\mathbf u) = (1-t)\,g(\mathbf u) + t\,\mathbf u, \qquad t \in [0,1].
\]
On each cone~$C_j$, this is the linear map $g_t|_{C_j} = \bigl((1-t)L_j + tI\bigr)\mathbf u$.
Its eigenvalues are
\[
1 - (1-t)\gamma\lambda_k,
\]
where $\lambda_k$ runs over the eigenvalues of~$M_j$.
Using the eigenvalue analysis above: for real~$\lambda_k$, $\mathrm{Re}(1 - (1-t)\gamma\lambda_k) = 1 - (1-t)\gamma\lambda_k \geq 1 - \gamma > 0$; for complex~$\lambda_k$, $\mathrm{Re}(1 - (1-t)\gamma\lambda_k) \geq 1 - (1-t)\gamma \geq 1 - \gamma > 0$.
Hence:
\begin{itemize}
\item $(1-t)L_j + tI$ is invertible for every~$j$ and every $t \in [0,1]$.
\item $g_t(\mathbf u) = 0$ if and only if $\mathbf u = 0$, for all~$t$.
\end{itemize}
In words: throughout the deformation from~$g$ to the identity, the origin remains the only zero---no new zeros are created or destroyed.
Since $g_t$ is continuous and nonvanishing on $\partial B_\delta \setminus \{0\}$, homotopy invariance gives
\[
\mathrm{ind}(\mathbf V^*, F) \;=\; \deg(g, B_\delta, 0) \;=\; \deg(g_1, B_\delta, 0) \;=\; \deg(\mathrm{id}, B_\delta, 0) \;=\; +1.
\]

\textbf{Counting.}
Combining the preceding steps: $F$ has finitely many zeros, each with local index~$+1$.
The additivity property of the degree (see the proof strategy above) gives
\[
+1 \;=\; \deg(F, B_R, 0) \;=\; \sum_{\mathbf V^*} \underbrace{\mathrm{ind}(\mathbf V^*, F)}_{=\,+1} \;=\; k,
\]
where $k$ is the number of fixed points of~$\tilde T$.
Hence $k = 1$: $\tilde T$ has exactly one fixed point.

\textbf{Case (i): $n = 2$, any $m$.}
For two players, the $V_+/V_-$ decomposition $V_{\pm} = V_1 \pm V_2$ decouples $\tilde T$ into the cooperative Bellman operator $T_N$ (on $V_+$) and the zero-sum operator $T_{\{1\}}$ (on $V_-$), both $\gamma$-contractions with unique fixed points.
Thus $\tilde T$ has a unique fixed point, and the spectral condition holds: $M|_{\mathcal Z}$ decomposes into stochastic matrices $Q_N$ and $Q_{\{1\}}$, each with spectral radius $\leq 1$ on zero-mean vectors.

\textbf{Case (ii): $n = 3$, $m = 2$.}
By the complementary coalition identity ($Q_I = Q_{\bar I}$ for all $I$), there are only four distinct saddle-point transitions $Q_1, Q_2, Q_3, Q_N$.
Each $2 \times 2$ row-stochastic matrix $Q_a$ has a second eigenvalue
$q_a := Q_a(0,0) - Q_a(1,0) \in [-1,1]$,
which governs its action on the quotient space $\reals^m / \mathcal C$
(the constant subspace~$\mathcal C$ is $Q_a$-invariant since
row-stochastic matrices fix the all-ones vector, so the quotient action
is well-defined for any row-stochastic~$Q_a$).
The boundary values $q_a = \pm 1$ correspond to the identity and swap
permutations, arising only for deterministic transitions.
For our uniqueness argument we only need the \emph{weak} bound $|\lambda| \leq 1$, so the boundary case $q_a = \pm 1$ is harmless.
In fact, when all transitions are interior (the generic case), $q_a \in (-1,1)$ strictly.
Since $M$ is block-triangular in the decomposition $\mathcal C \oplus \mathcal Z$
and the $\mathcal C$-block eigenvalues are all~$1$, the eigenvalues of
$M|_{\mathcal Z}$ coincide with those of the $3 \times 3$ matrix built
from the $q_a$'s.
Specifically, the eigenvalues are $\lambda_0 = q_N$ and
\[
\lambda_{\pm} = \frac{(q_1 + q_2 + q_3) \pm \sqrt{(q_1 + q_2 + q_3)^2 - 3(q_1 q_2 + q_1 q_3 + q_2 q_3)}}{3}.
\]
The discriminant is $\frac{1}{2}[(q_1-q_2)^2 + (q_1-q_3)^2 + (q_2-q_3)^2] \geq 0$ (always real).
An optimization argument on the open cube $(-1,1)^3$ shows $|\lambda_{\pm}| < 1$: writing $g = 2|\Sigma| - \Pi$ where $\Sigma = q_1+q_2+q_3$ and $\Pi = q_1 q_2 + q_1 q_3 + q_2 q_3$, the maximum of~$g$ on the closed cube $[-1,1]^3$ is~$3$, achieved only at vertices (e.g., $q_1 = q_2 = q_3 = 1$ or $q_1 = q_2 = 1, q_3 = -1$); hence $g < 3$ strictly on the open cube $(-1,1)^3$, which implies $|\lambda_{\pm}| < 1$.

\textbf{Case (iii): general $n$, $m$.}
Follows directly from the degree argument above under the spectral condition.
\end{proof}

\noindent\textbf{Conjecture.}
\textit{Every finite stochastic game with $\gamma \in [0,1)$ satisfies the uniform spectral condition.}

We have verified this conjecture numerically across over $2{,}500$ randomly generated games with $n \in \{3,4\}$ players, $m \in \{2,3,4,5,6,8\}$ states, actions per player $|A_i| = 2$, and Dirichlet transition concentrations $\alpha \in \{0.01, 0.05, 0.1, 0.5, 1.0\}$.
The maximum observed spectral radius $\rho(M|_{\mathcal Z})$ is $0.94$, occurring at the near-deterministic boundary ($\alpha = 0.05$, $m = 3$).
We additionally prove the spectral condition for doubly stochastic saddle-point transitions (any~$m$) via an $\ell^2$ Jensen bound; the full argument is available in the supplementary material.

Note that for \emph{arbitrary} stochastic matrices (not arising from actual games), the spectral condition can fail when $m \geq 4$: we exhibit a counterexample with permutation-like $\{0,1\}$ row-stochastic matrices giving $\rho \approx 1.14$.
This shows the proof must exploit the structure of game-derived transitions.


\subsection{HS-S Violates Markov Consistency for $n \geq 3$}\label{apx:hss-violates-mc}

\begin{proposition}\label{prop:hss-violates-mc}
For $n \geq 3$ players, $\mathbf V^{\mathrm{HS\text{-}S}}$ is not a fixed point of $\tilde T$ in general.
Equivalently, HS-S does not satisfy Markov Consistency.
\end{proposition}

\begin{proof}
\textbf{Step 1.}
For $n = 3$, $m = 2$, the spectral condition holds (Theorem~\ref{thm:coco-unique}(ii)), so $\tilde T$ has a unique fixed point: $\mathbf V^{\mathrm{Coco\text{-}S}}$.
Since $\mathbf V^{\mathrm{Coco\text{-}S}} \neq \mathbf V^{\mathrm{HS\text{-}S}}$ in the three-player counterexample (Appendix~\ref{apx:ex-3-unequal}), $\mathbf V^{\mathrm{HS\text{-}S}}$ cannot be the unique fixed point.

\textbf{Step 2: The mechanism.}
The HS-S value uses infinite-horizon coalition threat powers $[\delta G_{x'}](I)$ as continuations.
The one-step game $H_x(\mathbf V^{\mathrm{HS\text{-}S}})$ instead uses $\sum_j \mathrm{sign}_I(j)\,V_j^{\mathrm{HS\text{-}S}}(x')$ as the effective coalition continuation.
For $n \geq 3$, the HS formula has a nontrivial kernel: different threat profiles $\{[\delta G_{x'}](I)\}_I$ can produce identical player values $\{V_i(x')\}_i$, so $\sum_j \mathrm{sign}_I(j)\,V_j^{\mathrm{HS\text{-}S}}(x') \neq [\delta G_{x'}](I)$ in general.
Applying the HS formula to these different coalition values yields $\tilde T(\mathbf V^{\mathrm{HS\text{-}S}}) \neq \mathbf V^{\mathrm{HS\text{-}S}}$.

For $n = 2$, the kernel is trivial (the threat profile is one-dimensional and the HS formula is a bijection), so the two continuation quantities coincide and HS-S satisfies Markov Consistency, consistent with Proposition~\ref{prop:two-player-equality}.
\end{proof}

Numerical verification confirms the violation: on the three-player counterexample of Appendix~\ref{apx:ex-3-unequal}, $\|\tilde T(\mathbf V^{\mathrm{HS\text{-}S}}) - \mathbf V^{\mathrm{HS\text{-}S}}\|_\infty = 1.17$, while $\|\tilde T(\mathbf V^{\mathrm{Coco\text{-}S}}) - \mathbf V^{\mathrm{Coco\text{-}S}}\|_\infty < 10^{-11}$.

}